\documentclass[twocolumn, times, trackchanges]{aastex63_bren}

\usepackage[ruled,vlined]{algorithm2e}
\usepackage{graphicx}
\usepackage{bbm}
\usepackage{float}
\usepackage[hypertexnames=false]{hyperref}
\usepackage{subfiles} % Best loaded last in the preamble
\hypersetup{linkcolor=red,citecolor=green,filecolor=cyan,urlcolor=magenta}
\defcitealias{baron19}{B19}

\newcommand{\halpha}{H$\alpha$}
\newcommand{\hbeta}{H$\beta$}
\newcommand{\oiiifull}{$\text{[O\,{\sc iii}]}\lambda \, 5007\mathrm{\AA}$}
\newcommand{\oiii}{$\text{[O\,{\sc iii}]}$}
\newcommand{\oifull}{$\text{[O\,{\sc i}]}\lambda \, 6300\mathrm{\AA}$}
\newcommand{\oi}{$\text{[O\,{\sc i}]}$}
\newcommand{\niifull}{$\text{[N\,{\sc ii}]}\lambda \, 6584\mathrm{\AA}$}
\newcommand{\nii}{$\text{[N\,{\sc ii}]}$}
\newcommand{\siifull}{$\text{[S\,{\sc ii}]}\lambda\lambda \, 6717\mathrm{\AA}+6731\mathrm{\AA}$}
\newcommand{\sii}{$\text{[S\,{\sc ii}]}$}

\newcommand{\oiiihbeta}{$\log (\text{[O\,{\sc iii}]}/\text{H}\beta )$}
\newcommand{\niihalpha}{$\log (\text{[N\,{\sc ii}]}/\text{H}\alpha )$}
\newcommand{\siihalpha}{$\log (\text{[S\,{\sc ii}]}/\text{H}\alpha )$}
\newcommand{\oihalpha}{$\log (\text{[O\,{\sc i}]}/\text{H}\alpha )$}

\newcommand{\cofull}{$\mathrm{^{12}CO(2-1)}\,$}
\newcommand{\mic}{$\mathrm{\mu m}$}

\shorttitle{PHANGS-ML}
\begin{document}

\title{PHANGS-ML: dissecting multiphase gas and dust in nearby galaxies using machine learning}

\author[0000-0003-4974-3481]{Dalya Baron}
\email{dalyabaron@gmail.com}
\altaffiliation{Carnegie-Princeton Fellow}
\affiliation{The Observatories of the Carnegie Institution for Science. 813 Santa Barbara Street, Pasadena, CA 91101, USA}

\author[0000-0002-4378-8534]{Karin M. Sandstrom}
\affiliation{Department of Astronomy \& Astrophysics, University of California, San Diego. 9500 Gilman Drive, La Jolla, CA 92093, USA}

\author[0000-0002-5204-2259]{Erik Rosolowsky} 
\affiliation{Department of Physics, University of Alberta, Edmonton, Alberta, T6G 2E1, Canada}

\author[0000-0002-4755-118X]{Oleg V. Egorov}
\affiliation{Astronomisches Rechen-Institut, Zentrum f\"{u}r Astronomie der Universit\"{a}t Heidelberg, M\"{o}nchhofstra{\ss}e 12-14, 69120 Heidelberg, Germany}

\author[0000-0002-0560-3172]{Ralf S. Klessen} 
\affiliation{Universit\"{a}t Heidelberg, Zentrum f\"{u}r Astronomie, Institut f\"{u}r Theoretische Astrophysik, Albert-Ueberle-Stra{\ss}e 2,69120 Heidelberg, Germany}
\affiliation{Universit\"{a}t Heidelberg, Interdisziplin\"{a}res Zentrum f\"{u}r Wissenschaftliches Rechnen, Im Neuenheimer Feld 205, 69120 Heidelberg, Germany}

\author[0000-0002-2545-1700]{Adam K. Leroy}
\affiliation{Department of Astronomy, Ohio State University, 180 W. 18th Ave, Columbus, Ohio 43210}
\affiliation{Center for Cosmology and Astroparticle Physics, 191 West Woodruff Avenue, Columbus, OH 43210, USA}

\author[0000-0003-0946-6176]{M\'ed\'eric Boquien} 
\affiliation{Universit\'e C\^ote d'Azur, Observatoire de la C\^ote d'Azur, CNRS, Laboratoire Lagrange, 06000, Nice, France}

\author[0000-0002-3933-7677]{Eva Schinnerer}
\affiliation{Max Planck Institute for Astronomy, K\"{o}nigstuhl 17, D-69117, Germany}

\author[0000-0002-2545-5752]{Francesco Belfiore}
\affiliation{INAF -- Osservatorio Astrofisico di Arcetri, Largo E. Fermi 5, I-50157, Firenze, Italy}

\author[0000-0002-9768-0246]{Brent~Groves}
\affiliation{International Centre for Radio Astronomy Research, University of Western Australia, 35 Stirling Highway, Crawley, WA 6009, Australia}

\author[0000-0002-5235-5589]{J\'er\'emy Chastenet}
\affil{Sterrenkundig Observatorium, Ghent University, Krijgslaan 281-S9, 9000 Gent, Belgium}

\author[0000-0002-5782-9093]{Daniel~A.~Dale}
\affiliation{Department of Physics and Astronomy, University of Wyoming, Laramie, WY 82071, USA}

\author[0000-0003-4218-3944]{Guillermo A. Blanc}
\affiliation{Observatories of the Carnegie Institution for Science, 813 Santa Barbara Street, Pasadena, CA 91101, USA}
\affiliation{Departamento de Astronom\'{i}a, Universidad de Chile, Camino del Observatorio 1515, Las Condes, Santiago, Chile}

\author[0000-0002-6972-6411]{Jos\'e E. M\'endez-Delgado}
\affiliation{Astronomisches Rechen-Institut, Zentrum f\"{u}r Astronomie der Universit\"{a}t Heidelberg, M\"{o}nchhofstra{\ss}e 12-14, 69120 Heidelberg, Germany}

\author[0000-0001-9605-780X]{Eric W. Koch}
\affiliation{Center for Astrophysics $\mid$ Harvard \& Smithsonian, 60 Garden St., 02138 Cambridge, MA, USA}

\author[0000-0002-3247-5321]{Kathryn~Grasha}
\altaffiliation{ARC DECRA Fellow}
\affiliation{Research School of Astronomy and Astrophysics, Australian National University, Canberra, ACT 2611, Australia}   
\affiliation{ARC Centre of Excellence for All Sky Astrophysics in 3 Dimensions (ASTRO 3D), Australia}   

\author[0000-0002-5635-5180]{M\'elanie Chevance}
\affiliation{Zentrum f\"{u}r Astronomie der Universit\"{a}t Heidelberg, Institut f\"{u}r Theoretische Astrophysik, Albert-Ueberle-Str. 2, 69120 Heidelberg}
\affiliation{Cosmic Origins Of Life (COOL) Research DAO, coolresearch.io}

\author[0000-0002-8528-7340]{David A. Thilker}
\affiliation{Center for Astrophysical Sciences, Johns Hopkins University, 3400 N. Charles Street, Baltimore, MD 21218, USA}

\author[0000-0001-6498-2945]{Dario Colombo}
\affiliation{Argelander-Institut f\"{u}r Astronomie, Universit\"{a}t Bonn, Auf dem H\"{u}gel 71, 53121 Bonn, Germany}\affiliation{Max-Planck-Institut f\"{u}r Radioastronomie, Auf dem H\"{u}gel 69, 53121 Bonn, Germany}

\author[0000-0002-0012-2142]{Thomas~G.~Williams}
\affiliation{Sub-department of Astrophysics, Department of Physics, University of Oxford, Keble Road, Oxford OX1 3RH, UK}

\author[0000-0003-2721-487X]{Debosmita Pathak}
\affiliation{Department of Astronomy, Ohio State University, 180 W. 18th Ave, Columbus, Ohio 43210}

\author[0000-0002-9183-8102]{Jessica Sutter}
\affiliation{Department of Astronomy \& Astrophysics, University of California, San Diego, 9500 Gilman Drive, La Jolla, CA 92093}

\author[0000-0003-1845-0934]{Toby Brown}
\affiliation{Herzberg Astronomy and Astrophysics Research Centre, National Research Council of Canada, 5071 West Saanich Rd, Victoria, BC, V9E 2E7, Canada}

\author[0000-0002-5077-881X]{John F. Wu}
\affiliation{Space Telescope Science Institute, 3700 San Martin Dr, Baltimore, MD 21218}
\affiliation{Department of Physics \& Astronomy, Johns Hopkins University, 3400 N Charles St, Baltimore, MD 21218}

\author[0000-0003-4797-7030]{Josh E. G. Peek}
\affiliation{Space Telescope Science Institute, 3700 San Martin Drive, Baltimore, MD, 21218}
\affiliation{Department of Physics \& Astronomy, Johns Hopkins University, 3400 N. Charles Street, Baltimore, MD 21218}

\author[0000-0002-6155-7166]{Eric Emsellem}
\affiliation{European Southern Observatory, Karl-Schwarzschild-Straße 2, 85748, Garching, Germany}
\affiliation{University Lyon, Univ Lyon1, ENS de Lyon, CNRS, Centre de Recherche Astrophysique de Lyon UMR5574, 69230, Saint-Genis-Laval, France}

\author[0000-0003-3917-6460]{Kirsten~L.~Larson}
\affiliation{AURA for the European Space Agency (ESA), Space Telescope Science Institute, 3700 San Martin Drive, Baltimore, MD 21218, USA}

\author[0000-0002-3289-8914]{Justus Neumann}
\affiliation{Max-Planck-Institut f\"{u}r Astronomie, K\"{o}nigstuhl 17, D-69117 Heidelberg, Germany}

\suppressAffiliations



\begin{abstract}

The PHANGS survey uses ALMA, HST, VLT, and JWST to obtain an unprecedented high-resolution view of nearby galaxies, covering millions of spatially independent regions. 
The high dimensionality of such a diverse multi-wavelength dataset makes it challenging to identify new trends, particularly when they connect observables from different wavelengths.
Here we use unsupervised machine learning algorithms to mine this information-rich dataset to identify novel patterns. We focus on three of the PHANGS-JWST galaxies, for which we extract properties pertaining to their stellar populations; warm ionized and cold molecular gas; and Polycyclic Aromatic Hydrocarbons (PAHs), as measured over 150 pc-scale regions. 
We show that we can divide the regions into groups with distinct multiphase gas and PAH properties. In the process, we identify previously-unknown galaxy-wide correlations between PAH band and optical line ratios and use our identified groups to interpret them. 
The correlations we measure can be naturally explained in a scenario where the PAHs and the ionized gas are exposed to different parts of the same radiation field that varies spatially across the galaxies. This scenario has several implications for nearby galaxies: (i) The uniform PAH ionized fraction on 150 pc scales suggests significant self-regulation in the ISM, (ii) the PAH 11.3/7.7 \mic~ band ratio may be used to constrain the shape of the non-ionizing far-ultraviolet to optical part of the radiation field, and (iii) the varying radiation field affects line ratios that are commonly used as PAH size diagnostics. Neglecting this effect leads to incorrect or biased PAH sizes.

\end{abstract}


\section{Introduction}\label{sec:intro}

Over the past several decades, astronomy has been going through a data revolution. It was pioneered by the Sloan Digital Sky Survey, which imaged roughly one-third of the sky in five photometric bands, providing measured photometry for billions of objects and spectroscopy for millions (\citealt{york00, eisenstein11, dawson13}). Since then, past and ongoing surveys have been producing massive datasets that include millions to billions of objects with astrometric, photometric, spectroscopic, or time-series observations (e.g., Pan-STARRS; ZTF; Gaia; APOGEE; MANGA; DESI; \citealt{kaiser10, bellm14, bundy15, desi_colab_16, gaia16, majewski17, abdurrouf22, desi_edr_23, gaia23}). These observations, along with numerous derived products from them, were made publicly accessible through efficient and convenient interfaces and changed the way astronomers interact with observations, marking the beginning of the big data era in astronomy. In the near future, surveys by the Vera Rubin Observatory, Roman Space Telescope, Euclid, SDSS-V, and the Square Kilometre Array (e.g., \citealt{dewdney09, kollmeier17, ivezic19, euclid22}), to name a few, are expected to make another order-of-magnitude increase in data volume.

\begin{figure*}
	\centering
\includegraphics[width=1\textwidth]{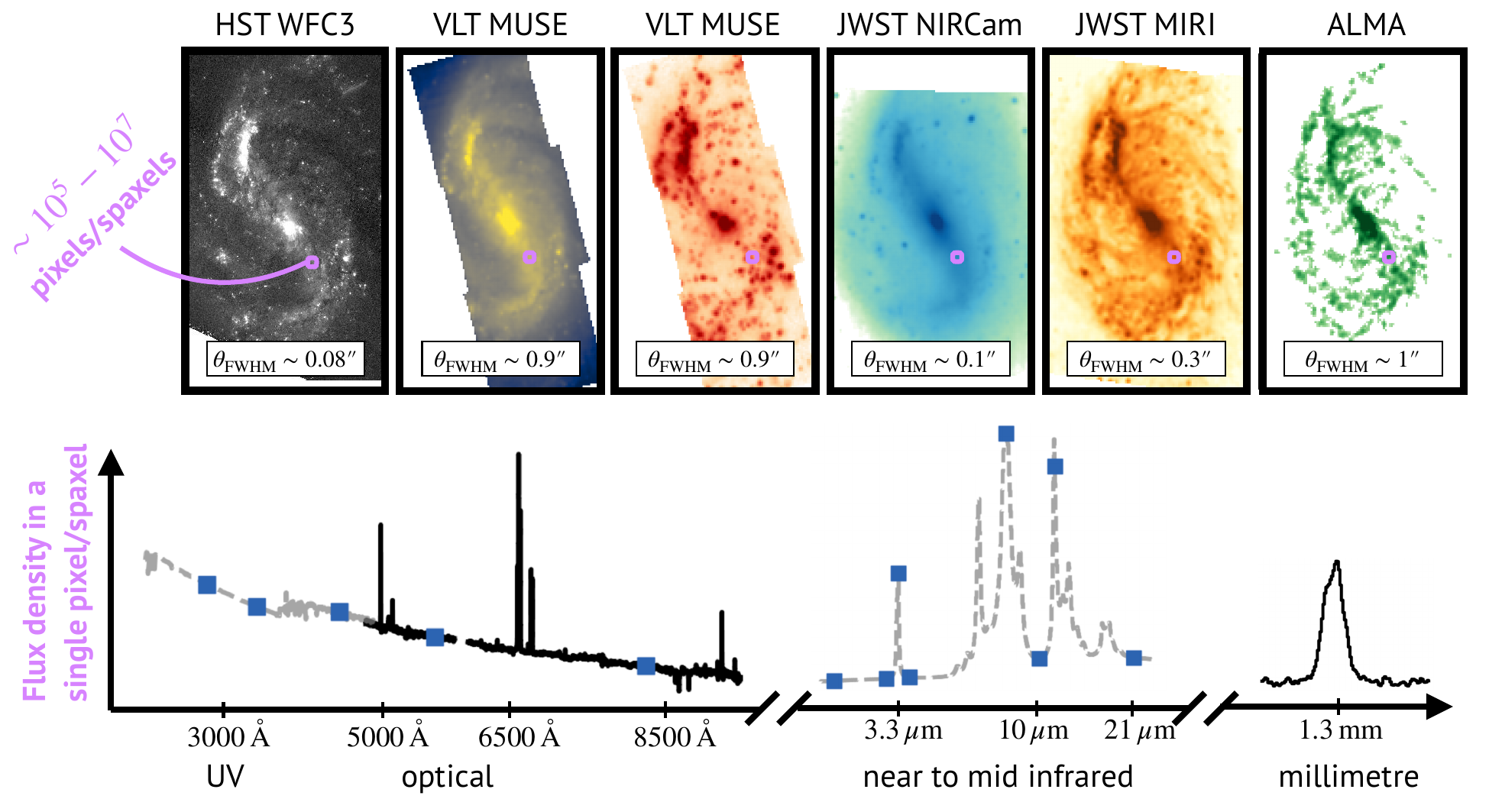}
\caption{\textbf{The information content in a single PHANGS galaxy.} The diagram uses data obtained for NGC~7496 to illustrate the multi-scale multi-wavelength information available as part of the PHANGS survey. The top row shows images of the galaxy obtained using: HST-WFC3 (F336W filter), VLT-MUSE (stellar continuum emission and \halpha~ surface brightness), JWST-NIRCam (F200W filter), JWST-MIRI (F770W filter), and ALMA. The different images have different spatial resolutions, ranging from $\sim 0.08''$ to $\sim 1''$, as indicated at the bottom of each image. Each PHANGS galaxy has $10^{5}$--$10^{7}$ pixels/spaxels. The multi-wavelength information available in each pixel/spaxel is shown in the bottom row, and it includes several UV-optical photometric bands by HST, a full optical spectrum between 4800 \AA\, and 9300 \AA\, by MUSE, eight JWST near and mid-infrared photometric bands, and a spectrum of the CO line reconstructed from millimeter interferometric observations by ALMA. The black solid lines represent observed spectra, blue rectangles represent photometry, and gray dashed lines represent the expected underlying spectrum. For presentation purposes, the flux densities in each wavelength region have been stretched vertically to cover the entire panel. In this work, we use a set of physically-motivated properties measured from these observations.}
\label{f:intro_fig}
\end{figure*}

The big data era in astronomy is not only characterized by an increase in data volume, related to the total number of observed sources, but is also characterized by an increase in \emph{data complexity}, which is related to an increased information content of a single astronomical source. With numerous surveys conducted using different telescopes and instruments, astronomical sources today often have multi-wavelength observations, from radio to X-ray, and in some wavebands, also as a function of time. In the nearby Universe, over the past several decades, surveys have mapped tens to hundreds of nearby galaxies from ultraviolet (UV) to radio using imaging and spectroscopy (e.g., SINGS; KINGFISH; THINGS; xCOLD GASS; \citealt{kennicutt03, dale07, walter08, moustakas10, kennicutt11, saintonge11, dale17, saintonge17}). Examples of galaxy clusters include the multi-wavelength mapping of the Virgo cluster galaxies (e.g., \citealt{cote04, boselli11, boselli18, brown21}). Outside the local universe, there are several multi-wavelength surveys that mapped galaxies at different redshifts in deep fields (e.g., GOODS; COSMOS; \citealt{ferguson00, giavalisco04, scoville07}).

The increase in data complexity raises some challenges, but also presents some new opportunities. On the one hand, it raises the question of how to incorporate the different types of observations, each with different sensitivities, spatial and spectral resolutions, and noise properties, within the same framework, in a sufficiently general manner to be applied to a variety of astronomical objects. In addition, the high-dimensionality of the data makes it challenging to identify trends, especially when they tie observables from different wavelengths or instruments. On the other hand, the increase in information content offers the unique opportunity to use the data itself to form novel hypotheses, a core approach in the field of data science. 

In the more traditional, model-driven or physics-driven approach, a scientific study starts with a hypothesis. Observations are planned and conducted to test the hypothesis, and their analysis leads to new insights, often resulting in new hypotheses. In data science, various statistical tools are used to visualize and dissect the high-dimensional space spanned by  the dataset. When the information content of the data is large, such tools may uncover previously-unknown trends or groups of objects, allowing one to form hypotheses directly from the data. Since this process relies less on a physical model or on prior knowledge, it may lead to unexpected discoveries. 

With the advent of multi-waveband opportunities, surveys have been producing larger and more complex datasets. The Physics at High Angular resolution in Nearby GalaxieS (PHANGS; \citealt{leroy21a, emsellem22, lee22, lee23}) survey is an example of a modern astronomical survey that pushes the limits of data complexity. With the goal of constraining the physics near or at the molecular cloud scale, the survey has been making high-resolution observations of nearby galaxies across the electromagnetic spectrum, utilizing various telescopes and instruments. Nineteen of the PHANGS galaxies have high-resolution maps obtained with the following telescopes: ALMA (mm interferometry; \citealt{leroy21a}), HST (UV and optical imaging; \citealt{lee22}), JWST (near-infrared and mid-infrared imaging; \citealt{lee23}), and VLT (optical integral field spectroscopy with MUSE; \citealt{emsellem22}). Each of the galaxies has $10^{5}$--$10^{7}$ independent spatial resolution elements\footnote{The number of independent resolution elements depends on the spatial resolution, with $\sim 10^{5}$ spaxels for MUSE and ALMA, $\sim 10^{6}$ pixels for JWST MIRI, and $\sim 10^7$ pixels for HST and JWST NIRCam.}, with each pixel/spaxel probing the conditions of multiphase gas, dust, and stars on scales of 5--100 pc (see figure \ref{f:intro_fig}). The unique combination of spectral coverage and high spatial resolution makes the information content of a single PHANGS galaxy comparable to that of the Legacy SDSS spectroscopic survey when considering the number of spectra and spectral resolution elements. 

The high information content of the PHANGS galaxies makes it an ideal dataset for applications of data-science tools. In this pilot study, our goal is to test whether unsupervised machine learning algorithms can be used to identify previously unknown trends or groups in the data. Such tools have been applied to astronomical datasets in various contexts (see reviews and references in \citealt[hereafter B19]{baron19}; and \citealt{fluke20}), with the most relevant and recent examples being (i) identification of underlying correlations in the high-dimensional data of molecular cloud populations constructed from multi-wavelength observations from PHANGS (\citealt{sun22}), and (ii) the application of a clustering algorithm to JWST observations of PAH emission in the Orion Bar (\citealt{pasquini23}). 

In this work, we estimate properties related to the stellar populations, multiphase gas conditions, and dust properties, on scales of 150 pc. We then use dimensionality reduction and clustering algorithms to divide these 150 pc-sized regions into groups. In the process, we identify new relations between PAHs and the warm ionized gas, and use our defined clusters to interpret these relations. This work is therefore complementary to recent studies that explicitly study the connection between PAHs, the ionized gas, and the radiation field, in the PHANGS-JWST galaxies, using a more physics-driven approach. In particular, \citet{egorov23} study the PAH-to-total dust fraction and its relation to the strength of the radiation field, parameterized using the gas ionization parameter, in HII regions. \citet{dale23} constrain different PAH properties, such as their size and charge distribution, in stellar clusters, while exploring the impact of changing radiation fields. \citet{chastenet23a} and Sutter et al. (in prep.) study how the PAH fraction depends on local and global conditions, including the phase of the interstellar medium, specific star formation rate, metallicity, and stellar mass. 

In section \ref{sec:data} we describe the data we use in our analysis. In section \ref{sec:methods} we describe our approach, which consists of three main steps: feature construction (section \ref{sec:methods:features}), dimensionality reduction (section \ref{sec:methods:UMAP}), and clustering (section \ref{sec:methods:clustering}). The resulting clusters, their interpretation, and the new relations found between PAHs and the warm ionized gas are presented in section \ref{sec:results}. The results section stands on its own and does not require a deep understanding of the methods. Therefore, readers who are interested in the results may skip section \ref{sec:methods} and go directly to section \ref{sec:results}. In section \ref{sec:discussion} we discuss possible extensions and generalizations of our methodology. We summarize and conclude in section \ref{sec:summary}.

 \floattable
\begin{deluxetable}{ccCCClC rrr}
\tablecaption{PHANGS-2D galaxy properties\label{tab:galaxy_properties}}
\tablecolumns{10}
\tablenum{1}
\tablewidth{0pt}
\tablehead{
\colhead{(1)}    & \colhead{(2)}    & \colhead{(3)}   & \colhead{(4)}                             & \colhead{(5)}                         & \colhead{(6)}  & \colhead{(7)}      & \colhead{(8)} & \colhead{(9)} & \colhead{(10)} \\
\colhead{Galaxy} & \colhead{D}      & \colhead{i}     & \colhead{SFR}                             & \colhead{$\log M_{*}$}                & \colhead{AGN?} & \colhead{FWHM}     & \colhead{N$_{\mathrm{pix, in}}$} & \colhead{N$_{\mathrm{pix, samp}}$} & \colhead{N$_{\mathrm{pix, mask}}$}  \\
\colhead{}       & \colhead{[Mpc]}  & \colhead{[deg]} & \colhead{[$\mathrm{M_{\odot}\,yr^{-1}}$]} & \colhead{[$\log \mathrm{M_{\odot}}$]} & \colhead{}     & \colhead{[arcsec]} & \colhead{}                       & \colhead{}                         & \colhead{}
}
\startdata
NGC~0628 & 9.84  & 9  & 1.75  & 10.2 & no  & 3.14\arcsec & 2\,560\,000 & 40\,000 & 6\,387 \\
NGC~1365 & 19.57 & 55 & 16.90 & 10.8 & yes & 1.58\arcsec & 346\,710    & 38\,523 & 12\,565\\
NGC~7496 & 18.72 & 36 & 2.26  & 9.8  & yes & 1.65\arcsec & 99\,000     & 24\,750 & 5\,055\\
\enddata

%\tablenotetext{a}{Galaxy distances from \citet{lee23}.}
%\tablenotetext{b}{Effective spatial resolution of the convolved multi-wavelength maps.}
%\tablenotetext{c}{Number of pixels in the initial ALMA WCS of the galaxy.}
%\tablenotetext{d}{Number of pixels in the RA-DEC coordinate grid after resampling.}
\tablecomments{(1)-(5) Galaxy properties from \citet{lee23}: name, distance, inclination, star formation rate, and stellar mass. (6) Indicator of AGN presence. (7) Effective spatial resolution of the convolved multi-wavelength maps. (8) Number of pixels in the initial ALMA WCS grid of the galaxy. (9) Number of pixels in the grid after down-sampling the ALMA grid to have at least 2 pixels per 150 pc. (10) Number of pixels that are not masked out in the pixel mask. The machine learning algorithms are applied to these sets of pixels.
}
\end{deluxetable}
%\vspace{5mm}

\section{Data}\label{sec:data}

To study the interplay between ISM gas and dust, star formation, and the observed stellar populations, we use multi-wavelength observations by ALMA, VLT-MUSE, JWST-NIRCam, and JWST-MIRI, obtained as part of the PHANGS survey \citep{leroy21a, emsellem22, lee23}. The first PHANGS-JWST data release includes fully-reduced and calibrated NIRCam and MIRI broad-band images of the three galaxies: NGC~0628, NGC~1365, and NGC~7496 (e.g., \citealt{lee23} and references therein). The broad-band images have been extensively tested and analyzed in a series of recent works (e.g., \citealt{belfiore23, chastenet23b, dale23, egorov23, leroy23, sandstrom23a, sandstrom23b}), making them an ideal benchmark for applications of machine learning algorithms. We therefore focus on these three galaxies here. 

Since our goal is to combine information from different instruments (ALMA, MUSE, and JWST NIRCam and MIRI), each with a different spatial resolution, and to consider pixels from different galaxies within the same analysis, we have to ensure that the pixels trace information originating from the same physical scale. We therefore use PHANGS data products that are based on maps convolved to a resolution of 150 pc, as described in each of the subsections below. We list the effective angular resolution (\arcsec) of the convolved maps for each of the galaxies in table \ref{tab:galaxy_properties}. 

We project all the convolved maps into the world coordinate system (WCS) defined by the ALMA observations using {\sc reproject.exact} by {\sc astropy} \citep{reproject_2020, astropy22}. We then downsample the RA-DEC coordinate grid to have two pixels per resolution element of 150 pc in each of the galaxies\footnote{The pixels in the downsampled grid therefore trace distances of 76.3 pc, 85.4 pc, and 89.6 pc, for NGC~0628, NGC~1365, NGC~7496, respectively. We ensured that using only one pixel per resolution element of 150 pc results in a comparable a low-dimensional embedding to that we obtain with two pixels.}. We list the initial number of pixels in the ALMA observation grid and the number of pixels after the downsampling in table \ref{tab:galaxy_properties}. 

\begin{figure*}
	\centering
\includegraphics[width=1\textwidth]{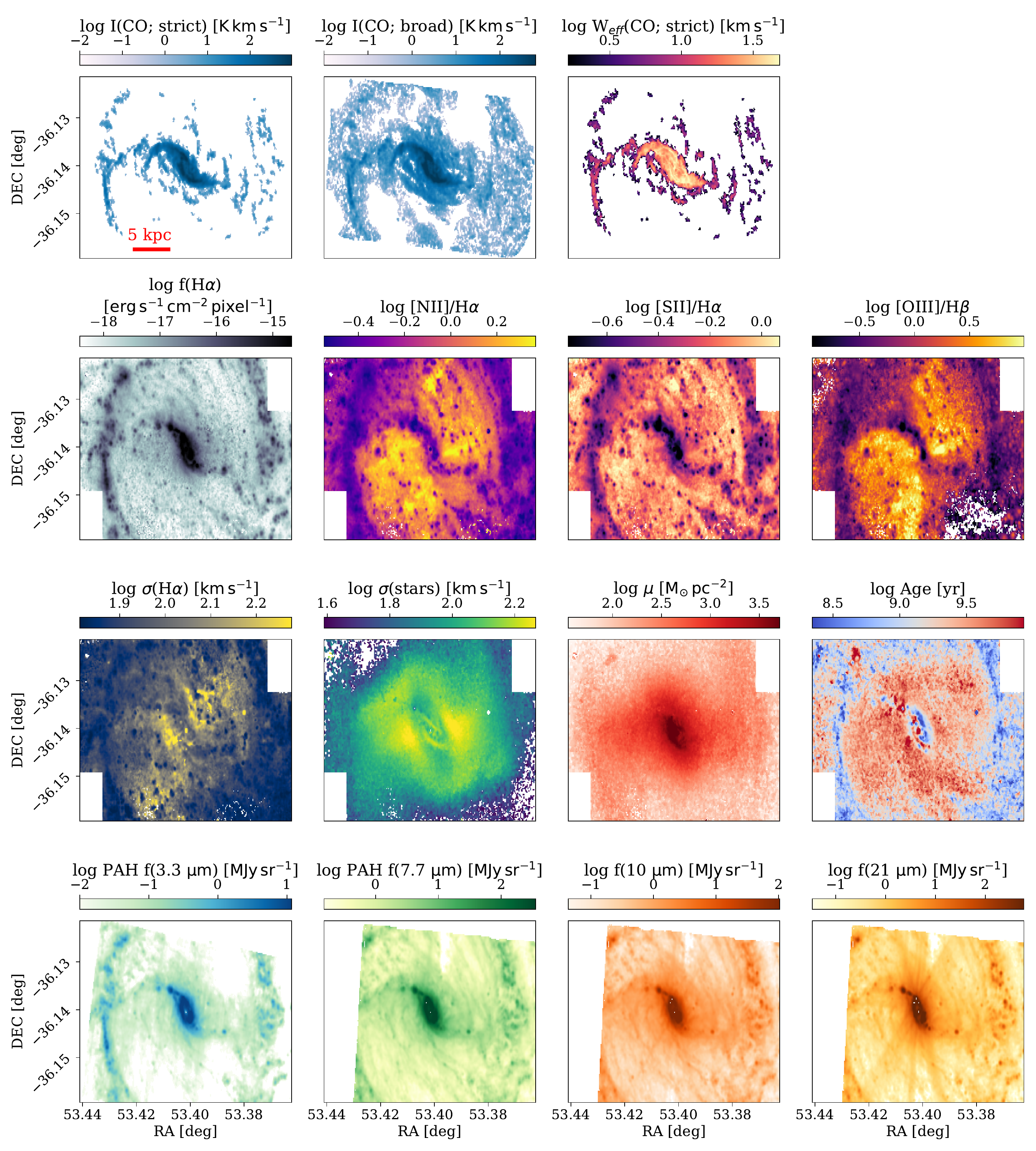}
\caption{\textbf{Different properties derived from the ALMA, MUSE, and JWST observations of NGC~1365.} The first row shows cold molecular gas properties from ALMA: \cofull moment 0 (intensity) derived using the ``strict" (lower noise but less complete) and ``broad" (noisier and more complete) masks, and the CO effective width. The second and third rows show gas and stellar population properties from MUSE: dust-corrected \halpha~ surface brightness, the line ratios \niihalpha, \siihalpha, and \oiiihbeta, \halpha~ velocity dispersion, stellar velocity dispersion, stellar mass surface density, and the age of the stellar population. The fourth row shows dust grain properties from JWST: PAH 3.3 \mic~ and 7.7 \mic~ emission features, and broad-band mid-infrared emission from hot dust at 10 \mic~ and 21 \mic. The 21 \mic~ image is saturated in the center of the galaxy and shows noticeable diffraction spikes. All the maps are convolved to a common resolution of 150 pc.}
\label{f:NGC1365_initial_feature_display}
\end{figure*}

As an illustration, we show different properties of NGC~1365 derived from the ALMA, MUSE, and JWST maps in figure \ref{f:NGC1365_initial_feature_display}. These include properties pertaining to the cold molecular gas, warm ionized gas, stellar populations, PAHs, and large dust grains. In the rest of the section we describe the data products we use and their analysis.

\subsection{ALMA}\label{sec:data:alma}

To trace the cold molecular gas properties, we use the PHANGS-ALMA survey \citep{leroy21a, leroy21b}. The survey mapped the \cofull (CO hereafter) line emission at a spatial resolution of $\sim 1'' \sim 100$ pc in 90 nearby galaxies. After imaging, the cubes are convolved to a succession of physical resolutions. We use the PHANGS-ALMA data products obtained after convolving the cubes to a spatial resolution of 150 pc.

The catalog includes two main types of products. The ``strict" mask products include two-dimensional maps generated from the data cubes after applying stringent signal identification criteria. Since these require a detection of the signal with high confidence, these maps have low noise, but they include less of the total CO flux and are thus somewhat incomplete. The ``broad" mask products include two-dimensional maps generated using all the sight lines where signal is identified at any resolution, making them more complete than the ``strict" mask products. However, since these include regions with faint emission, they appear noisier and can contain false positives. In figure \ref{f:NGC1365_initial_feature_display} we show the CO moment 0 (total intensity) derived using the ``strict" and ``broad" masks, as well as the CO effective width ($W_{eff}$, a line width measure) calculated using the ``strict" masks. Since our analysis requires high completeness and can tolerate some additional noise, we use the CO flux derived using the ``broad" mask in our feature extraction (see details in section \ref{sec:methods:features}).

\subsection{MUSE}\label{sec:data:muse}

To trace the warm ionized gas conditions and the stellar population properties, we use the PHANGS-MUSE survey \citep{emsellem22}. This survey mapped 19 of the PHANGS-ALMA galaxies with the integral field spectrograph MUSE at a spatial resolution of $\sim 1''$. The data analysis pipeline of the survey includes (i) mosaicking and homogenization of individual MUSE pointings for a given galaxy, (ii) performing spatial binning of the spectra to reach sufficient signal to noise ratio (SNR) for stellar population synthesis modeling, (iii) fitting the stellar continuum to extract the stellar kinematics, reddening, and stellar populations, and (iv) fitting the optical emission lines to extract gas kinematics and line fluxes. In this work, we use various properties (see below) derived from the MUSE cubes after they have been convolved to a resolution of 150 pc (PHANGS DR2.2 release).

For the warm ionized gas properties, we use the surface brightness maps of the Balmer lines \hbeta~ and \halpha, and the following emission lines: \oiiifull, \oifull, \niifull, and \siifull~ (\oiii, \oi, \nii, and \sii~ hereafter). These are estimated by fitting the spectra in the convolved cubes with a set of Gaussians (see details in \citealt{emsellem22}). To estimate the dust-corrected \halpha~ surface brightness, we assume case-B recombination ($T=10^{4}$ K), a dusty-screen, and the \citet{cardelli89} extinction law, with the color excess given by:
\begin{equation}\label{eq:reddening}
	{\mathrm{E}(B-V) = \mathrm{2.33 \times log\, \Bigg[ \frac{(H\alpha/H\beta)_{obs}}{2.86} \Bigg] \, \mathrm{mag} }},
\end{equation}
where $\mathrm{(H\alpha/H\beta)_{obs}}$ is the observed \halpha/\hbeta~ surface brightness ratio. We then correct the observed \halpha~ surface brightness using the derived $\mathrm{E}(B-V)$ values. 

In our analysis, we consider the dust-corrected \halpha~ surface brightness, the \halpha~ gas velocity dispersion, and the line ratios \oiiihbeta, \niihalpha, \siihalpha, and \oihalpha, which are typically used to constrain the main source of ionizing radiation. These line ratios are based on lines close in wavelength, so they are nearly reddening independent. We thus do not correct the lines for reddening prior to the line ratio estimation. We do correct the \halpha~ surface brightness for reddening since our derived features in section \ref{sec:methods:features} include ratios of \halpha~ to CO or PAH emission. We propagate the surface brightness uncertainties to the feature uncertainties, and pixels in which these properties are measured with SNR $<3$ are masked out (see discussion about feature missingness in section \ref{sec:discussion:missing_features}).

For the stellar population properties, we use maps of the stellar velocity dispersion, stellar mass density, mass-weighted stellar age, mass-weighted stellar metallicity, and reddening towards the stars. These properties were derived using stellar population synthesis fits of binned stellar spectra, using Voronoi bins of the convolved MUSE cubes (\citealt{emsellem22, pessa23}).

\subsection{JWST}\label{sec:data:jwst}

To trace PAH properties and dust-reprocessed stellar or AGN light, we use the Cycle 1 PHANGS-JWST survey data \citep{lee23, williams24}. It is a JWST treasury survey aimed at collecting imaging data in eight bands from 2 to 21 \mic~ of the 19 galaxies observed as part of PHANGS-ALMA, PHANGS-MUSE, and PHANGS-HST. The broad band images include the four NIRCam bands F200W, F300M, F335M, and F360M, probing near-infrared emission at 2 \mic, 3 \mic, 3.35 \mic, and 3.6 \mic~ respectively, and the four MIRI bands F770W, F1000W, F1130W, and F2100W, probing mid-infrared emission at 7.7 \mic, 10 \mic, 11.3 \mic, and 21 \mic. These filters are designed to cover different PAH emission bands, which are expected to be sensitive to different grain size and charge distribution, to capture dust continuum emission, and the silicate 9.7 \mic~ absorption feature.

We use the same version of the surface brightness maps presented and described in the PHANGS-JWST Cycle 1 Focus Issue\footnote{\url{https://iopscience.iop.org/collections/2041-8205_PHANGS-JWST-First-Results}} (version 0.8; see e.g., \citealt{belfiore23, chastenet23b, dale23, lee23, sandstrom23a, sandstrom23b}). These maps are convolved to a spatial resolution of 150 pc using the approach described in \citet{aniano11}. \citet{williams24} describes the full data reduction pipeline.

To study the PAH properties, we use the MIRI F770W and F1130W broad-band images, which are expected to be dominated by the 7.7 and 11.3 \mic~ PAH features. We also use the NIRCam bands F300M, F335M, and F360M, and follow the procedure outlined by \citet{sandstrom23a} to estimate the 3.3 \mic~ PAH flux (F335M$\mathrm{_{PAH}}$ hereafter). This procedure uses the F300M and F360M bands to subtract the starlight contribution from the F335M band, thus isolating the flux from the 3.3 \mic~ PAH feature. 

For the dust continuum emission, we considered both the MIRI F1000W and F2100W filters, though both suffer from different limitations. The F1000W filter shows strong correlations with the PAH-tracing filters F770W and F1130W, and a weaker correlation with the hot dust-tracing filter F2100W, suggesting that the filter is in fact dominated by PAH emission in a large fraction of the pixels (e.g., \citealt{belfiore23, leroy23}). The F2100W filter traces only dust continuum emission, but it is saturated in the central pixels of NGC~1365 and NGC~7496, presenting significant diffraction spikes (see figure \ref{f:NGC1365_initial_feature_display} and \citealt{chastenet23b}). Since using the F2100W filter would require us to mask out the saturated galaxy centers and the diffraction spikes, we do not include it in our feature construction. We do include the F1000W filter, but note that it traces PAH emission much more than the hot dust continuum. We use both F1000W and F2100W in our interpretation of the resulting clusters and trends. 

Following the papers in the Focus Issue, we use the convolved maps of NGC~7496 to estimate the noise level in all the relevant bands (F335M$\mathrm{_{PAH}}$, F770W, F1130W, F1000W, F2100W)\footnote{Our estimated noise levels of the convolved, lower-resolution, maps are consistent with those estimated by \citet{lee23} and \citet{chastenet23b} for the high-resolution maps given the convolution kernel size.}. NGC~7496 is the only target with sufficiently empty space that is not contaminated by emission from the source. The estimated noise root mean square (RMS) levels, in units of $\mathrm{MJy\,sr^{-1}}$, are 0.0071, 0.021, 0.023, 0.013, and 0.082, for the F335M$\mathrm{_{PAH}}$, F770W, F1130W, F1000W, F2100W bands, respectively. We use these values to mask out pixels in which the measured flux is lower than 3 times the noise RMS in all three galaxies. \vspace{1cm}

\section{Methods}\label{sec:methods}

In this section, we apply unsupervised machine learning algorithms to the maps constructed from the ALMA, MUSE, and JWST observations. These algorithms are used to divide the pixels from the different galaxies into groups according to their multi-wavelength properties, where pixels in a given group show distinct values or correlations between their stellar, gas, and dust properties, from pixels in other groups. This allows us to explore the complex multi-wavelength PHANGS dataset without an initial model-driven hypothesis. Instead, we form data-driven hypotheses by inspecting the resulting groups and identifying previously unknown trends and correlations within them.

In our analysis, each pixel from each of the galaxies is considered as a separate \emph{object} with a set of measured \emph{features}. The measured features are constructed from the multi-wavelength maps, and they trace the stellar, gas, and dust properties within each of the pixels. The features do not include information related to the galaxy a pixel belongs to, or its location within the galaxy, though both are used when interpreting the results. Once a final list of objects with measured features is constructed, we apply a dimensionality reduction algorithm to this data. The output of the dimensionality reduction is an embedding of the high-dimensional data into a two-dimensional space, where every object (pixel) is represented by two numbers. We then apply a clustering algorithm to the distribution of the objects in the two-dimensional space, which allows us to assign a class to each of the objects. Therefore, this three-step procedure divides the PHANGS pixels into groups according to their multi-wavelength observations.

We start by describing our feature construction scheme in section \ref{sec:methods:features}, which includes our definition of features, and their scaling and normalization. Since the data contains a non-negligible fraction of non-detections, the section also describes our adopted pixel masking strategy aimed at minimizing missing feature values, and our strategy to handle the remaining missing values. In section \ref{sec:discussion:missing_features} we discuss the issue of missing values more generally, and propose methods to handle non-detections and upper limits in future works. In section \ref{sec:methods:UMAP} we apply the dimensionality reduction algorithm {\sc umap} and obtain a two-dimensional embedding of the input dataset. We also discuss the impact of different hyper-parameter choices on the resulting two-dimensional embedding. In section \ref{sec:methods:clustering} we apply several different clustering algorithms to the two-dimensional embedding and define the final clusters that will be used to group the pixels. 

\subsection{Feature Extraction}\label{sec:methods:features}

We use the convolved, reprojected, and downsampled maps obtained from the ALMA, MUSE, and JWST observations described in section \ref{sec:data}. Since our goal is to study the interplay between stellar, gas, and dust properties, we require $> 3\sigma$ detection of the \halpha~ and CO emission lines, the PAH bands, and the dust continuum emission\footnote{The stellar population properties are measured using high-SNR binned spectra, and their measured uncertainties are generally smaller than 3 times the measured value. Therefore, the stellar properties are quite complete throughout the field of view.}. For that requirement, we construct a pixel mask map for each of the galaxies. In each pixel mask map, a pixel value is set to $1$ if this pixel is not masked out in \emph{every one} of the following maps: dust-corrected \halpha~ surface brightness, CO intensity using the ``broad" mask, F335M$\mathrm{_{PAH}}$, F770W, and F1000W. Otherwise, the pixel value is set to $0$. Since the \halpha~ surface brightness is dust corrected, requiring a $3\sigma$ detection ensures the detection of the weaker \hbeta~ line. While this requirement does not ensure the detection of the \oiii, \nii, \sii, and \oi~ lines, their detection fraction is quite high (see missing fractions in table \ref{tab:features}).

\begin{figure*}
	\centering
\includegraphics[width=1\textwidth]{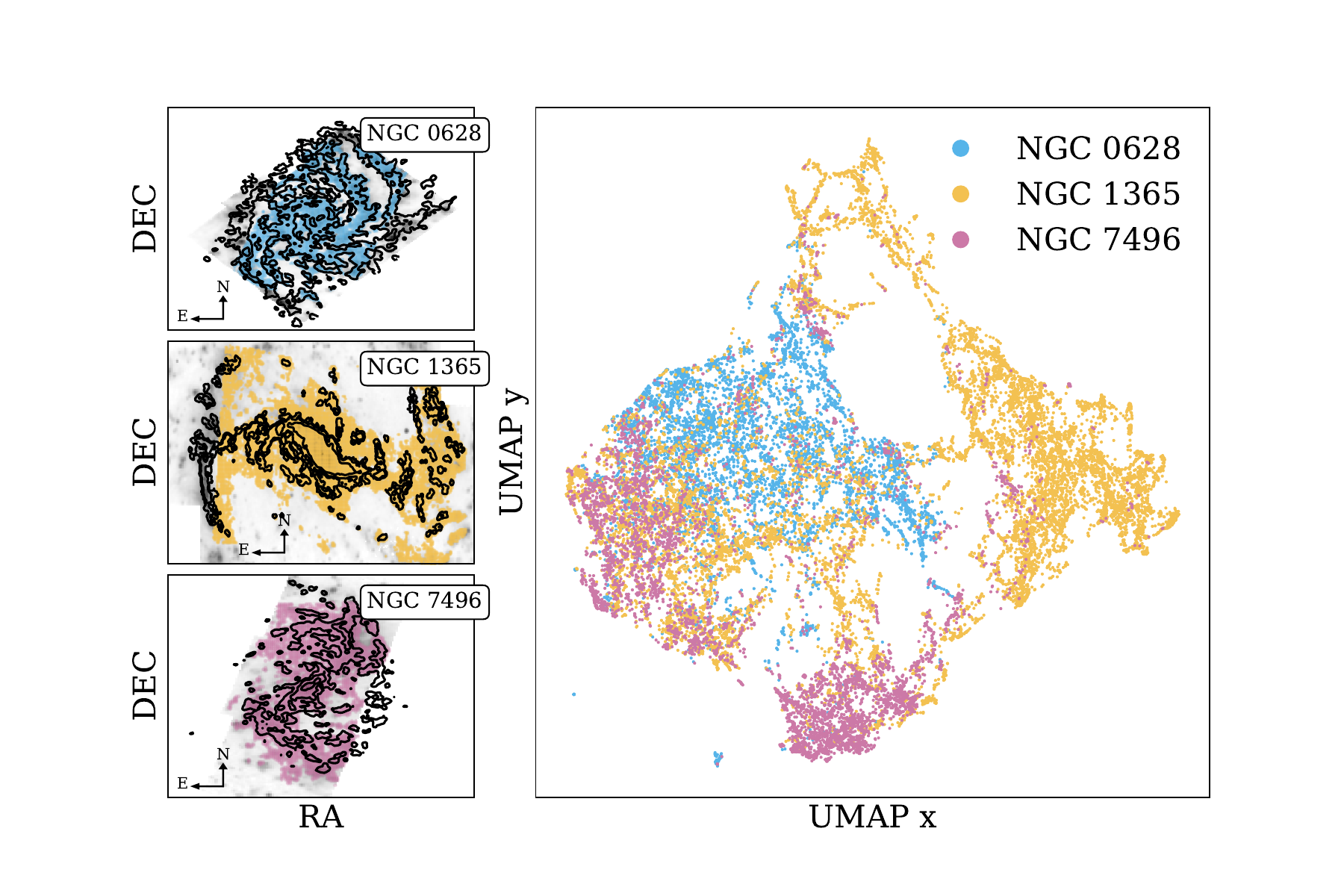}
\caption{\textbf{{\sc umap} dimensionality reduction of the PHANGS multi-wavelength pixels.} The left panels show the three galaxies NGC~628, NGC~1365, and NGC~7496, where pixels that are included in the analysis are marked with colors. In these panels, the grayscale background represents the \halpha~ surface brightness, and the black contours represent the CO intensity. The right panel shows our adopted {\sc umap} embedding. Every point corresponds to an object (pixel) from our input dataset, which includes 24\,007 objects with 13 features each (features trace 150 pc-size regions). The two-dimensional embedding was obtained using the following hyper-parameters: \texttt{metric=correlation}, \texttt{n\_neighbors=10}, and \texttt{min\_dist=0}, though we show in \ref{app:umap_parameters} that the global structure in the two-dimensional space remains stable when changing the metric and the \texttt{n\_neighbors} parameter. Each pixel is colored according to the galaxy it belongs to, information that is not included in the input dataset. The embedding shows several over-dense regions that may be interpreted as clusters, connected through filamentary structures of points, suggesting that the features form continuous relations in the complex high-dimensional space.}
\label{f:UMAP_with_galaxies}
\end{figure*}

We list the number of pixels in the pixel mask map in table \ref{tab:galaxy_properties}. The included pixels are also marked with colors in figure \ref{f:UMAP_with_galaxies} in the results. They constitute only 15\%, 32\%, and 20\%, of all the pixels in the downsampled maps of NGC~0628, NGC~1365, and NGC~7496, respectively. There are two main factors contributing to the large number of masked-out pixels. The first is the incomplete overlap between the fields of view of the different instruments (see e.g., figure 1 in \citealt{lee23}). The second factor is our requirement of CO and PAH detection, which masks out around 30\% of the pixels (see e.g., figure \ref{f:UMAP_with_galaxies}). These pixels are masked out because the CO and PAH 3.3 \mic~ emission are below the sensitivity limit. 

Among the maps we consider in our masking, the dust-corrected \halpha~ and the 7.7 \mic~ and 10 \mic~ surface brightness maps are quite complete, with a small fraction of masked-out pixels. The CO and 3.3 \mic~ maps have a larger fraction of masked-out pixels, and they typically have the same pixels masked-out in both. Since we are interested in studying multiphase gas and dust, which cannot be done without a CO and 3.3 \mic~ PAH emission detection\footnote{We require a detection of the 3.3 \mic~ PAH feature as this band is particularly useful when constraining the PAH size distribution. See section \ref{sec:results:PAH_optical_lines_ratios_corr} for additional details.}, we choose to exclude such pixels from the analysis. Our choice to use the CO emission identified using the ``broad" mask is motivated by the CO detection requirement, since the CO line is detected in a larger fraction of the pixels compared to that of the ``strict" mask. If instead we had used the CO intensity identified using the ``strict" mask, we would have had to mask out approximately 80-90\% of the pixels in each galaxy, excluding most of the diffuse ISM. In section \ref{sec:discussion:missing_features} we discuss the implications of excluding such pixels from the analysis and propose possible methods to include upper limits and non-detections in future analyses of this kind.

\floattable
\begin{deluxetable}{llR}
\tablecaption{Selected features\label{tab:features}}
\tablecolumns{3}
\tablenum{2}
\tablewidth{0pt}
\tablehead{
\colhead{Feature} & \colhead{Unit} & \colhead{missing fraction after pixel mask}
}
\startdata
$\log \mathrm{f(H\alpha)/I(CO)}$ & $\log ( \mathrm{erg\,s^{-1}\,cm^{-2}\,pixel{^{-1}}/K\,km\,s^{-1}} )$ & - \\
$\log \mathrm{f(H\alpha)/f(10\, \mu m)}$ & $\log (\mathrm{erg\,s^{-1}\,cm^{-2}\,pixel{^{-1}}/MJy\,sr^{-1}})$ & - \\
$\log \mathrm{I(CO)/f_{PAH}(7.7\, \mu m)}$ & $\log (\mathrm{K\,km\,s^{-1}/MJy\,sr^{-1}})$ & - \\
$\log \mathrm{f([N\, II])/f(H\alpha)}$ & & 0.033\% \\
$\log \mathrm{f([S\, II])/f(H\alpha)}$ & & - \\
$\log \mathrm{f([O\, I])/f(H\alpha)}$ & & 0.36\% \\
$\log \mathrm{f([O\, III])/f(H\beta)}$ & & 1.53\% \\
$\log \mathrm{Age}$ & $\log (\mathrm{yr})$ & - \\ 
gas $\mathrm{E}(B-V)$ & $\mathrm{mag}$ & - \\ 
$\log \mathrm{f_{PAH}(3.3\,\mu m)/f_{PAH}(7.7\,\mu m)}$ & & - \\ 
$\log \mathrm{f_{PAH}(3.3\,\mu m)/f_{PAH}(11.3\,\mu m)}$ & & - \\ 
$\log \mathrm{f_{PAH}(11.3\,\mu m)/f_{PAH}(7.7\,\mu m)}$ & & - \\ 
$\log \mathrm{\Big(f_{PAH}(7.7\,\mu m) + f_{PAH}(11.3\,\mu m) \Big)/f(10\, \mu m)}$ & & - \\ 
\hline
$\log \mathrm{\sigma(H\alpha)}$ $^{*}$ & $\log (\mathrm{km\,s^{-1}})$ & 0.066\% \\ 
$\log \mathrm{\mu}$ $^{*}$ & $\log (\mathrm{M_{\odot}\,pc^{-2}})$ & 0.12\% \\ 
$\log \mathrm{\sigma(H\alpha)/\sigma(stars)}$ $^{**}$ & & 12.4\% \\ 
\enddata
\tablecomments{Summary of the selected features, their units, and their missing fraction \textit{after} applying the pixel mask. The data was first scaled, then clipped, and then normalized. All features except the gas $\mathrm{E}(B-V)$ are scaled logarithmically. Then all features are clipped to be between the 0.5th and 99.5th percentiles of the distribution. All the features are then normalized by subtracting the mean and diving by the standard deviation of the clipped distribution. The last three features were excluded from the data after some initial tests: $^{*}$: these features dominated the two-dimensional embedding and resulted in a trivial embedding with three clusters corresponding to the three different galaxies, $^{**}$: high-fraction of missing values and little impact on the resulting embedding.}
\end{deluxetable}

Table \ref{tab:features} summarizes the features we considered, most of which are surface brightness \textit{ratios}. Although the machine learning algorithms we use can in principle be applied to measured surface brightness values directly, we choose to work with ratios as we find them more easy to interpret. The first three features, \halpha~ to CO, \halpha~ to 10 \mic, and CO to 7.7 \mic~ PAH, trace the interplay between star formation (traced by \halpha), different gas phases (traced by \halpha~ and CO), and dust grains. We also consider the warm ionized gas line ratios \niihalpha, \siihalpha, \oihalpha, and \oiiihbeta~ as features, as these are known to be sensitive to the source of ionizing radiation through standard BPT diagrams (\citealt{baldwin81, veilleux87, kewley01, kauff03a}), as well as to metallicity, ionization parameter, and more (e.g., \citealt{kewley19}). 
%Since we consider ratios as our features, many of the features are correlated with each other

We consider properties pertaining to the dynamics and stellar population traced by the MUSE observations: the ionized gas velocity dispersion measured with the \halpha, the gas-to-stellar velocity dispersion ratio, the stellar mass surface density, the age of the stellar population, and the reddening towards the line-emitting gas (see equation \ref{eq:reddening}). Three of these features were later excluded from the analysis -- (i) the \halpha~ velocity dispersion, (ii) and the stellar mass surface density, since they dominated the dimensionality reduction and resulted in a trivial embedding where the pixels were divided into three clusters according to the galaxy they belong to\footnote{The clipping and normalization we perform (see below) to the features should have, in principle, ensured that the feature values do not differ significantly between the different galaxies. In practice, however, these two features showed non-Gaussian distributions with significant tails, which affected the normalization and resulted in non-standard distributions that are different for the different galaxies. Since these tails correspond to robust measurements that represent extreme dynamical environments, we do not clip them. To include such features in future studies, it might be necessary to perform histogram equalization prior to the clipping and normalization.}, even after correcting for galaxy inclination, and (iii) the gas to stellar velocity dispersion ratio, which had a large fraction of missing values and had little impact on the low-dimensional embedding. 

We include several flux ratios that trace different PAH properties. The 3.3 \mic~ PAH feature primarily traces small and neutral PAHs, while the 7.7 \mic~ feature traces larger and positively-charged ions, and the 11.3 \mic~ feature represents grains that are larger and neutral (e.g., \citealt{boersma16, boersma18, maragkoudakis20, draine21, rigopoulou21, maragkoudakis22}). Therefore, the 3.3/11.3 \mic~ and 3.3/7.7 \mic~ PAH ratios are sensitive to the PAH size distribution, though they are also sensitive to the shape of the FUV-optical radiation field (see e.g., \citealt{draine21} and Appendix \ref{app:pah_models}). The 11.3/7.7 \mic~ PAH ratio is primarily sensitive to the PAH charge distribution, though also to the shape of the radiation. To trace the PAH abundance, several recent studies defined $\mathrm{R_{PAH}}= \mathrm{(F770W + F1130W)/F2100W}$, which is a ratio of PAH to dust-continuum flux (e.g., \citealt{chastenet23a, chastenet23b, egorov23}; Sutter et al., in prep). Since the F2100W is saturated in the centers of NGC~1356 and NGC~7496, we use an alternative feature, (F770W + F1130W)/F1000W, where F1000W is used to trace the hot dust continuum. However, we found that the (F770W + F1130W)/F1000W feature does not correlate with $\mathrm{R_{PAH}}$, most-likely since the F1000W band is dominated by PAH emission, rather than by hot dust continuum emission. In addition, the band may also be affected by 9.7 \mic~ silicate absorption (e.g., \citealt{smith07}).

Most of the features we consider are distributed over several orders of magnitude. In such a case, any dimensionality reduction algorithm that starts by measuring pairwise distances between the features will be dominated by features with larger values (see e.g., \citetalias{baron19}). To give even weights to small and large feature values, we use a logarithmic scaling of the features. That is, the \halpha~ to CO flux ratio feature is represented by $\log [ \mathrm{f(H\alpha)/I(CO) ]}$ rather than $\mathrm{f(H\alpha)/I(CO)}$. The only exception is the feature $\mathrm{E}(B-V)$, which is not distributed over several orders of magnitude.

We then apply clipping and normalization to the scaled features. For each feature, we clip its values to be between the 0.5th and 99.5th percentiles of the distribution. The clipping ensures that catastrophic outliers do not have a significant impact on the normalization of the feature, which is the next and final stage of the feature construction. These outliers are mostly the result of problems in observations, their reduction, or the estimation of a feature value from them. By definition, the fraction of objects with clipped feature values is very low, and thus the clipping of their values does not have an impact on the two-dimensional embedding of the rest of the objects. Indeed, we found the two-dimensional embedding by {\sc umap} to be stable to different clipping schemes, ranging from (0.1, 99.9)\% to (1, 99)\%. However, even a low fraction of catastrophic outliers can have a significant impact on the estimated standard deviation of a feature, which can affect the normalization of the whole feature, and thus the {\sc umap} embedding. It is therefore essential to perform some clipping before the normalization. 

Finally, each scaled and clipped feature $x$ is normalized as $(x - \mu_{x})/\sigma_{x}$, where $\mu_{x}$ is the average feature value and $\sigma_{x}$ is the standard deviation. The normalization is done to ensure that the dimensionality reduction will not be dominated by features with a large dynamical range (e.g., see discussion in \citetalias{baron19}).

Since we exclude pixels where the CO and/or 3.3 \mic~ PAH are not detected, our analysis is complete in HII regions and much of the dense ISM as these show brighter CO and mid-infrared emission, but is incomplete in the most diffuse part of the ISM, where CO and mid-infrared emission are much fainter. Therefore, our results may not be applicable to the most diffuse parts of local galaxies. In table \ref{tab:features} we list the fraction of missing values in the features we consider after applying the pixel mask. Due to our pixel mask, the fraction of missing values is very low. We compared two different imputation methods to replace these missing values (see additional details in section \ref{sec:discussion:missing_features} in the discussion): KNN search \citep{hruschka03, jonsson04} and Random Forest Regression \citep{stekhoven11, shah14} by {\sc sklearn} \citep{pedregosa11}. The two methods gave comparable results, and neither had a significant impact on the resulting two-dimensional embedding. The results shown in section \ref{sec:results} are based on a dataset where the missing values have been replaced using the KNNImputer\footnote{\url{https://scikit-learn.org/stable/modules/generated/sklearn.impute.KNNImputer.html}}.\vspace{1cm}

\subsection{Dimensionality Reduction with UMAP}\label{sec:methods:UMAP}

In this section we use the dimensionality reduction algorithm {\sc umap} (Uniform Manifold Approximation and Projection; \citealt{mcinnes18}) to embed our input dataset into a two-dimensional space. 

{\sc umap} is a non-linear dimensionality reduction algorithm that aims to represent high-dimensional data in a lower-dimensional space while preserving the underlying structure and relationship among data points. It operates on the assumption that the data lie on a manifold, which is a low-dimensional curved surface embedded within the high-dimensional space. It attempts to learn an approximation of this manifold and then project the data onto the lower-dimensional space. {\sc umap} has been shown to preserve local as well as global structures in the data, and has been widely used in a variety of fields (see e.g., \citealt{becht18, ali19, cao19, carter19, packer19}). 

The two main use cases of {\sc umap} are: (i) visualization of complex high-dimensional datasets, and (ii) dimensionality reduction prior to the application of clustering algorithms. The latter, which is also the use case in this work, is done because many clustering algorithms do not scale well with a large number of features (e.g., \citealt{xu15}), and thus {\sc umap} is used as an intermediate stage to reduce the initial dimensions of the dataset. This intermediate step also improves the interpretability of the final clustering output, as the clusters and their properties can be easily visualized in the two-dimensional space given by {\sc umap}.

{\sc umap} has several hyper-parameters, and setting different values of the hyper-parameters can change the resulting embedding significantly. The first, and probably most important, hyper-parameter is \texttt{metric}, which defines the distance metric to be used when estimating distances between the objects in the high-dimensional space. Different metrics are sensitive to different aspects and scales in the feature space. As a result, while one metric may suggest proximity between two objects, another may suggest a considerable separation (see discussion in \citetalias{baron19}). Given that {\sc umap} relies on pairwise distances between objects for the embedding process, the selection of an appropriate metric becomes crucial. 

The second most important hyper-parameter of {\sc umap} is \texttt{n\_neighbors}, which controls how the algorithm balances local versus global structure when learning the manifold of the data (see examples in \citealt{mcinnes18}). Setting a low value of \texttt{n\_neighbors} will result in an embedding that is more sensitive to local structure of the data, sometimes at the expense of accurately representing the global structure. On the other hand, increasing the value of \texttt{n\_neighbors} expands the neighborhoods considered when estimating the manifold, which may result in the loss of fine-grained details. 

\begin{figure*}
	\centering
\includegraphics[width=1\textwidth]{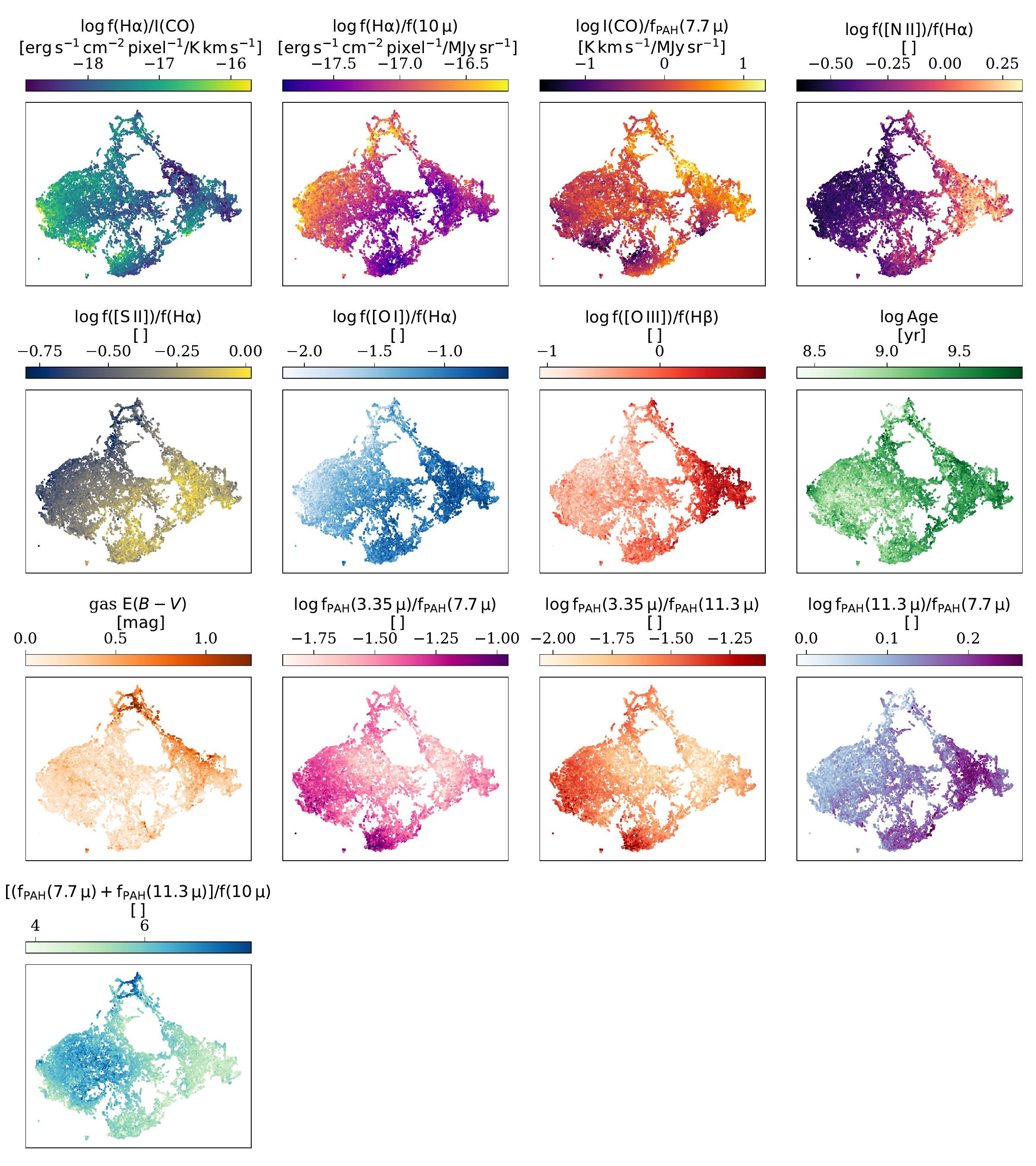}
\caption{Our adopted two-dimensional {\sc umap} embedding color-coded by the features in the input dataset.}
\label{f:UMAP_colorcoded_by_features}
\end{figure*}

The third hyper-parameter, \texttt{min\_dist}, controls how tightly points can be packed in the low-dimensional space. Larger values of \texttt{min\_dist} will force even close neighbors to be separated in the low-dimensional embedding, while lower values will result in clumpier embeddings. 

Following the feature construction scheme in section \ref{sec:methods:features}, the input dataset has 24\,007 objects with 13 features each. Each object represents a pixel in one of the three galaxies we consider, and the 13 features (listed in Table \ref{tab:features}) trace different properties related to the stellar population, gas, dust, and star formation, over a 150 pc scale. 

We applied {\sc umap} to our input dataset while exploring a wide range of hyper-parameter choices. In particular, we examined the two-dimensional embedding using 11 different distance metrics, and using a wide range of \texttt{n\_neighbors} values (see \ref{app:umap_parameters})\footnote{One could, in principle, apply {\sc umap} to reduce the dimensions to 3 or 4, and then apply clustering to the lower-dimensional embedding. We find it challenging to visualize and interpret the low-dimensional embedding for 3 or 4 dimensions, and to identify a suitable set of hyper-parameters for the clustering algorithms. We therefore only explore two-dimensional embeddings.}. We find that the global structure of the data in the two-dimensional space remains stable when changing the metric and/or \texttt{n\_neighbors}. We reach a similar conclusion for the \texttt{min\_dist} parameter, which primarily changes the density of points. This suggests that the clusters that would have been identified in the two-dimensional embeddings by {\sc umap} with different hyper-parameters should be roughly the same. Therefore, the results presented in section \ref{sec:results} should not depend significantly on the assumed hyper-parameters. 

For the rest of the paper, we adopt the two-dimensional {\sc umap} embedding obtained using \texttt{metric=correlation}, \texttt{n\_neighbors=10}, and \texttt{min\_dist=0}.  This set of hyper-parameters was adopted after a visual inspection of the resulting embeddings, where we selected an embedding where we expect cluster identification to be less challenging technically, as we describe below. However, since different sets of hyper-parameters result in embeddings that satisfy the criteria described below, the selection of this particular set of hyper-parameters is somewhat arbitrary. 

To select an embedding where identifying clusters is expected to be less challenging, we favor embeddings where clusters that we identify by eye are more separated from each other, and are separated by roughly the same distances one from the other. In addition, we prefer embeddings where the density of points in different clusters does not vary significantly, and try to avoid embeddings with a large number of filamentary structures, as these tend to result in clusters that do not match our visual perception (many clustering algorithms are designed to work effectively in a flat geometry, and filamentary structures are a strong departure from this assumption). 

In figure \ref{f:UMAP_with_galaxies} we show our adopted two-dimensional {\sc umap} embedding, where every pixel is colored according to the galaxy it belongs to. In figure \ref{f:UMAP_colorcoded_by_features} we show our adopted embedding color-coded by the features in the input dataset, where strong gradients can be seen in many of the features throughout the embedding.

\subsection{Clustering}\label{sec:methods:clustering}

This section includes the final phase of our three-step process, where we apply a clustering algorithm to the two-dimensional embedding by {\sc umap} to divide the PHANGS pixels into groups. To partition the two-dimensional distribution, we examine several different clustering algorithms: K-means, Birch, OPTICS, DBSCAN, and Hierarchical Clustering (see \citealt{xu15} and \citetalias{baron19} for reviews about clustering algorithms). The different algorithms have different optimization objectives and different stopping criteria, making them sensitive to different aspects of the data. For example, by construction, K-means identifies evenly-sized clusters and can only operate in a flat geometry, while OPTICS and DBSCAN may identify unevenly-sized clusters and can operate in a non-flat geometry. The output of Hierarchical Clustering depends on the assumed linkage method, which defines how clusters are linked to different clusters. We considered the four linkage methods: Ward, complete, average, and single, each being sensitive to different types of structures (see figure 11 in \citetalias{baron19} and related text).

Each of the clustering algorithms has different hyper-parameters, and changing the values of these parameters can have a significant impact on the resulting clusters (see e.g., \citetalias{baron19}). For example, in K-means, Birch, and Hierarchical Clustering, the number of clusters is a hyper-parameter of the algorithm and is set by the user. When applying clustering algorithms directly to high-dimensional datasets, it may be challenging to select the ideal number of clusters\footnote{There are various suggested techniques to select a suitable number of clusters automatically, but these techniques are usually tailored to a specific clustering algorithm, and cannot be applied broadly and generally with all considered clustering algorithms (e.g., \citealt{xu15}).}, since it is not straightforward to visualize the detected clusters in the high-dimensional space. In our case, since the clustering algorithms are applied to the two-dimensional embedding by {\sc umap}, we were able to visualize the resulting clusters, color-coded by different features (figure \ref{f:UMAP_colorcoded_by_features}), and select a suitable number of clusters. The chosen numbers, which were selected to not be larger than $\sim$10, so we can inspect the clusters manually and compare between their properties, and not smaller than $\sim$4, so that objects with different properties will not be grouped together, are 6 for K-means and Hierarchical Clustering, and 7 for Birch. 

\begin{figure*}
	\centering
\includegraphics[width=1\textwidth]{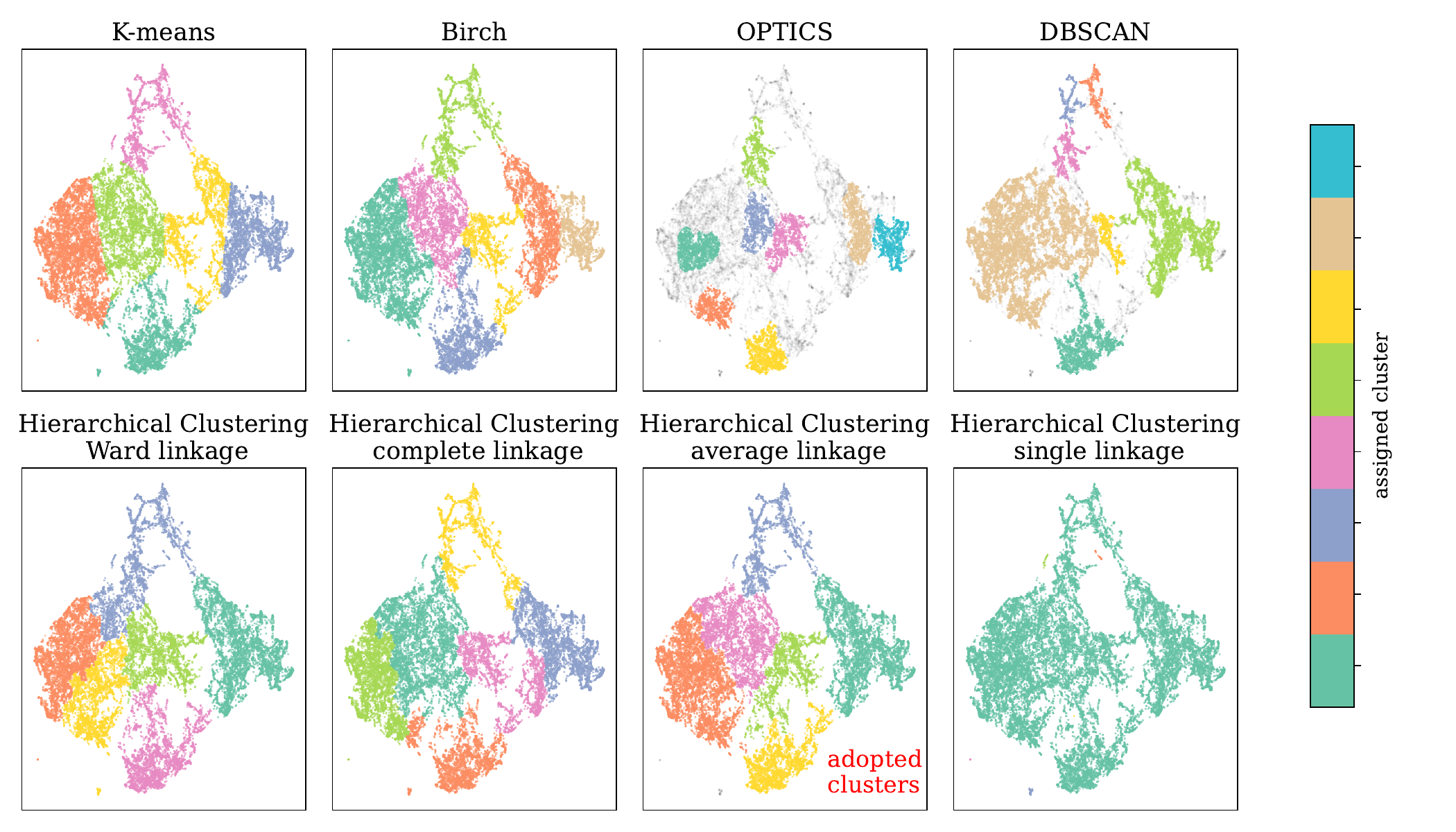}
	\caption{\textbf{Different clustering algorithms applied to the two-dimensional embedding of the PHANGS pixels.} Each panel shows the application of a different clustering algorithm to our adopted two-dimensional embedding by {\sc umap}. In each panel, the points are color-coded according to their assigned cluster (all points for K-means, Birch, and Hierarchical Clustering), or are marked with light gray if they are not clustered (in the case of OPTICS and DBSCAN). The figure shows that while some regions in the two-dimensional map are always considered as separate and well-defined clusters (e.g., the clusters at the bottommost and at the topmost), others are more ambiguous, with different clustering algorithms dividing the objects into different groups. We adopt the clusters identified using Hierarchical clustering with the average linkage.}
\label{f:clustering_alg_variation}
\end{figure*}

While K-means, Birch, and Hierarchical Clustering divide all the points into clusters, OPTICS and DBSCAN may divide only some points into clusters, leaving others unclustered. The hyper-parameters of OPTICS and DBSCAN primarily control the minimum cluster membership and neighborhood size (see e.g., \citealt{xu15}), and we set these parameters to result in roughly 6--8 clusters. For OPTICS, we used the hyper-parameters (\texttt{min\_samples}, \texttt{xi}, \texttt{min\_cluster\_size}) to be (300, 0.001, 0.04). For DBSCAN, we set the hyper-parameters (\texttt{min\_samples}, \texttt{min\_cluster\_size}) to be (10, 300). We used the python {\sc sklearn} library to apply these different clustering algorithms \citep{pedregosa11}.

Figure \ref{f:clustering_alg_variation} shows the different clustering algorithms applied to our adopted two-dimensional embedding by {\sc umap}. One of the methods, Hierarchical Clustering with the single linkage, can be excluded immediately as it clusters the majority of the points into a single cluster, and marks five very small outlier groups as the remaining clusters. We also exclude OPTICS and DBSCAN as they leave a significant fraction of points unclustered, and we wish to include as many objects as possible in the analysis. Next, we exclude K-means, Birch, and Hierarchical clustering with the complete linkage as they cluster together groups that appear separate upon visual inspection (yellow group in K-means, yellow group in Birch, and pink and orange groups in Hierarchical clustering). This leaves Hierarchical clustering with Ward linkage versus the average linkage, both of which roughly agree on three of the clusters (top-most, right-most, and bottom-most), but disagree about the division of points in the left-most over-density of points. Our choice to adopt the average linkage was motivated by the distributions of features seen in figure \ref{f:UMAP_colorcoded_by_features}, where it is apparent that the average linkage divides the points in the left-most region into points with lower PAH 3.3/11.3 \mic~ ratios and points with higher ratios. The results reported in section \ref{sec:results} are therefore based on the clusters identified with Hierarchical clustering with the average linkage method.

\section{Results}\label{sec:results}

\begin{figure*}
	\centering
\includegraphics[width=1\textwidth]{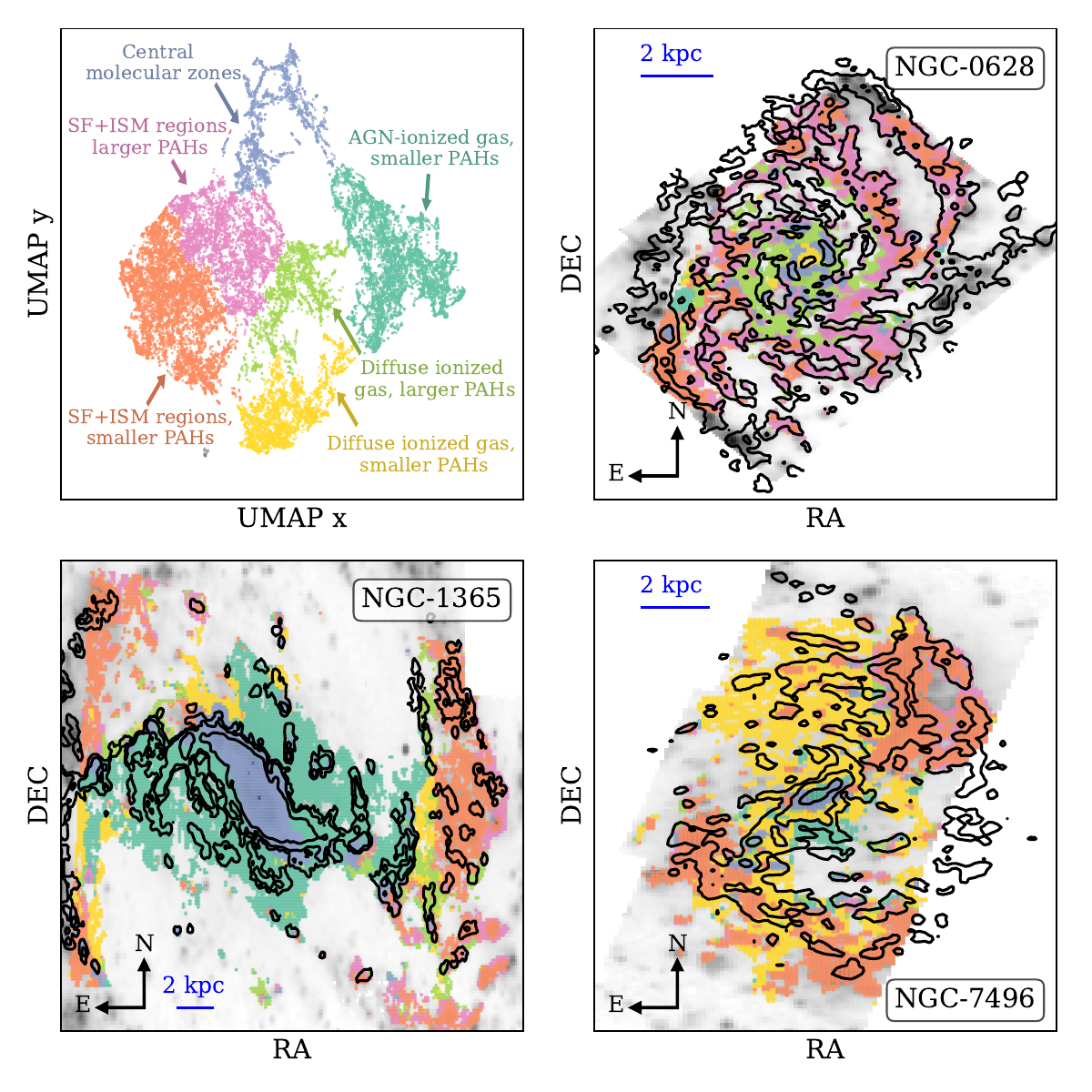}
	\caption{\textbf{PHANGS multi-wavelength pixels partitioned into 6 groups using dimensionality reduction and clustering algorithms.} The top left panel shows the two-dimensional embedding by {\sc umap} of the input dataset, which includes 24\,007 PHANGS pixels with 13 measured features that trace the stellar population, gas, dust, and star formation properties. Each point in the two-dimensional space represents a pixel from one of the galaxies we consider. The objects are divided into groups using the Hierarchical Clustering algorithm with the average linkage, and each point is color-coded according to its assigned group. The other three panels show the spatial distribution of the six groups, which map to large-scale coherent structures within the galaxies. In these panels, the grayscale background represents the \halpha~ surface brightness, and the black contours represent the CO emission. }
\label{f:classes_on_galaxies}
\end{figure*}

We apply dimensionality reduction and clustering algorithms to the PHANGS multi-wavelength dataset of NGC~0628, NGC~1365, and NGC~7496, and divide the pixels into six groups. This dataset is constructed from maps extracted from ALMA, MUSE, and JWST observations, and it traces the properties of the stellar population, multiphase gas, dust, and star formation, on a scale of 150 pc in these galaxies. 

The top left panel of figure \ref{f:classes_on_galaxies} shows our adopted two-dimensional embedding by {\sc umap} (section \ref{sec:methods:UMAP}), where every point represents a pixel from one of the three galaxies. The distribution of points in the two-dimensional space shows several regions with an over-density of points, which may be interpreted as separate clusters, connected to each other through filamentary structures. This highly connected filamentary structure\footnote{We observe the same highly-connected filamentary structure in the two-dimensional embedding when using one pixel per spatial resolution element (150 pc) instead of two (section \ref{sec:data}). This suggests that this structure is not driven by our pixels sub-sampling the 150 pc resolution.}, seen for different {\sc umap} hyper-parameter choices (see \ref{app:umap_parameters}), indicates that the multi-wavelength features of the PHANGS pixels form continuous relations in the high-dimensional space they span. It suggests that the PHANGS pixels do not represent different entities with distinct physical properties (e.g., the difference between a star and a quasar), but rather the same entity with varying physical conditions (e.g., optical spectra of stars with different temperatures). This is not surprising given that the features we constructed trace different physical properties of gas and dust, averaged over a 150 pc scale.

The points in the two-dimensional space are color-coded according to their assigned group using our adopted clustering algorithm (section \ref{sec:methods:clustering}). Since the points do not form well-separated clusters in the two-dimensional space, the resulting groups are somewhat arbitrary, with different clustering algorithms and different hyper-parameter choices changing the resulting groups. Figure \ref{f:clustering_alg_variation} and the top left panel of figure \ref{f:classes_on_galaxies} demonstrate the arbitrary nature of partitioning objects that form a continuous sequence into distinct groups -- there is more than one way to divide the objects into groups, each resulting in groups that are distinct in different aspects. Despite this ambiguity, dividing objects into groups is a common practice in science and in astronomy in particular, with examples ranging from classifying the stellar sequence into distinct stellar types, (O, B, A, F, G, K, M), the classification of core collapse supernovae according to their light-curves or spectra, classification of AGN into type I and II, and more. The practice of dividing objects, even when they form a sequence, into groups is useful as it allows one to compare the properties of objects in different groups, and by that, gain insight into the physics that drive the continuous variation in their properties. The difference of this work is the use of statistical tools to dissect the high-dimensional space, rather than using predefined physics-motivated properties, such as line ratios, metallicity, mass, luminosity, temperature, etc.

In this section, we describe the properties of the adopted groups, showing that they each have distinct gas ionization and PAH properties (section \ref{sec:results:cluster_properties}). We then present newly identified galaxy-wide correlations between PAH band and optical line ratios and use the adopted groups to interpret them (section \ref{sec:results:PAH_optical_lines_ratios_corr}).

\subsection{Distinct gas ionization and PAH properties in the different clusters}\label{sec:results:cluster_properties}

In this section we interpret the six groups using different observables. We use the spatial distributions of pixels in different groups, as well as the feature values and their relation to other features. In particular, we use optical line diagnostic diagrams, PAH band ratios, and the relations between the \halpha, CO, and 10 \mic~ emission.

Figure \ref{f:classes_on_galaxies} shows the spatial distribution of the adopted groups. Although the input dataset did not include information regarding the galaxy a region belongs to, nor information regarding the relative location of a region within the galaxy, the groups map onto large-scale coherent structures within the galaxies.

\begin{figure*}
	\centering
\includegraphics[width=0.9\textwidth]{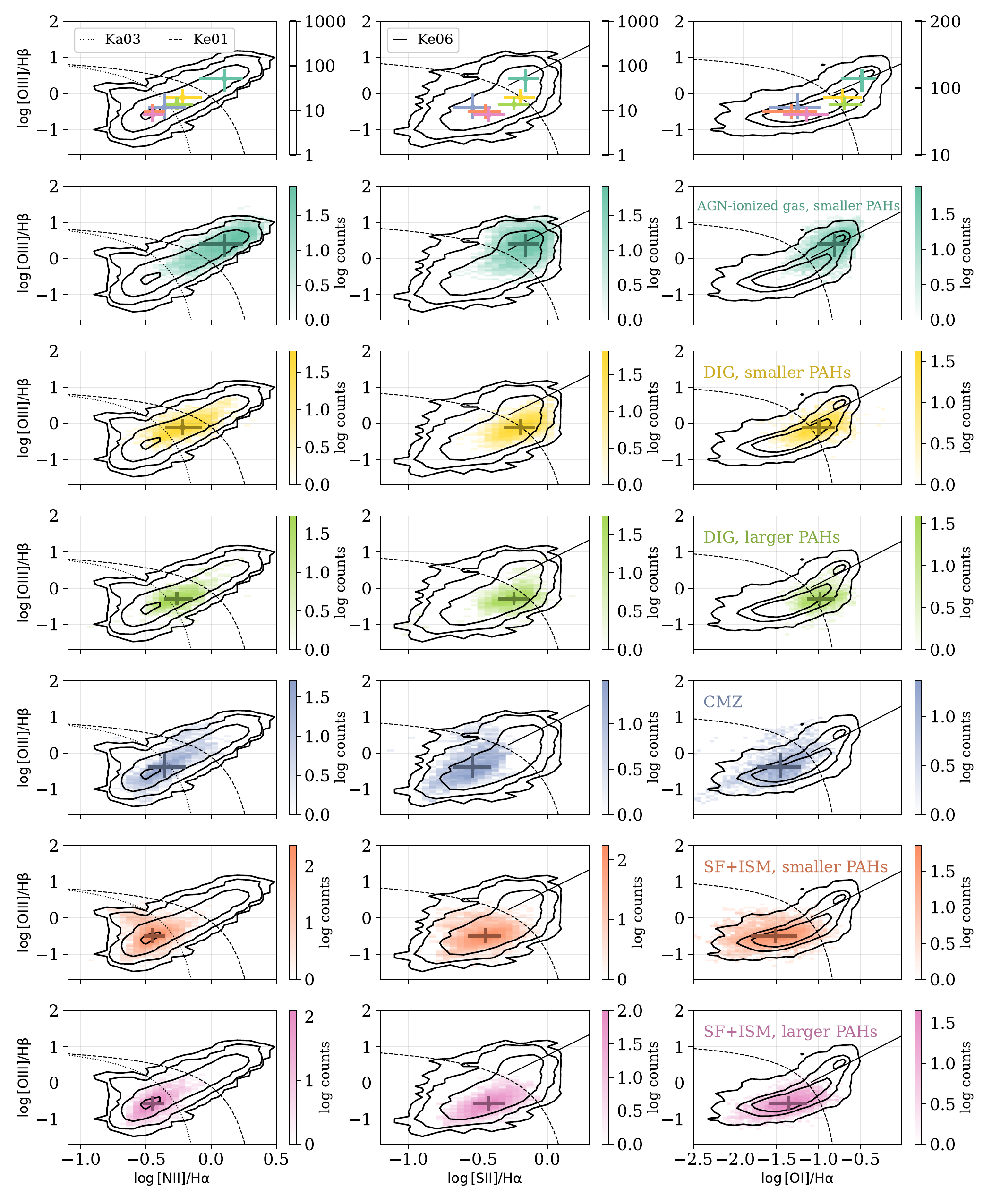}
\caption{\textbf{Location of the identified groups in optical line diagnostic diagrams.} Each row shows the optical line ratios on standard diagnostic diagrams (e.g., \citealt{baldwin81, veilleux87, kewley01}): \oiiihbeta~ versus \niihalpha, \siihalpha, \oihalpha. These diagrams are used to constrain the main source of ionizing radiation. In the left panel, we show the separating criteria by \citet{kewley01} and \citet{kauff03a}, that are used to separate ionization by young massive stars from AGN. In the middle and right panels, we show the LINER-Seyfert separating criteria by \citet{kewley06}. The black contours represent the distribution of line ratios in all the pixels we considered in our analysis. The colormaps represent 2D histograms of the line ratios for a given group, in logarithmic scale. The crosses represent the 16th, 50th, and 84th percentiles of the distributions in each of the band ratios. The different groups show different distributions in these diagnostic diagrams, suggesting different sources of ionizing radiation, as described in the text. Details on the classification of each group can be found in section \ref{sec:results:cluster_properties}.}
\label{f:BPT_diagram_of_clusters}
\end{figure*}

In Figure \ref{f:BPT_diagram_of_clusters} we show the distribution of the groups in standard optical line diagnostic diagrams (BPT diagrams; \citealt{baldwin81, veilleux87, kewley01}). These diagrams are used to constrain the main source of ionizing radiation, by classifying a set of emission lines into one of the three classes: (i) HII regions, where the ratios are consistent with ionization by O and early B-type stars, (ii) Seyfert, where the ratios are consistent with ionization by AGN, and (iii) LINER/LIER, where the ratios may be consistent with either AGN ionization, photoionization by hot and evolved stars, or shock-excited gas (e.g., \citealt{kewley01, kauff03a, kewley06, allen08, cidfernandes10, rich11, rich15, belfiore22}). 

\begin{figure*}
	\centering
\includegraphics[width=0.95\textwidth]{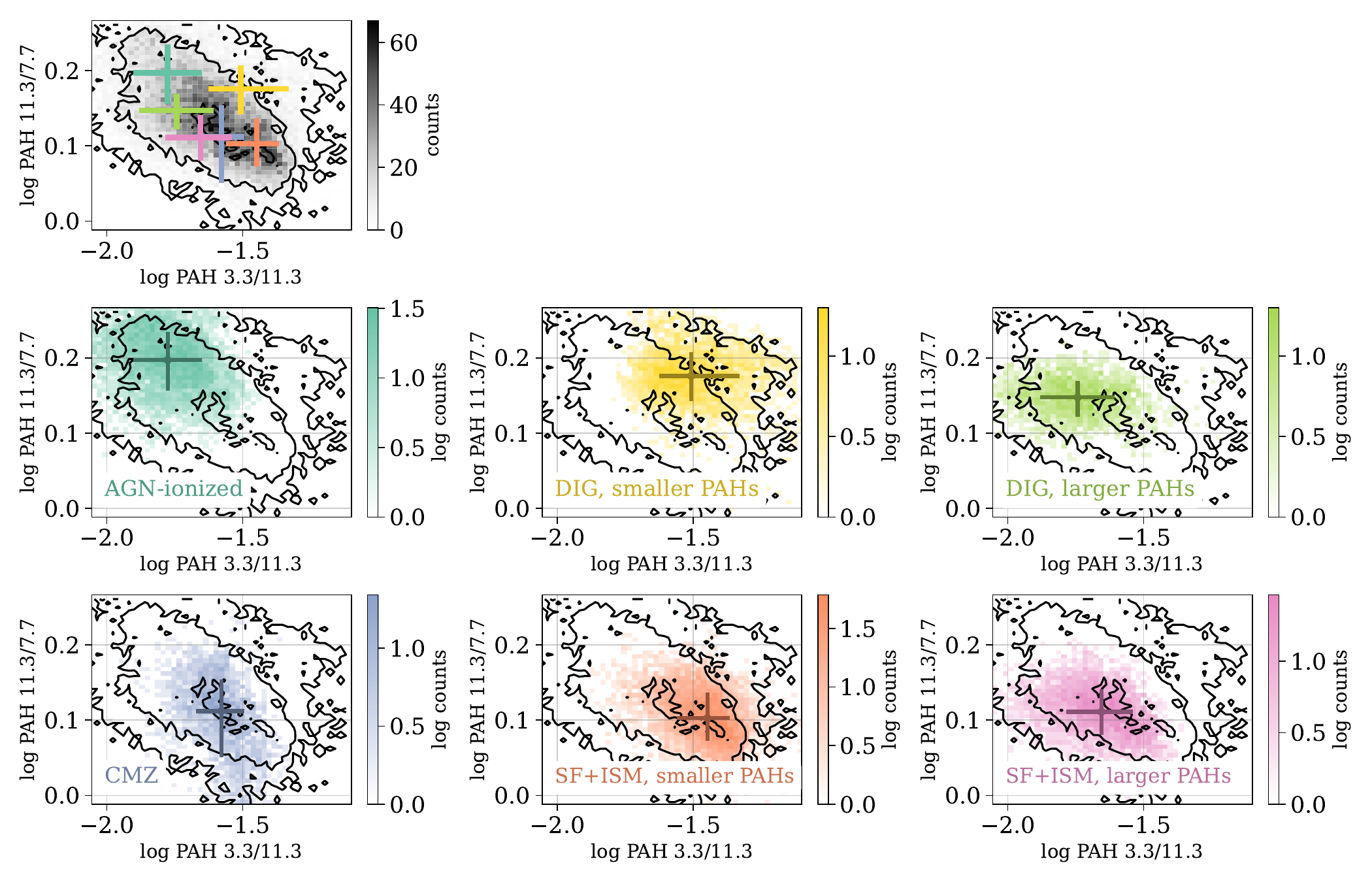}
\caption{\textbf{Distribution of the identified groups in the PAH band ratios plane.} The panels show the distribution of pixels in the PAH 11.3/7.7 \mic~ versus 3.3/11.3 \mic~ plane. The top left panel shows the distribution of all the pixels we consider using gray-scale color-coding, and the rest of the panels show the distributions in each individual group. While the gray-scale colormap on top represents the 2D histogram counts in a linear scale, the individual group panels show the counts in a logarithmic scale. The black contours represent the distribution of all the pixels we consider, and they are the same in the different panels. The crosses represent the 16th, 50th, and 84th percentiles of the distributions in each of the band ratios for each group. Details on the classification of each group can be found in section \ref{sec:results:cluster_properties}.}
\label{f:log_PAH_11p3-7p7_versus_PAH_3p35-11p3}
\end{figure*}

In figure \ref{f:log_PAH_11p3-7p7_versus_PAH_3p35-11p3} we show the distribution of the groups in the PAH 11.3/7.7 \mic~ versus 3.3/11.3 \mic~ plane (hereafter PAH 11.3/7.7 and 3.3/11.3 ratios). These bands ratios are sensitive to the ionized fraction of PAHs, the PAH size distribution, and the shape of the incident FUV-optical radiation (e.g., \citealt{draine21, rigopoulou21} and section \ref{app:pah_models}).

Figures \ref{f:BPT_diagram_of_clusters} and \ref{f:log_PAH_11p3-7p7_versus_PAH_3p35-11p3} show that the different groups have distinct ionized gas and PAH properties, as probed by the different optical line and PAH band ratios. Below, we describe these general properties\footnote{The order in which we describe these clusters also corresponds to the order in which they are presented in all the figures from this point forward.}: 

 \textbf{(1) AGN-photoionized gas (turquoise group)}: These pixels can primarily be found in NGC~1365 and NGC~7496\footnote{Some of the pixels of this group belong to NGC~628. These pixels represent the very few pixels that are located in HII regions in the BPT diagram that were assigned to this cluster. In the two-dimensional embedding, they are located at the intersection with the CMZ cluster described below. Their properties are more in line with those of pixels in the CMZ cluster, and we believe that they should not belong to the AGN group.}, both of which host AGN in their center (e.g., \citealt{morganti99}). The optical line ratios place most of the pixels in the Seyfert/LINER region in the BPT diagram, with \oiiihbeta~ ratios too high to be powered solely by hot and evolved stars (e.g., \citealt{cidfernandes10, byler19, belfiore22}). The pixels are located primarily along the AGN ionization cones, suggesting that the gas is photoionized by the AGN. Some of the pixels in this group are in the bars of NGC~1365, and these pixels show properties intermediate between the AGN-photoionized group and the CMZ group below. Interestingly, other clustering algorithms (e.g., BIRCH and OPTICS in figure \ref{f:clustering_alg_variation}) divide this group into two -- one that corresponds to pixels in the bars and the other to pixels within the AGN ionization cones. From here forward, when referring to this group, we will focus on the pixels that are within the AGN ionization cones, located on kpc scales, in the bulge that is dominated by an older stellar population. The pixels in this group show the highest PAH 11.3/7.7 band ratio, and the lowest PAH 3.3/11.3 band ratio. Importantly, while the AGN seems to be dominating the ionizing radiation, resulting in the Seyfert-like line ratios, the old stellar population in the bulge probably dominates the FUV-optical radiation, and is thus responsible for the PAH heating.

\noindent \textbf{(2+3) Diffuse ionized gas (yellow and green groups)}: These pixels are seen in all three galaxies, and they primarily probe regions with faint \halpha~ and CO emission (e.g., figures \ref{f:log_Ha_10mic_versus_f_Ha} and \ref{f:log_Ha_CO_versus_I_CO} in the appendix). The optical line ratios are consistent with LINER/LIER-like emission. In addition, these pixels spatially-coincide with regions identified by \citet{belfiore22} as regions dominated by diffuse ionized gas, where the gas is ionized by a combination of radiation leaking from HII regions along with emission from hot and evolved stars, called the HOLMES mixing sequence. Both of the groups show quite high 11.3/7.7 PAH band ratio. The two groups differ in their 3.3/11.3 PAH band ratio, with the yellow group showing significantly larger values than those of the green group.

\noindent \textbf{(4) Central molecular zone (CMZ; slate blue group)}: Most of the pixels in this group belong to the central part of NGC~1365, a region known for hosting extreme star formation that is powered by a massive molecular gas reservoir (see \citealt{schinnerer23}; and an overview by \citealt{henshaw23}). Interestingly, some of the central pixels of NGC~0628 and NGC~7496 are also assigned to this group. The optical line ratios place the pixels in the HII region of the BPT diagram, suggesting ionization by young massive stars. This group differs from the two SF+ISM groups below primarily due to its very bright CO emission (figure \ref{f:log_Ha_CO_versus_I_CO} in the appendix), in particular with respect to the observed \halpha, and significant dust extinction. It shows quite a low 11.3/7.7 PAH band ratio, consistent with those observed in the SF+ISM groups below, and an intermediate 3.3/11.3 ratio, in between those of the SF+ISM groups. Some of the pixels of this group deviate from the strong correlation between 3.3/7.7 and 3.3/11.3 (see figure \ref{f:log_PAH_3p35-7p7_versus_PAH_3p35-11p3} in the appendix), which may suggest the presence of Silicate 9.7 \mic~ absorption that reduces the observed flux in F1130W.

\begin{figure*}[ht]
	\centering
\includegraphics[width=1\textwidth]{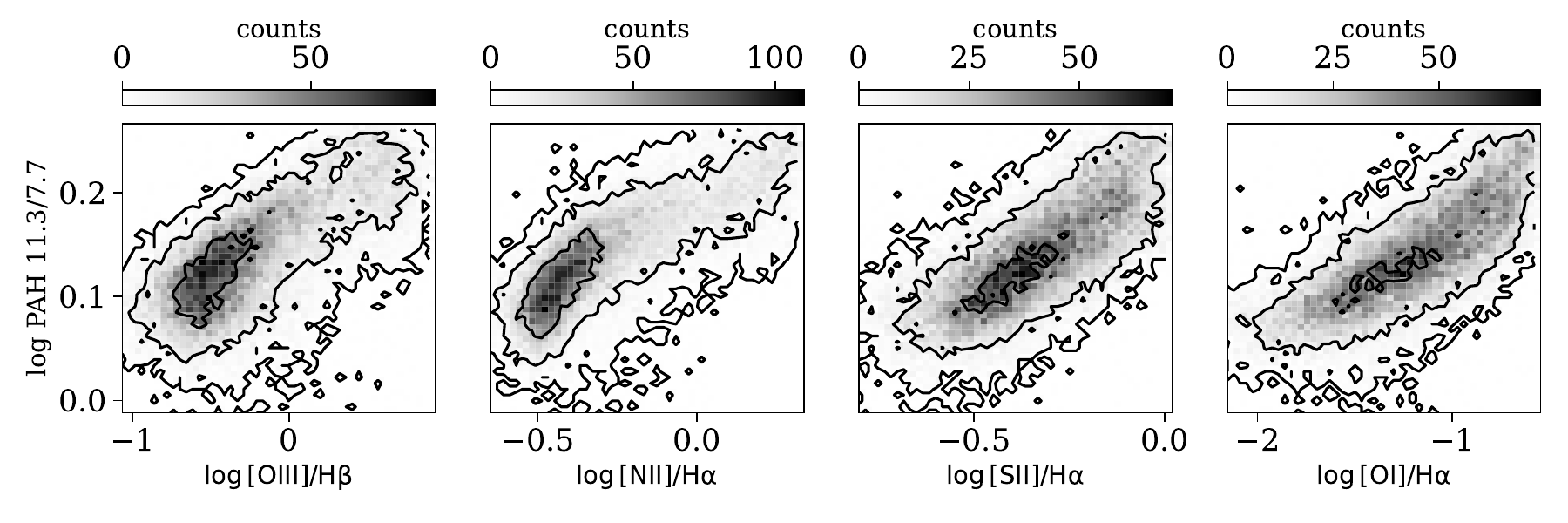}
\caption{\textbf{PAH band ratio $\log (11.3/7.7)$ versus optical line ratios.} The different panels show the distribution of the PAH band ratio $\log (11.3/7.7)$ versus the optical line ratios \oiiihbeta, \niihalpha, \siihalpha, and \oihalpha, for all the PHANGS pixels.}
\label{f:PAH_11p3_7p7_versus_optical_line_ratios_all}
\end{figure*}

\begin{figure*}[ht]
	\centering
\includegraphics[width=1\textwidth]{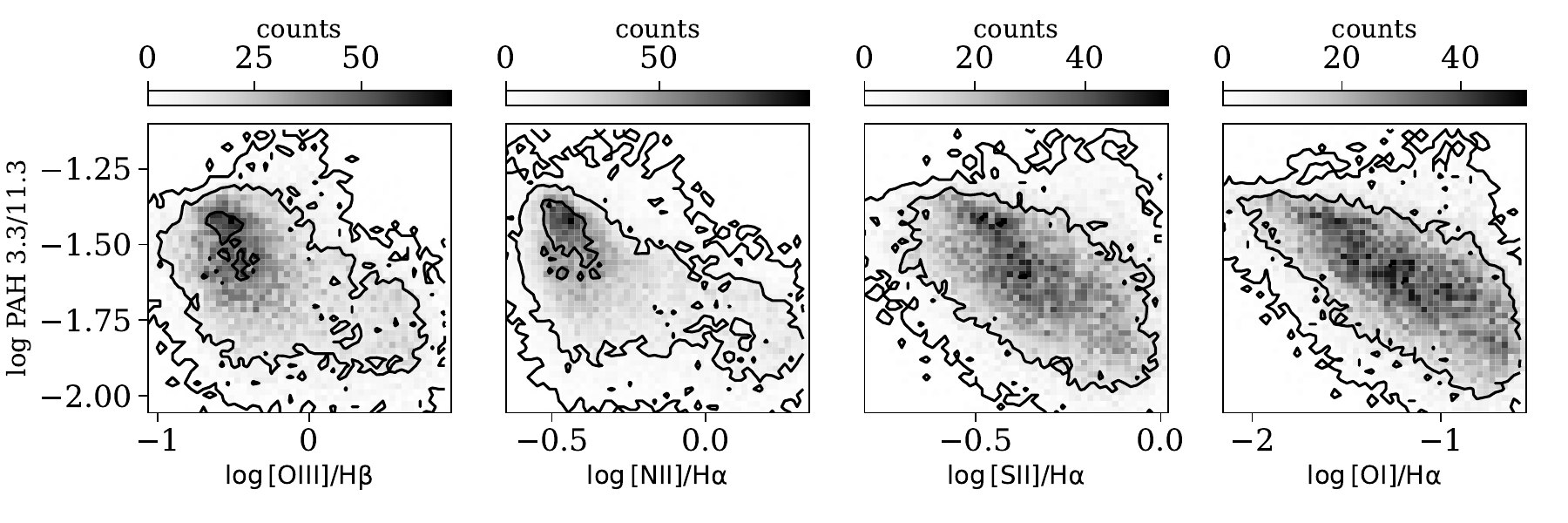}
\caption{\textbf{PAH band ratio $\log (3.3/11.3)$ versus optical line ratios.} The different panels show the distribution of the PAH band ratio $\log (3.3/11.3)$ versus the optical line ratios \oiiihbeta, \niihalpha, \siihalpha, and \oihalpha, for all the PHANGS pixels.}
\label{f:PAH_3p3_11p3_versus_optical_line_ratios_all}
\end{figure*}

\noindent \textbf{(5+6) Star-forming regions and ISM (orange and pink groups)}: These pixels are primarily seen in the spiral arms of the three galaxies. Their \halpha, CO, and 10 \mic~ emission show strong correlations, suggesting standard star formation that is powering the observed \halpha~ and 10 \mic~ emission. The optical line ratios are consistent with ionization by young massive stars. Similarly to the CMZ group, they show comparable and low 11.3/7.7 PAH band ratios. The first difference between the two groups is in their 3.3/11.3 PAH band ratio, which is significantly larger in the orange group than in the pink group. In addition, the orange group has a higher \halpha/CO ratio than the pink group, which may suggest that it corresponds to more dispersed and evolved clouds compared to those of the pink group. 

The three galaxies we consider dominate different regions of the two-dimensional embedding by {\sc umap} (see figure \ref{f:UMAP_with_galaxies}). This can be explained in the context of the groups we identify -- (i) NGC~1365 is unique in having a large number of pixels that trace the CMZ, the bar, and AGN-photoionized gas on kpc scales. It therefore dominates two out of the six groups.  (ii) The diffuse ionized gas in NGC~7496 differs from that of NGC~0628 and NGC~1365 in PAH band ratios, thus dominating one of the groups. (iii) The star forming and ISM regions of NGC~0628 differ from from those NGC~1365 and NGC~7496 in PAH band ratios. It therefore dominates another group. At this point, it is unclear whether adding the rest of the 19 PHANGS galaxies would fill-in the space, making pixels that originate from a given galaxy indistinguishable from pixels of other galaxies. For that, it may be necessary to correct some of the features for inclination. We plan to address this question in a future publication where we plan to include all 19 PHANGS galaxies. \vspace{2cm}

\subsection{Close connection between the heating of PAHs and the ionization of the warm ionized gas}\label{sec:results:PAH_optical_lines_ratios_corr}

We identify significant and tight correlations between different PAH band and optical line ratios (see the full correlation matrix in Figure \ref{f:correlation_matrix_with_labels} in the appendix). These correlations are seen across the entire dataset, extending from the star-forming regions and the ISM, through the diffuse ionized gas, to the AGN-photoionized gas. The correlations are also detected in individual groups in which the dynamical range is large enough. Ionizing radiation is expected to destroy PAHs, and observations suggest much weaker PAH emission in regions dominated by ionized gas, such as HII regions (e.g., \citealt{chastenet19, chastenet23a, chown23, egorov23, lee23, peeters23}; and reviews \citealt{tielens08, li20}), though at our spatial resolution, the PHANGS pixels include contributions from both. These correlations suggest a strong connection between the heating of PAHs and the ionization of the warm ionized gas on 150 pc scales.

In Figure \ref{f:PAH_11p3_7p7_versus_optical_line_ratios_all} we show the PAH band ratio 11.3/7.7 versus the optical line ratios \oiii/\hbeta, \nii/\halpha, \sii/\halpha, and \oi/\halpha, for the PHANGS pixels considered in our analysis. The 11.3/7.7 band ratio shows strong correlations with all of them. In figure \ref{f:PAH_3p3_11p3_versus_optical_line_ratios_all} we show the PAH band ratio 3.3/11.3 versus the optical line ratios. The correlations show a larger scatter than those of the 11.3/7.7 band ratio, but they extend over twice as large a dynamical range. In addition, there is a clear difference in the relation seen for the lower ionization transitions traced by \sii/\halpha~ and \oi/\halpha~ compared to the higher ionization transitions \nii/\halpha~ and \oiii/\hbeta, with the former showing stronger correlations with the 3.3/11.3 band ratio.

Since the PAH band ratios are based on the broad band filter ratios F1130W/F770W and F335M$\mathrm{_{PAH}}$/F1130W, we first ensure that these correlations are not due to varying contributions of dust continuum emission to the F1130W filter flux. Since the F1000W filter likely has a large contribution from PAHs under most conditions in the PHANGS galaxies (e.g., \citealt{leroy23}), we use the F2100W filter flux to trace the dust continuum emission. If the observed correlations are due to a varying contribution of hot dust emission, we expect to find significant correlations between F2100W/F770W and F335M$\mathrm{_{PAH}}$/F2100W and the optical line ratios (F2100W replaces F1130W). We find no such correlations. Therefore, in what follows, we assume that the change in F1130W/F770W and F335M$\mathrm{_{PAH}}$/F1130W band ratios trace changes in PAH emission. 

Various PAH band ratios, including those we consider here, have been used to place constraints on the PAH size and charge distribution in different environments (see reviews by \citealt{tielens08, li20}). Theoretical calculations and laboratory measurements suggest that neutral PAHs have significantly different infrared spectra than ionized PAHs, with the former showing strong 3.3 \mic~ and 11.3 \mic~ bands compared to the 7.7 \mic~ band, and the latter showing stronger 7.7 \mic~ band compared to the 3.3 \mic~ and 11.3 \mic~ bands (e.g., \citealt{allamandola99, tielens08}). Therefore, the 11.3/7.7 PAH band ratio has been used extensively as a PAH ionization diagnostic, and to a lesser extent, as a PAH size diagnostic (see below; e.g., \citealt{kaneda05, smith07, galliano08, diamond_stanic10, vega10, lai22, chastenet23b, dale23}). Since smaller PAHs have smaller heat capacities than larger PAHs, stochastic heating by single photon absorption raises their peak temperature to higher values (e.g., \citealt{draine01, draine11}), leading to higher luminosity in shorter wavelength bands such as 3.3 \mic~ compared to longer wavelength bands such as 11.3 \mic. Therefore, PAH band ratios such as 3.3/11.3, and to a lesser extent, the 11.3/7.7, have been used as PAH size diagnostics (e.g., \citealt{smith07, galliano08, lai23, chastenet23b, dale23, ujjwal24}).

\begin{figure*}
	\centering
\includegraphics[width=1\textwidth]{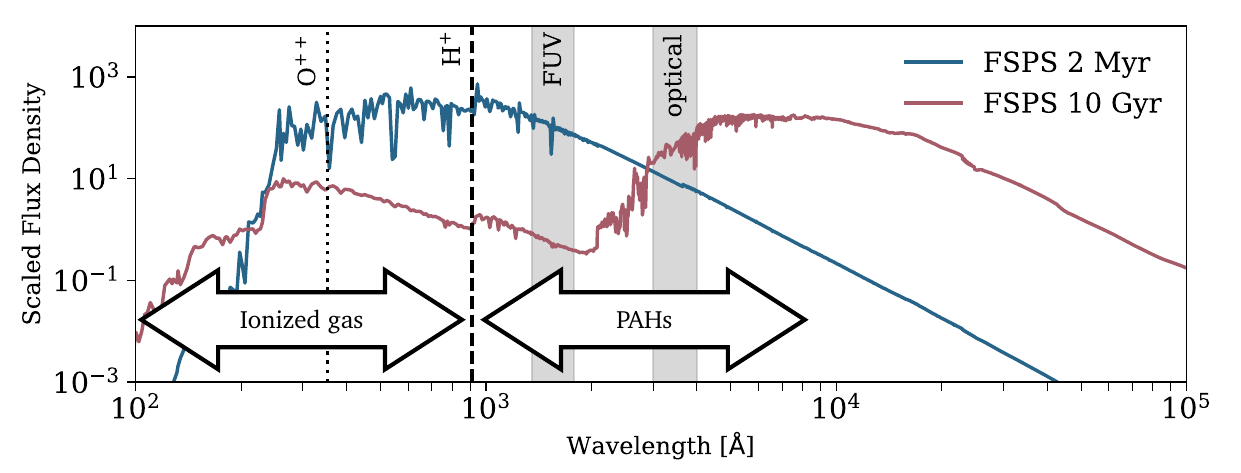}
\caption{\textbf{PAHs and ionized gas are sensitive to different parts of the illuminating radiation field.} The lines represents two example SEDs calculated using the flexible stellar population synthesis code by \citet{conroy09}. They correspond to single stellar populations of age 2 Myr and 10 Gyr (see Appendix \ref{app:seds} for additional details), where the flux density of the 10 Gyr star has been scaled to higher values for representational purposes. The black vertical lines represent the wavelengths that correspond to photon energies that can ionize Hydrogen (H$^{+}$; dashed) and Oxygen twice (O$^{++}$; dotted). The gray bands correspond to the wavelength ranges 1350--1780 \AA\, and 3000-4000 \AA, which are our adopted definitions for FUV and optical luminosities. The conditions in the warm ionized gas primarily depend on the flux densities at $\lambda \leq 912$ \AA. The PAHs are exposed to non-ionizing radiation, and their heating depends on the shape of the FUV-optical radiation field, which we parametrize using $\nu L_{\nu}(\mathrm{FUV})/\nu L_{\nu}(\mathrm{optical})$ (see Appendix \ref{app:pah_models} for details).}
\label{f:SED_general_explanation_edit}
\end{figure*}

\citet{draine21} and \citet{rigopoulou21} showed that various PAH band ratios are also sensitive to the shape of the incident radiation field. The same population of PAHs may show significantly different 11.3/7.7 and 3.3/11.3 band ratios when heated by radiation that is dominated by young versus older stars. Harder far-ultraviolet (FUV)-optical radiation field, typical of young and massive stars, is expected to lead to hotter PAHs, resulting in higher luminosity in the 3.3 \mic~ feature, and to a lesser extent, the 7.7 \mic~ feature, compared to the 11.3 \mic. Since the PHANGS 150 pc-sized pixels trace a variety of radiation fields, this effect must be taken into account when interpreting PAH band ratios. \citet{dale23} demonstrated that, in the PAH band ratio plane available for the PHANGS galaxies (3.3/11.3 \mic~ versus 3.3/7.7 \mic; see their figure 3), the impact of varying PAH size distribution and a varying radiation field are degenerate with each other. In this work, we use the observed optical line ratios to break this degeneracy.

\subsubsection{A varying radiation field as the main driver of the correlations}\label{sec:results:PAH_optical_lines_ratios_corr:interp}

To interpret the observed correlations, we use our adopted groups, as well as the PAH emission models by \citet{draine21} and a range of spectral energy distributions (SEDs) of the incident radiation field (see details in \ref{app:PAH_gas_models_interp}). \citet{draine21} calculated the infrared emission spectra of PAHs for different illuminating radiation SEDs, radiation intensities, and PAH size and charge distributions. To remove the underlying continuum emission from starlight or other hot, small grains, they employed a clipping scheme that focuses on strong PAH emission features. Using this clipping scheme, they studied the sensitivity of PAH band ratios to the SED of the illuminating radiation and the PAH size and charge distribution. 

We use the infrared spectra predicted by \citet{draine21}. In particular, we consider models calculated using the small, standard, and large PAH size distributions, and with low, standard, and high ionization distributions. For the illuminating radiation SED, we consider 12 different templates, covering a range of stellar population ages and FUV luminosities. The PAHs are believed to be located in regions that are shielded from ionizing radiation (e.g., \citealt{chastenet19, chastenet23a, egorov23, lee23}; and reviews \citealt{tielens08, li20}), and their heating is dominated by non-ionizing FUV-optical radiation (see figure \ref{f:SED_general_explanation_edit}). To parameterize the dependence of PAH band ratios on the incident radiation field, we define the FUV-to-optical luminosity ratio to be $\nu L_{\nu}(1350-1780\,\mathrm{\AA}) / \nu L_{\nu} (3000-4000\,\mathrm{\AA})$, as illustrated in figure \ref{f:SED_general_explanation_edit}. 

\begin{figure*}
	\centering
\includegraphics[width=1\textwidth]{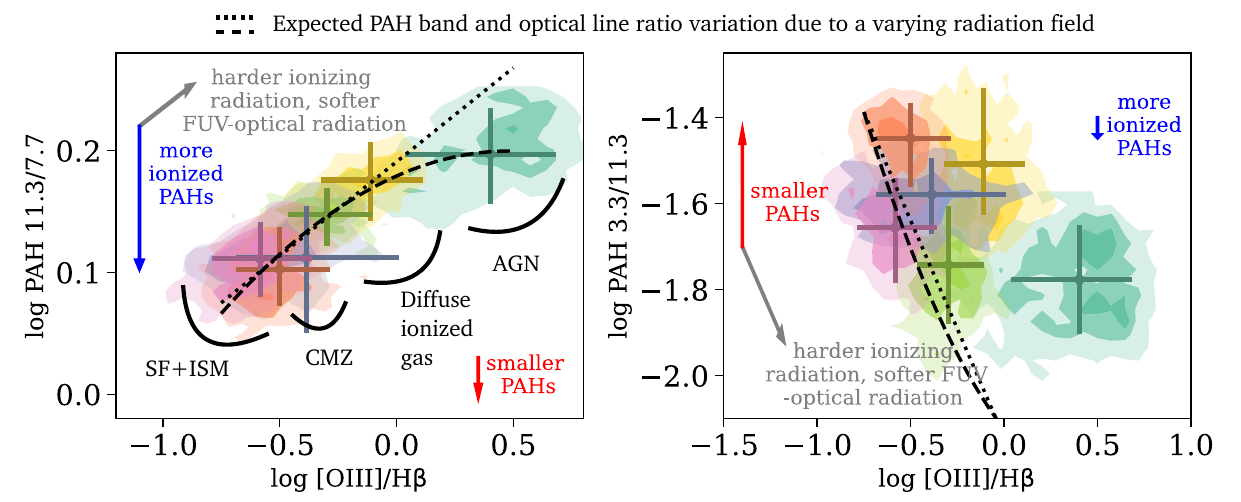}
\caption{\textbf{Interpreting the relation between PAH band ratios and \oiiihbeta.} The panels show the PAH band ratios $\log (11.3/7.7)$ and $\log (3.3/11.3)$ versus \oiiihbeta, where each group is plotted separately. The contours represent the 2D distribution for each group, and the crosses represent the 16th, 50th, and 84th percentiles. The blue and red arrows represent the expected change in PAH band ratios when changing the PAH size and charge distribution (see Appendix \ref{app:pah_models}). The gray arrows represent the expected change in PAH band and \oiiihbeta~ ratios assuming that the PAHs and the ionized gas are exposed to different parts of the same varying radiation field. The varying radiation field is characterized by harder ionizing radiation and softer FUV-optical slope, a scenario that is expected under a wide range of assumptions regarding what powers it. The relation on the left panel is consistent with the ``common varying radiation field'' interpretation, without the need to invoke PAHs with different charge distributions. The dashed and dotted black lines in the left panel represent two polynomial fits to the 11.3/7.7 versus \oiii/\hbeta~ relation. These best fits are then propagated to the expected relations between 3.3/11.3 and \oiii/\hbeta~ (right panel), assuming the ``common varying radiation field'' interpretation and the PAH models by \citet{draine21}. Clusters that deviate from these lines in the right panel may represent PAHs with different size distributions. The dotted and dashed lines are given by the following equations; Left panel ($y = \log $ PAH 11.3/7.7, $x = \log $ \oiii/\hbeta): $y = 0.15x + 0.19$, $y = -0.087x^2 + 0.084x + 0.18$. Right panel ($y = \log $ PAH 3.3/11.3, $x = \log $ \oiii/\hbeta): $y = -x -2.14$, $y = 0.57x^2 -0.55 -2.12$. }
\label{f:PAH_optical_interpretation_oiii_hbeta}
\end{figure*}

We use the clipping scheme by \citet{draine21} to estimate the PAH band ratios 11.3/7.7 and 3.3/11.3, and parameterize in Appendix \ref{app:pah_models} how they vary for different FUV-to-optical ratios, PAH size distributions, and PAH charge distributions. Since the JWST photometry includes contributions from non-PAH emission sources, and since the photometric bands do not coincide with the clipping points defined by \citet{draine21}, we do not expect the absolute PAH band ratios to match those predicted by the models. Instead, we use the clipping scheme only to compare the observed \textit{trends in ratios} with those predicted by the models. \citet{dale23} integrated the \citet{draine21} models under the JWST filter bandpasses, allowing them to compare the absolute values of the PAH band ratios to those predicted by the model. The observed PAH band ratios in the regions they consider (for the same three PHANGS galaxies) is within the ranges spanned by the models. In particular, they find that the PAH band ratios are more aligned with models of larger and ionized PAHs in compact stellar clusters and stellar associations.

In figure \ref{f:PAH_optical_interpretation_oiii_hbeta} we show the PAH band ratios 11.3/7.7 and 3.3/11.3 versus \oiii/\hbeta~ ratio, using our adopted groups. We indicate on the figure the expected changes in the PAH band ratios when changing the PAH size or charge distribution. Since these changes affect only the PAH bands and have no effect on the optical line ratios, they form arrows in the vertical direction only. The relative sizes of the arrows represent the expected change in PAH band ratios when changing to size/ionization from low to standard, or from standard to large/high. We also show the expected mutual variation in the PAH band and optical line ratios under the assumption that the PAHs and ionized gas are exposed to different parts of the same spatially varying radiation, as described later in this section. 

\noindent \textbf{``Non-varying radiation field" interpretation:} Under the assumption that the radiation field spectrum heating the PAHs is not varying, figure \ref{f:PAH_optical_interpretation_oiii_hbeta} suggests that PAHs are more ionized in regions with low optical line ratios. In particular, using our groups, PAHs are more ionized in star-forming regions and ISM, and are more neutral in the diffuse ionized gas and in the AGN-photoionized gas. The right panel of the figure shows a negative correlation between 3.3/11.3 and the optical line ratio, suggesting PAHs are smaller in regions with low optical line ratios. This would suggest smaller PAHs in the star-forming and ISM regions, and larger PAHs in the diffuse ionized gas. These are in line with the conclusions of \citet{ujjwal24} for the three PHANGS galaxies we consider. However, the 3.3/11.3 shows a strong positive correlation with the \halpha/CO ratio (see left panel of figure \ref{f:interp_PAH_sizes_wrt_Ha_CO_edit}. According to this interpretation, it would suggest that PAHs are smaller in regions with larger fraction of ionized-to-molecular gas. This is the opposite of what would be expected if smaller PAHs are destroyed by ionizing radiation more efficiently than larger PAHs. 

\noindent \textbf{``Common varying radiation field'' interpretation:} The left panel of figure \ref{f:PAH_optical_interpretation_oiii_hbeta} shows that the tight correlation between the PAH band ratio 11.3/7.7 and the optical line ratios is in fact a sequence in ionized gas conditions, with the different groups occupying distinct ranges within the correlation. It may suggest that the PAHs and the ionized gas are exposed to \textit{different parts} of the same radiation field. That is, while the ionized gas is exposed to the entire radiation field, and its properties are set by the \textit{ionizing part} of the radiation, the PAHs are located in regions shielded from the ionizing radiation, and are exposed only to the \textit{non-ionizing FUV-optical part} of the radiation (see figure \ref{f:SED_general_explanation_edit}). Under this interpretation, the radiation field varies over kpc scales and its variation drives the changes in both PAH band and optical line ratios. The PHANGS pixels show significant variations in their stellar and/or AGN properties, with some regions dominated by young stellar populations, while others dominated by old stars or AGN radiation. Therefore, we argue that this scenario is \emph{unavoidable}. The only question is its impact on the PAH band ratios compared to those of varying PAH size and charge distributions. 

\begin{figure*}
	\centering
\includegraphics[width=1\textwidth]{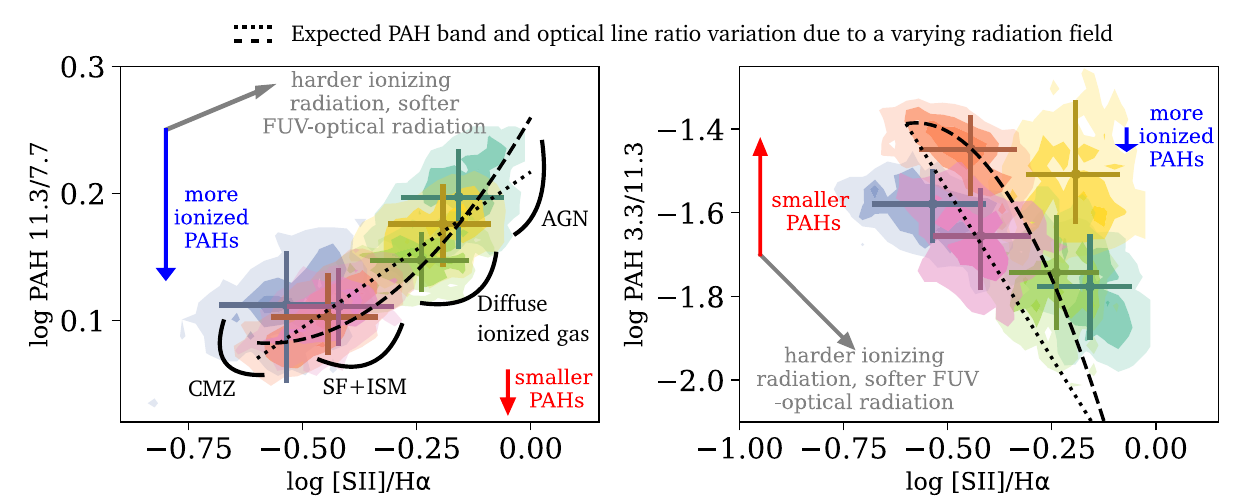}
\caption{\textbf{Interpreting the relation between PAH band ratios and \siihalpha.} Similar to figure \ref{f:PAH_optical_interpretation_oiii_hbeta}, but using the lower ionization \siihalpha~ line ratio.
The dotted and dashed lines are given by the following equations; Left panel ($y = \log $ PAH 11.3/7.7, $x = \log $ \sii/\halpha): $y = 0.22x + 0.21$, $y = 0.53x^2 + 0.61x + 0.26$. Right panel ($y = \log $ PAH 3.3/11.3, $x = \log $ \sii/\halpha): $y = -1.6x -2.34$, $y = -3.5x^2 -4.02 -2.54$. }
\label{f:PAH_optical_interpretation_sii_halpha}
\end{figure*}

The gray arrows in figure \ref{f:PAH_optical_interpretation_oiii_hbeta} represent the expected relations between the PAH band and optical line ratios under the ``common varying radiation field'' interpretation. The expected relation depends on several factors: (i) the dependence of PAH bands on the FUV-to-optical luminosity ratio, (ii) the SED shape, in particular, the connection between the FUV-optical and ionizing parts of the radiation field, and (iii) the dependence of optical line ratios on the shape of the ionizing radiation.

We explore the dependence of PAH bands on the FUV-to-optical luminosity ratio using the \citet{draine21} models in Appendix \ref{app:pah_models}. In particular, we find that varying the radiation field from old (1--10 Gyr) to young (2 Myr) stellar populations increases the FUV-to-optical luminosity ratio by 2.5 dex, and at the same time, changes the PAH band ratios 11.3/7.7 and 3.3/11.3 by -0.1 dex and 0.55 dex respectively. Interestingly, these are quite similar to the ranges spanned by 11.3/7.7 and 3.3/11.3 in the PHANGS pixels.

We explore the connection between the FUV-optical and ionizing parts of the radiation field in Appendix \ref{app:seds}. We use several different stellar libraries, covering single stellar populations with ages of 2 Myr to 10 Gyr, while varying the metallicity, stellar isochrones, and stellar templates. Since figure \ref{f:PAH_optical_interpretation_oiii_hbeta} suggests a sequence that covers HII regions, the diffuse ionized gas, and AGN-photoionized gas, we also consider models that are constructed to reproduce the optical line ratios in the diffuse ionized gas and AGN-photoionized gas. In particular, we consider the HOLMES mixing sequence introduced by \citet{belfiore22} to explain the observed optical line ratios in the diffuse ionized gas in PHANGS galaxies. In this model, the SED is a combination of radiation from a young stellar population leaking from HII regions and from hot and evolved stars, where the latter is used as it contributes the hard ionizing radiation required to power LINER-like emission line ratios. We also construct AGN+SF mixing sequences. We use a standard \citet{shakura73} accretion disk SED with several improvements (general relativistic corrections and radiative transfer in the disk atmosphere; e.g., \citealt{slone12}), which results in Seyfert-like line ratios in standard line diagnostic diagrams (see Appendix A in \citealt{baron19b}). Similarly to the HOLMES mixing sequence, the AGN SED is mixed with a young (2 Myr) stellar template by varying the relative contribution of each. 

Using these different SEDs, in Appendix \ref{app:seds} we confirm that in all of these cases, i.e., stellar age sequence from young to old; HOLMES mixing sequence with varying contribution of old versus young stars; and AGN+SF mixing sequence with varying contributions from the stars versus the AGN, the general observed trend is a harder ionizing slope for a softer FUV-optical slope, matching the direction required to explain the relation in the left panel of figure \ref{f:PAH_optical_interpretation_oiii_hbeta}. Moreover, we find that a decrease of 2.5 dex in $\nu L_{\nu}(\mathrm{FUV})/\nu L_{\nu}(\mathrm{optical})$ results in an increase of $\sim$1 dex in $\nu L_{\nu}(\mathrm{O^{++}})/\nu L_{\nu}(\mathrm{H^{+}})$, depending on the mixing sequence used. Therefore, the expected slope as indicated by the gray arrow is roughly consistent with the observed slope in the left panel of figure \ref{f:PAH_optical_interpretation_oiii_hbeta}.

The exact slope expected under the ``common varying radiation field'' interpretation also depends on the relation between the ionizing radiation and optical line emissivities. In particular, the optical line ratios we consider are measured with collisionally excited transitions whose line emissivity depends on the gas temperature (e.g., \citealt{osterbrock06}). Harder ionizing radiation leads to higher electron temperature (e.g., \citealt{garnett92, byler17}). Both of these are expected to increase the expected optical line ratios, where the line emissivity depends linearly on ionic abundances (O$^{++}$, S$^{+}$, etc) and exponentially on the electron temperature (e.g., \citealt{osterbrock06}). Indeed, we find equally-strong correlations between 11.3/7.7 and the low-ionization transitions \nii/\halpha, \sii/\halpha, and \oi/\halpha~ (figures \ref{f:PAH_11p3_7p7_versus_optical_line_ratios_all} and \ref{f:PAH_optical_interpretation_sii_halpha}), which can naturally be explained in a case where a change in temperature drives the change in the \oiii, \nii, \sii, and \oi~ line emissivities. These optical line ratios depend on the $\nu L_{\nu}(\mathrm{O^{++}})/\nu L_{\nu}(\mathrm{H^{+}})$ ratio through the relation between the hardness of the ionizing radiation and the electron temperature. 
We intend to use detailed photoionization models to further investigate this scenario in a future work. 

\underline{To summarize}, the observed correlations between the PAH band 11.3/7.7 ratio and the optical line ratios \oiii/\hbeta, \nii/\halpha, \sii/\halpha, and \oi/\halpha~ can be naturally explained in a scenario where the PAHs and ionized gas are exposed to different parts of the same spatially varying radiation field, without the need to invoke PAHs with different charge distributions. Since the PHANGS pixels trace regions with widely varying radiation SEDs that are a combination of young stars, old stars, and/or AGN, a variation of PAH band ratios due to the changing radiation field SED is \emph{unavoidable}. We use the PAH models by \citet{draine21} and a wide range of assumptions about the variation of the incident radiation field to show that the expected slope of the relation is roughly consistent with that observed. 

\underline{The main implications} of this scenario are: 
(I) The very small scatter ($\sim 0.03$ dex) in the relation between 11.3/7.7 and \oiii/\hbeta~ (or any other optical line ratio) implies that the ionized PAH fraction is quite uniform on scales of 150 pc across different environments in local galaxies. Since the fraction of ionized PAHs is set by the balance between ionization (by photons or collisions) and recombination, this uniformity suggests a strong self-regulation of the ISM that limits variations in gas temperature and electron density. In particular, the property $U \sqrt{T} / n_e$, where $U$ is dimensionless intensity parameter, could, in principle, vary by a factor of a few (see \citealt{tielens05, galliano08, draine21}). According to our interpretation, this property can vary by no more than $\sim$100\% across different environments (SF+ISM and the diffuse ionized gas). In a future study, we plan to use all 19 PHANGS galaxies to place constrains on $U \sqrt{T} / n_e$ in different environments. 
(II) The 11.3/7.7 PAH band ratio may potentially be used to trace the shape of the FUV-optical parts of the radiation field across nearby galaxies. The combination of the 11.3/7.7 and optical line ratios may therefore be used to constrain simultaneously the ionizing and non-ionizing parts of the radiation field. 
(III) The varying radiation field is expected to impact PAH band ratios that are typically used as PAH size indicators (like the 3.3/11.3 band ratio). To constrain the PAH size distribution, the variation of the radiation SED must be accounted for. 

\begin{figure*}
	\centering
\includegraphics[width=1\textwidth]{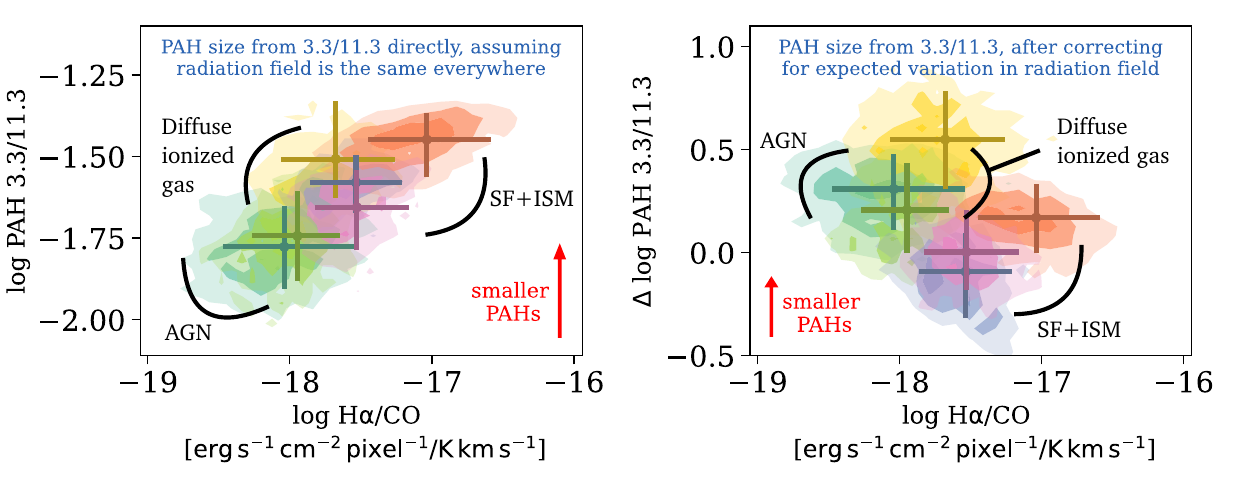}
\caption{\textbf{Different PAH heating scenarios lead to opposite interpretations regarding PAH sizes.} The two panels show the variation of the 3.3/11.3 PAH band ratio versus the \halpha/CO ratio. The left panel shows the measured 3.3/11.3 PAH band ratio. Assuming the ``Non-varying radiation field" interpretation, the 3.3/11.3 can be used directly to constrain the PAH size distribution. The left panel therefore suggests that regions with higher fractions of ionized gas (traced by larger \halpha/CO ratios) are associated with smaller PAHs. This seems to be in tension with the common picture that smaller PAHs are more efficiently destroyed by ionizing radiation than larger PAHs. The right panel shows $\Delta $3.3/11.3, which is the measured 3.3/11.3 PAH band ratio after correcting for the expected variation due to changing SED (subtracting the black dotted line in figure \ref{f:PAH_optical_interpretation_sii_halpha} from the measured 3.3/11.3 values). Contrary to the left panel, the right panel suggests that regions with higher fractions of ionized gas host larger PAHs, in line with the common picture that larger PAHs survive in harsher environments. The yellow and orange clusters, which predominantly come from NGC~7496, are above this relation and show significantly smaller PAH size distributions (see section \ref{sec:results:PAH_optical_lines_ratios_corr:size} for additional details). }
\label{f:interp_PAH_sizes_wrt_Ha_CO_edit}
\end{figure*}

Finally, it is worth noting that previous Spitzer-based studies found surprisingly high 11.3/7.7 PAH band ratios in AGN-dominated systems and in elliptical galaxies (e.g., \citealt{kaneda05, smith07, kaneda08, galliano08, diamond_stanic10, vega10}, which in some cases, were found outside the range of ratios predicted by models. There are two main differences between the variation of the 11.3/7.7 PAH band ratio found here and those reported by the studies mentioned above. First, while the Spitzer-based studies found a 11.3/7.7 PAH band ratio that may be larger by $\sim$one order of magnitude compared to the ratio observed in normal environments, the variation we observe is much smaller, of the order of 0.1 dex. While a change of 1 dex cannot be explained solely by a change of the hardness of radiation field, a change of 0.1 dex can. Secondly, our work focuses on 150 pc-sized regions, where the wealth of multi-wavelength observations allows us to constrain separately the dominant source of ionizing radiation and the dominant source of FUV--optical photons. In our case, although we classify the first group as "AGN-photoionized gas", these pixels are very different from the AGN-dominated systems presented and discussed by, e.g., \citet{smith07, diamond_stanic10, lai22}, since in our case, old stars probably dominate the PAH heating.

\subsubsection{Constraining the PAH size distribution}\label{sec:results:PAH_optical_lines_ratios_corr:size}

The right panels of figures \ref{f:PAH_optical_interpretation_oiii_hbeta} and \ref{f:PAH_optical_interpretation_sii_halpha} show the PAH band ratio 3.3/11.3 versus the optical line ratios \oiii/\hbeta~ and \sii/\halpha. Although the 3.3/11.3 had been used as a PAH size indicator (e.g., \citealt{lai23, ujjwal24}), it is also sensitive to the shape of the incident radiation. In particular, varying the stellar age from 2 Myr to 1 Gyr can change the ratio by 0.55 dex, similar to the entire range spanned by 3.3/11.3 in figures \ref{f:PAH_optical_interpretation_oiii_hbeta} and \ref{f:PAH_optical_interpretation_sii_halpha}. This has been pointed out by \citet{chastenet23b} when interpreting the PAH band ratios observed in the first three PHANGS-JWST galaxies, where, similarly to \citet{dale23}, they argued that the hardness of the radiation field and PAH size may both affect the observed 3.3/11.3 ratio. In this section, we use the observed correlations between the PAH band and optical line ratios to break this degeneracy and disentangle the impact of the radiation field from that of the PAH size distribution on the observed 3.3/11.3 PAH band ratio.

Under the ``common varying radiation field'' interpretation, both 11.3/7.7 and 3.3/11.3 are expected to change with the optical line ratios, with the two slopes (11.3/7.7 vs. optical line ratio) and (3.3/11.3 vs. optical line ratio) connected to each other through the shape of the radiation field. In section \ref{sec:results:PAH_optical_lines_ratios_corr:interp} above we suggest that the relation between the 11.3/7.7 and the optical line ratios can be explained entirely using a varying radiation field. We can use the \citet{draine21} models to propagate this relation to the expected relation between 3.3/11.3 and the optical line ratios.

In the left panels of figures \ref{f:PAH_optical_interpretation_oiii_hbeta} and \ref{f:PAH_optical_interpretation_sii_halpha}, we show two polynomial fits to the observed relation between 11.3/7.7 and the optical lines. Since the 11.3/7.7 is a function of the FUV-to-optical luminosity ratio of the radiation field, these best-fitting polynomials connect the FUV-to-optical luminosity ratio with the observed \oiii/\hbeta~ or \sii/\halpha~ line ratios. Using the relation between 3.3/11.3 and the FUV-to-optical luminosity ratio (Appendix \ref{app:pah_models}), we can therefore calculate the expected relation between 3.3/11.3 and \oiii/\hbeta~ or \sii/\halpha, under the ``common varying radiation field'' interpretation. We show these lines in the right panel of figures \ref{f:PAH_optical_interpretation_oiii_hbeta} and \ref{f:PAH_optical_interpretation_sii_halpha}.

The right panels of figures \ref{f:PAH_optical_interpretation_oiii_hbeta} and \ref{f:PAH_optical_interpretation_sii_halpha} suggest that a significant part of the observed 3.3/11.3 variation is in fact due to a varying radiation field, rather than a varying PAH size distribution. Significant deviations from the expected slopes may be attributed to changes in the PAH size distribution. For example, among the two SF+ISM groups, the orange group is consistent with hosting smaller PAHs. Among the two diffuse ionized gas groups, the yellow group corresponds to regions with smaller PAHs. Finally, the figures suggest smaller PAHs in the AGN-photoionized gas group.

\begin{figure*}
	\centering
\includegraphics[width=1\textwidth]{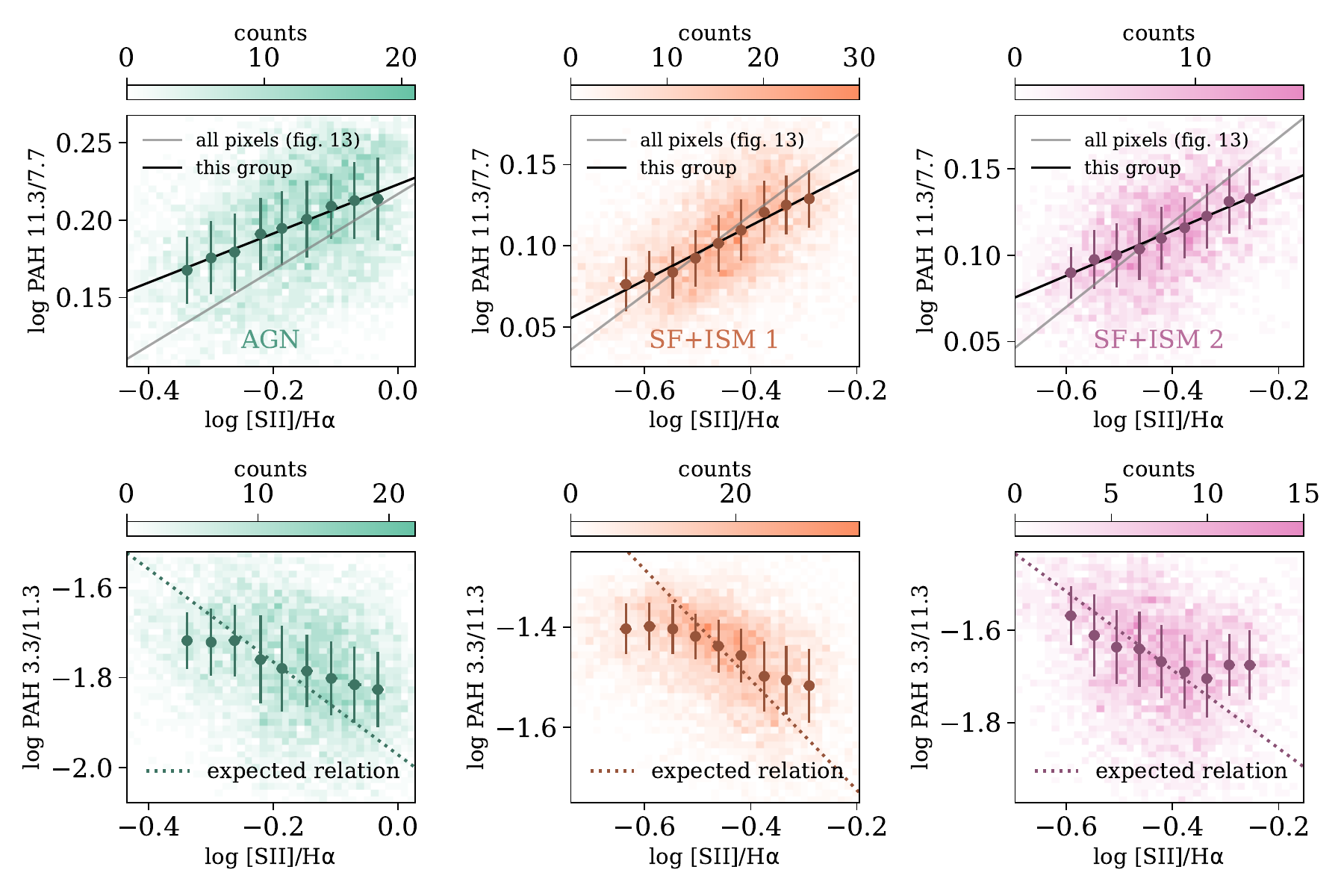}
\caption{\textbf{PAH band ratios versus \siihalpha~ for individual groups.} The figure shows three groups that show significant correlations between their PAH band and optical line ratios: AGN-photoionized gas (left), SF+ISM with smaller PAHs (middle), and SF+ISM with larger PAHs (right). The top panels show the PAH band ratio 11.3/7.7 and the bottom 3.3/11.3. The black lines at the top row represent the best linear fits to the observed relations. The light gray lines represent the best linear fit for all the groups together (figure \ref{f:PAH_optical_interpretation_sii_halpha}). The dashed lines at the bottom row represent the expected relation between 3.3/11.3 and \sii/\halpha, propagated from the best-fit relation between 11.3/7.7 and \sii/\halpha, using the PAH models by \citet{draine21}, and assuming that the PAHs and gas are exposed to different parts of the same varying radiation. The expected relations are steeper than the observed relations, which may suggest variations in PAH size distributions within each group.}
\label{f:PAH_versus_SIIHalpha_individual_clusters}
\end{figure*}

To further illustrate the impact of the different PAH heating interpretations on the derived PAH sizes, in figure \ref{f:interp_PAH_sizes_wrt_Ha_CO_edit} we show the 3.3/11.3 PAH band ratio versus the \halpha/CO ratio. The left panel shows the measured 3.3/11.3 ratio, and assuming that the radiation field does not vary, the ratio can be used directly to constrain PAH sizes. According to this ``Non-varying radiation field" interpretation, the 3.3/11.3 increases with the \halpha/CO ratio, suggesting that regions with a larger fraction of ionized gas tend to host smaller PAHs. The positive trend in the left panel is surprising, given that smaller PAHs are believed to be more easily destroyed by ionizing radiation compared to larger PAHs (reviews by \citealt{tielens08, li20}). The right panel of figure \ref{f:interp_PAH_sizes_wrt_Ha_CO_edit} shows $\Delta$3.3/11.3, which is the expected 3.3/11.3 after accounting for the variation due to the varying illuminating SED. We calculate $\Delta$3.3/11.3 by subtracting the black dotted line in the right panel of figure \ref{f:PAH_optical_interpretation_sii_halpha} from the measured 3.3/11.3 band ratio\footnote{We reach the same conclusion if we use the relation with the \oiii/\hbeta~ (figure \ref{f:PAH_optical_interpretation_oiii_hbeta}) instead.}. According to this ``common varying radiation field'' interpretation, PAHs are generally larger with increasing \halpha/CO, consistent with the idea that larger PAHs survive in regions with more ionizing radiation. 

The right panel of figure \ref{f:interp_PAH_sizes_wrt_Ha_CO_edit} shows a significant scatter that is dominated by the 3.3/11.3 ratios of the yellow and orange groups, both show significantly larger 3.3/11.3 ratios than those of their equivalent groups (the green and pink; diffuse ionized gas and SF+ISM respectively). Even if the ``common varying radiation field'' interpretation describes more accurately the observed PAH band ratios, the scatter in the diagram suggests that a simple picture of `larger PAHs in regions more dominated by ionizing radiation' is too simple to describe the observed variations in the 3.3/11.3 band in the PHANGS galaxies, and that additional factors may be at play. We plan to revisit this question in a future work, where we plan to apply this analysis to the full PHANGS+JWST sample of 19 galaxies.

\subsubsection{PAH bands vs. optical line ratios in individual groups}\label{sec:results:PAH_optical_lines_ratios_corr:individual_clusters}

We find significant correlations between the PAH band and optical line ratios in a few groups where the dynamical range is large enough (see table \ref{f:correlation_matrix_with_labels} in the appendix for the full correlation matrix). In figure \ref{f:PAH_versus_SIIHalpha_individual_clusters} we show these relations with the \sii/\halpha~ ratio for the three groups: AGN-photoionized gas, and the two SF+ISM groups. The top row shows the 11.3/7.7 PAH band ratio. The observed slopes of the 11.3/7.7 versus \sii/\halpha~ relations observed for the different groups are comparable to each other (0.16, 0.13, and 0.15 for AGN, SF+ISM 1, and SF+ISM 2, respectively) and to the best-fitting slope found for all PHANGS pixels in figure \ref{f:PAH_optical_interpretation_sii_halpha} (0.22).

We find that the observed trends between the 11.3/7.7 and \sii/\halpha~ ratio are all in line with the ``common varying radiation field'' interpretation, with different types of radiation dominating in each of the groups. For the AGN-ionized gas group, the radiation may be a sequence of varying contribution of AGN and stellar light. Importantly, while the AGN may dominate the ionizing radiation in this group, resulting in the Seyfert-like line ratios we observe in figure \ref{f:BPT_diagram_of_clusters}, it probably does not dominate the FUV-optical part of the radiation field. This is because these regions are at distances of a few kpc from the center, where the old stellar population probably dominates the PAH heating.

For the SF+ISM groups, the variation can be driven by a sequence in stellar ages (or equivalently, a mixing of younger and slightly-older stellar populations). Since these clusters are consistent with HII ionization in the BPT diagrams, the relevant stellar ages are 1--10 Myr. For gas metallicity of $\log (Z/Z_{\odot}) \sim -0.3$, which is roughly the metallicity in the three PHANGS galaxies we consider (e.g., \citealt{williams22}), a variation of the stellar age from 1 Myr to 10 Myr results in a variation of gas electron temperature from 12,000 to 7,000 K \citep{byler17}, which can explain the varying \sii~ emission. In particular, the radiation of a 1 Myr-old star has harder ionizing radiation and softer FUV-optical slope. The harder ionizing radiation results in higher electron temperature and thus brighter \sii~ emission \citep{byler17}. At the same time, the softer FUV-optical slope results in larger 11.3/7.7 PAH band ratio \citep{draine21}, which explains the positive relation between 11.3/7.7 and \sii/\halpha.

In the above discussion, we assume that the optical line ratios in the two SF+ISM groups vary due to a varying sequence in stellar ages. We disfavor the interpretation that the optical line ratios are driven by a varying ionization parameter, a property that is typically invoked to explain the optical line ratios in HII regions (see review by \citealt{kewley19}). For a varying ionization parameter, we expect a negative correlation between \oiii/\hbeta~ and \sii/\halpha, while for a varying stellar age we expect a positive one (e.g., \citealt{blanc15, byler17, kewley19}). Since we observe that \oiii/\hbeta~ increases with increasing \sii/\halpha, we favor the varying stellar age sequence. 

The bottom panel of figure \ref{f:PAH_versus_SIIHalpha_individual_clusters} shows the 3.3/11.3 PAH band ratio versus \sii/\halpha. Since the 3.3/11.3 band ratio is also sensitive to variations in the incident radiation, we use the best-fitting relation between 11.3/7.7 and \sii/\halpha~ and propagate it to the expected relation between 3.3/11.3 and \sii/\halpha. In all the three groups, figure \ref{f:PAH_versus_SIIHalpha_individual_clusters} shows that the observed relation between $\log (3.3/11.3)$ and \siihalpha~ is shallower than that expected. This may suggest small variations in the PAH size distribution within each of the clusters. We intend to further investigate it in future works.

\section{Extensions and generalizations of the methodology}\label{sec:discussion}

\subsection{Omitted observations: rationale, implocations, and future considerations}\label{sec:discussion:method_improvemnts:not_included}
% think about a better subsection name

For this pilot study, we choose to concentrate on a limited set of properties derived from the PHANGS observations, leaving out many potential properties of interest that could be included in future works. The choice to limit the number of considered properties simplifies the problem significantly but also limits the discovery space to that spanned by the properties we focus on. In this section we briefly discuss some of the omitted properties and the reasons for excluding them, and describe additional  properties that may be included in future analyses of this type.  

The current analysis is based on MUSE, JWST, and ALMA observations, and it does not use HST photometry. In particular, the PHANGS-HST galaxies have been observed with the filters F275W, F336W, and F438W, covering UV and blue optical wavelengths not observed by MUSE (\citealt{lee22}). Including these wavelengths may add information related to the young stellar populations, burstiness of star formation, star formation histories, and reddening towards the stars. 

For the ALMA data, we use the integrated CO intensity obtained through the ``broad" mask and leave out other properties, such as the width of the line. The decision to use only the integrated CO intensity, and not the line width, was motivated by our attempt to include as many pixels as possible in the analysis. Since the CO line width is available only in the ``strict" mask products, it would have required us to exclude $\sim$90\% of the pixels. Even with the ``broad" mask, the CO detection requirement (along with the 3.3 \mic~ PAH detection) excludes $\sim$30\% of the pixels, restricting our analysis to regions hosting molecular gas masses equivalent to those of giant molecular clouds within 1\arcsec. In a future work, we plan to explore the use of the ``flat" CO intensity maps presented by \citet{leroy23} and used by \citet{belfiore23}, where CO flux measurement is available in every pixel.

For the MUSE data, we use derived properties that are based on high-SNR observables such as the integrated stellar continuum and strong emission lines. Additional properties that can be derived from the MUSE observations include the gas metallicity, ionization parameter, electron density, and temperature in the warm ionized phase; and the neutral atomic column density, traced by NaID absorption. Some of these properties require the detection of weaker lines, making maps derived with them incomplete\footnote{Alternatively, the spaxels can be binned until the weaker lines are measured with sufficient SNRs. In that case, the maps will be complete but may have a much coarser resolution.}. In addition, the derivation of some of these properties is often based on physical models that may not be general enough to describe the variety of gas conditions in the PHANGS galaxies. For example, the metallicity is estimated using calibrations that are valid in HII regions, but not in the diffuse or AGN-photoionized gas (e.g., \citealt{pilyugin16, kreckel20, williams22}).

For the JWST data, our selected features trace primarily PAH emission, and we do not include features that trace the continuum emission from hot and larger dust grains (21 \mic). Several PHANGS-JWST studies show that the F2100W filter behaves differently from the JWST filters we consider, in that it traces both gas column density and heating, showing connection to the \halpha~ and CO emission (e.g., \citealt{belfiore23, hassani23, leroy23, pathak24}). In addition, by excluding the F2100W filter, we also exclude $\mathrm{R_{PAH}}$, a feature that traces the PAH-to-total dust mass, and shows galaxy-wide variations that are related to the ionization parameter and metallicity (e.g., \citealt{chastenet23a, egorov23}; Sutter et al. in prep.). In the first three PHANGS-JWST galaxies studied here, the F2100W filter is saturated in the centers of two (NGC~1365 and NGC~7496), showing significant diffraction spikes on kpc scales. Including this feature in the analysis would have required us to mask-out these pixels, leaving out the two galaxy centers that host AGN. We plan to revisit this choice in the future as more galaxies are included in the analysis. 

Finally, it is worth noting that unsupervised machine learning algorithms can, in principle, be applied directly to the raw data (illustrated in figure \ref{f:intro_fig}). In the PHANGS survey case, these include the MUSE spectra, the photometric bands by HST and JWST, and the clean CO spectra reconstructed from the ALMA interferometric observations. Working with the raw data allows a more general analysis that is not limited by our prior physical knowledge, and may have a bigger potential for unexpected discoveries. It also provides a clear way to account for non-detections. However, working with the raw data also presents some challenges, including (i) treatment of highly correlated features, (ii) different features carrying different amounts of information, requiring some physically-motivated weighting scheme, (iii) presence of catastrophic outliers due to problems in observations and reduction, and most importantly, (iv) the challenge of interpreting the output of the unsupervised learning algorithms when applied to highly-complex datasets (see discussion in \citetalias{baron19}). These challenges are also present when applying machine learning algorithms to derived features, but the usage of raw data can make them considerably more difficult to address.

\subsection{Treatment of missing measurements}\label{sec:discussion:missing_features}

In astronomy, as well as in other fields, it is typical to find missing values in datasets. The values could be missing due to different reasons. Values may be missing if observations were not conducted. This can occur, for example, when certain astronomical objects are included in one survey but not in another. Values may be derived from fitting observations with theoretical models, where missing values might occur if the fitting failed, or the observations are outside of the parameter space allowed by the theoretical models. Perhaps the most common type of missing values in astronomical datasets are non-detections, where observations were performed but the astronomical object was too faint to be detected. These are also referred to as upper limits, and their statistical properties are significantly different from those of features missing because observations were not conducted. It has been known for several decades that excluding objects with upper limits from a dataset may lead to significant biases in estimating summary statistics such as the mean, median, or standard deviation of a distribution, as well as biases in the output of regression and/or correlation analyses (e.g., \citealt{feigelson85, isobe86}).

In the machine learning and data science literature, excluding objects with missing values has also been shown to lead to significant biases in estimated parameters and to an increasing prediction error. Interestingly, studies suggest that using a noisier but more complete dataset leads to smaller prediction errors compared to using cleaner but less complete datasets (e.g., \citealt{niederhut18}). This has been shown specifically in the context of supervised machine learning, where a model is trained to predict a certain variable, and the prediction can be compared to some \emph{ground truth} value (regression or classification task). In unsupervised machine learning tasks, such as dimensionality reduction or clustering, the effect of excluding objects with missing values is more ambiguous, as there is typically no definitive reference point for comparison. Nevertheless, we choose to construct a noisier and more complete dataset by using the CO intensity derived using the ``broad" mask. 

Several approaches have been suggested to handle missing values in a dataset (e.g., \citealt{little02, schafer02, newman14, niederhut18}). The simplest one is to replace the missing values by the mean value within a given feature. However, this simple approach of imputing the mean has been shown to lead to biased results and to underestimate the prediction variance \citep{niederhut18}. Additional approaches include more sophisticated imputations based on K nearest neighbor (KNN) search, Random Forest Regressor, and multiple imputation (e.g., \citealt{hruschka03, jonsson04, buuren11, stekhoven11, shah14, niederhut18}), or approaches that are not based on imputation but rather on estimating distances between features with missing values, and then applying the algorithm to the distance matrix directly (e.g., \citealt{eirola13} and references therein). All of these approaches assume that the missing features are ``missing at random", which means that the likelihood of a particular measurement being missing is independent of the value it could have had. This in turn assumes that any object with a missing value is well-represented in the dataset, and by fitting a model to the features of other objects, the missing feature value can be predicted. Once the missing values are replaced by predicted values, standard machine learning algorithms can be applied to the data without the need to exclude objects with missing features. 

In section \ref{sec:methods:features} we examine two different methods for missing value imputation: KNN search and Random Forest Regression, both assume that the features are ``missing at random". The missing features are optical line ratios that are based on strong emission lines: \nii, \oi, and \oiii. Our manual inspection of the pixels shows that when one of the ratios is missing, the others are measured and can be used to predict the value of the missing ratio. In this particular case, the assumption of ``missing at random" approximately holds, since we are working with line ratios that have been well-measured throughout most of field of view. 

In many cases, astronomical upper limits cannot be considered as ``missing at random", since their missingness directly depends on their (low) value. Furthermore, objects with missing values are not necessarily well-represented by objects that were detected, as the two may represent different populations of objects (e.g., star-forming versus passive galaxies that show significantly different correlations between their properties). In a future work where we plan to include all 19 Cycle 1 PHANGS-JWST galaxies, we intend to examine whether the assumption of ``missing at random" may hold approximately. The Cycle 1 PHANGS-JWST galaxies were observed to the same depth, but they span a distance range of 5--20 Mpc (see table 1 in \citealt{lee23}). Therefore, convolving all the maps to the same physical scale of 150 pc, and working in units of luminosity per physical scale of 150 pc, would result in lower noise levels for the closer galaxies. Then, the CO and 3.3 \mic~ PAH luminosities in brighter regions corresponding to the closer galaxies, along with other measured features, may be used as an approximate model for the undetected CO and 3.3 \mic~ PAH luminosities in farther away galaxies. It may therefore be possible to build a regression model that predicts the undetected CO and 3.3 \mic~ luminosities from the detected \halpha, 7.7 \mic, 11.3 \mic, and 10 \mic~ luminosities using pixels where all the features are measured. This, of course, assumes that the CO luminosity distributions are similar, and that the physical properties of pixels with undetected CO and 3.3 \mic~ can be well-represented by pixels with detections. 

An alternative solution for including upper limits in the dataset is to represent all features as probability distribution functions (PDFs). Features that are not missing may be represented by a normal distribution whose mean is the measured value and its standard deviation is the measurement uncertainty. Upper limits may be represented by a step or a box function whose limits are defined by the survey depth and other physical considerations (e.g., flux is non-negative). Then, instead of estimating pairwise distances between numbers, distances can be estimated using metrics that estimate distances between PDFs, such as the Wasserstein distance (e.g., \citealt{rubner00, ramdas15}) or the \citet{kullback51} divergence. With the pairwise distances estimated, traditional unsupervised learning algorithms can be used without excluding objects with missing values. We have developed and tested such an approach for the Random Forest algorithm \citep{reis19}, though additional tests must be performed in the case of unsupervised learning. We intend to compare different solutions for including upper limits for the 19 PHANGS-JWST galaxies in a future work.

\section{Summary}\label{sec:summary}

The PHANGS survey has been making high-resolution observations of nearby galaxies across the electromagnetic spectrum (figure \ref{f:intro_fig}). This complex multi-wavelength dataset offers the opportunity to explore the interplay between different processes operating on scales as small as the molecular cloud scale. This interplay is what controls the baryon life cycle in galaxies as well as their star formation, and thus is key to their overall evolution. In this work, we use unsupervised machine learning algorithms to dissect the complex high-dimensional space spanned by the PHANGS multi-wavelength observations. With these tools, we identify groups and previously-unknown trends in the data, allowing us to form data-driven hypotheses from this rich dataset.

We extract properties of interest from the ALMA, MUSE, JWST NIRCam, and JWST MIRI observations of the three PHANGS galaxies: NGC~0628, NGC~1365, and NGC~7496 (section \ref{sec:data}), from $\sim$24\,000 pixels that are half of the 150 pc resolution. These properties pertain to the stellar populations; warm ionized and cold molecular gas properties; and PAH and dust conditions (section \ref{sec:methods:features} and table \ref{tab:features}); on scales of 150 pc. We apply the dimensionality reduction algorithm {\sc umap} to embed the high-dimensional dataset into a two-dimensional plane (section \ref{sec:methods:UMAP}). We then use the Hierarchical clustering algorithm to divide the pixels into groups (section \ref{sec:methods:clustering}), each group showing distinct properties. In the process, we identify novel galaxy-wide correlations across the different regions, and crucially, use our defined groups to interpret them. Our results and their broader implications are summarized below. 

\noindent \textbf{(I) The identified groups have distinct multiphase gas and PAH properties (section \ref{sec:results:cluster_properties} and figure \ref{f:classes_on_galaxies}).} The 150 pc-sized regions are divided into 6 groups, where each maps to large-scale ($\sim$kpc) and coherent structures within the galaxies. Using optical line ratios such as \oiii/\hbeta, \nii/\halpha, \sii/\halpha, and \oi/\halpha, the \halpha-to-CO flux ratio, the PAH band ratios 11.3/7.7 and 3.3/11.3, and additional properties, we interpret the groups, finding that they mostly differ in their multiphase gas and PAH properties. The different groups trace gas photoionized by the AGN; by standard star formation in spiral arms; extreme star formation fueled by galactic bars; and diffuse gas ionized by a combination of radiation leaking from HII regions with harder radiation from hot and evolved stars. The different groups also show different PAH properties, with some showing significantly smaller PAH size distributions than others.

\noindent \textbf{(II) There is a close connection between the heating of PAHs and the ionization of the warm ionized gas (section \ref{sec:results:PAH_optical_lines_ratios_corr}).} We identify significant and tight correlations between different PAH band and optical line ratios (figures \ref{f:PAH_11p3_7p7_versus_optical_line_ratios_all} and \ref{f:PAH_3p3_11p3_versus_optical_line_ratios_all}). These correlations are seen across the entire dataset, covering the star-forming regions and the ISM, the diffuse ionized gas, and the AGN-photoionized gas. They suggest a strong connection between the heating of PAHs and the ionization of the warm ionized gas on 150 pc scales. \vspace{-0.3cm}
\begin{itemize}
		\item The observed correlations between the PAH band 11.3/7.7 ratio and the optical line ratios \oiii/\hbeta, \nii/\halpha, \sii/\halpha, and \oi/\halpha~ can be naturally explained in a scenario where the PAHs and ionized gas are exposed to different parts of the same spatially varying radiation field, without the need to invoke PAHs with different charge distributions (section \ref{sec:results:PAH_optical_lines_ratios_corr:interp} and figure \ref{f:PAH_optical_interpretation_oiii_hbeta}). Since the PHANGS pixels trace regions with widely varying radiation fields that are a combination of young stars, old stars, and/or AGN, a variation of PAH band ratios due to the changing radiation SED is \emph{unavoidable}. We combine PAH models with a wide range of assumed radiation fields to show that the observed slope of the relation is roughly consistent with that observed. %\vspace{-0.3cm}
  
            \item The scatter in the PAH 11.3/7.7 - optical lines relations is small, $\sim$0.03 dex, and suggests that the fraction of ionized PAHs is quite uniform on 150 pc scales in nearby galaxies. It suggests a significant self-regulation in the ISM across different regions.%\vspace{-0.2cm}
            
		\item The 11.3/7.7 PAH band ratio may potentially be used to trace the shape of the non-ionizing far-ultraviolet to optical parts of the radiation field. Combining it with optical line ratios such as \oiii/\hbeta~ and \sii/\halpha~ may therefore serve as a powerful diagnostic of the local radiation field, including its ionizing and non-ionizing parts simultaneously.%\vspace{-0.2cm}
  
		\item The varying radiation field is expected to also impact PAH band ratios that are typically used as PAH size diagnostics (section \ref{sec:results:PAH_optical_lines_ratios_corr:size} and figure \ref{f:PAH_optical_interpretation_oiii_hbeta}). We show that the 3.3/11.3 PAH band ratio is strongly impacted by the varying radiation, and we combine empirical fits of the 11.3/7.7 vs. optical line ratios relations with PAH models to correct the 3.3/11.3 band ratio for the varying radiation field. Once the varying radiation field is accounted for, we find that PAHs tend to be smaller in regions with low \halpha/CO ratios, and larger in regions with high \halpha/CO ratios. This is in line with the picture that smaller PAHs are more easily destroyed by ionizing radiation than larger PAHs. We also show that using the 3.3/11.3 band ratio directly to infer the PAH size distribution, which implicitly assumes that the incident radiation field is constant throughout, leads to completely different conclusions regarding the PAH size distribution in different regions in the galaxies (figure \ref{f:interp_PAH_sizes_wrt_Ha_CO_edit}).%\vspace{-0.2cm}
		\item We identify the same correlations between PAH band and optical line ratios in individual groups where the dynamical range is large enough (section \ref{sec:results:PAH_optical_lines_ratios_corr:individual_clusters}). The observed slopes of the correlations are comparable between the groups and comparable to the slope we find in the galaxy-wide correlations. Our analysis suggests that the variation in the 11.3/7.7 PAH band can be completely attributed to a variation in the radiation field. On the other hand, the 3.3/11.3 PAH band ratio requires an additional component, for example, a small variation of the PAH size distribution within each group. 
\end{itemize}		

This analysis is based on the first three galaxies in the PHANGS-JWST survey, covering $\sim$24\,000 75 pc-sized pixels. We plan to apply a similar analysis to all the 19 PHANGS galaxies observed by JWST in Cycle 1, and to combine it with an extensive set of photoionization models to better explore the PAH-ionized gas correlations. 

In this era of big data in astronomy, the abundance of multi-wavelength opportunities leads to complex and high-dimensional datasets, where millions of objects are observed across the electromagnetic spectrum, sometimes also as a function of time. In many cases, we have only scratched the surface of what is possible to learn from these information-rich datasets. When the information content of the data is so large, the data itself can be used to form new hypotheses, an approach that is at at the heart of data science. This work illustrates that unsupervised machine learning algorithms can be used to mine for novel information in complex multi-wavelength datasets such as that of the PHANGS survey. To expedite the discovery process, we used simple, yet powerful, machine learning algorithms and applied them to a set of physically-motivated features. This allowed us to quickly interpret the results and focus on the scientific implications of the newly-discovered correlations. On the other hand, this decision limited the discovery space to that spanned by the features we chose to consider. In future works, we plan to examine more sophisticated algorithms that can be applied directly to the raw data. While these may be more challenging to interpret, they can also potentially lead to new unexpected discoveries. 
 
\acknowledgments{

D. Baron is grateful to L. Armus, T. Lai, and G. Rudie, for insightful discussions regarding the interpretation of the clusters and correlations presented in this work. 
D. Baron is supported by the Carnegie-Princeton fellowship.
This project started during the GALEVO-23 program: Building a Physical Understanding of Galaxy Evolution with Data-driven Astronomy. This research was supported in part by grant NSF PHY-1748958 to the Kavli Institute for Theoretical Physics (KITP). 

KS acknowledges funding support from JWST-GO-02107.006-A.
MB acknowledges support by the ANID BASAL project FB210003 and by the French government through the France 2030 investment plan managed by the National Research Agency (ANR), as part of the Initiative of Excellence of Universit\'e C\^ote d’Azur under reference number ANR-15-IDEX-01. 
G.A.B. acknowledges the support from the ANID Basal project FB210003. 
TB acknowledges support from the National Research Council of Canada via the Plaskett Fellowship of the Dominion Astrophysical Observatory.
KG is supported by the Australian Research Council through the Discovery Early Career Researcher Award (DECRA) Fellowship (project number DE220100766) funded by the Australian Government.
EWK acknowledges support from the Smithsonian Institution as a Submillimeter Array (SMA) Fellow and the Natural Sciences and Engineering Research Council of Canada.

Some of the data presented in this paper were obtained from the Mikulski Archive for Space Telescopes (MAST) at the Space Telescope Science Institute. The specific observations analyzed can be accessed via \dataset[https://doi.org/10.17909/ew88-jt15]{https://doi.org/10.17909/ew88-jt15}. STScI is operated by the Association of Universities for Research in Astronomy, Inc., under NASA contract NAS5–26555. Support to MAST for these data is provided by the NASA Office of Space Science via grant NAG5–7584 and by other grants and contracts.}\vspace{1.8cm}

\software{Astropy \citep{astropy13, astropy18, astropy22},
		  IPython \citep{perez07},
          scikit-learn \citep{pedregosa11}, 
          SciPy \citep{scipy01},
		  matplotlib \citep{hunter07},
		  reproject \citep{reproject_2020},
		  UMAP \citep{mcinnes18}}\vspace{2cm}

\bibliography{ref.bib}

%\onecolumngrid

\appendix

\section{UMAP hyper-parameter exploration}\label{app:umap_parameters}

We examine the two-dimensional embeddings by {\sc umap} for a range of distance metrics and \texttt{n\_neighbors} parameters. For the distance metric, we examine 10 metrics from the list of proposed metrics in the {\sc umap} python library\footnote{\url{https://umap-learn.readthedocs.io/en/latest/parameters.html\#metric}}: Euclidean, Manhattan, Chebyshev, Minkowski, Canberra, Braycurtis, Mahalanobis, WMinkowski, Cosine, and Correlation distance. In addition to these, we estimate the unsupervised Random Forest (URF) distance matrix, used in \citet{baron17} to perform anomaly detection on galaxy spectra. The motivation to examine the URF-based distance is that the Random Forest algorithm can be generalized to handle missing values, upper limits, and measurement uncertainties \citep{reis19}, and can thus be used to include objects with missing features in our future work of the full PHANGS-JWST sample. 

\begin{figure*}
	\centering
\includegraphics[width=0.95\textwidth]{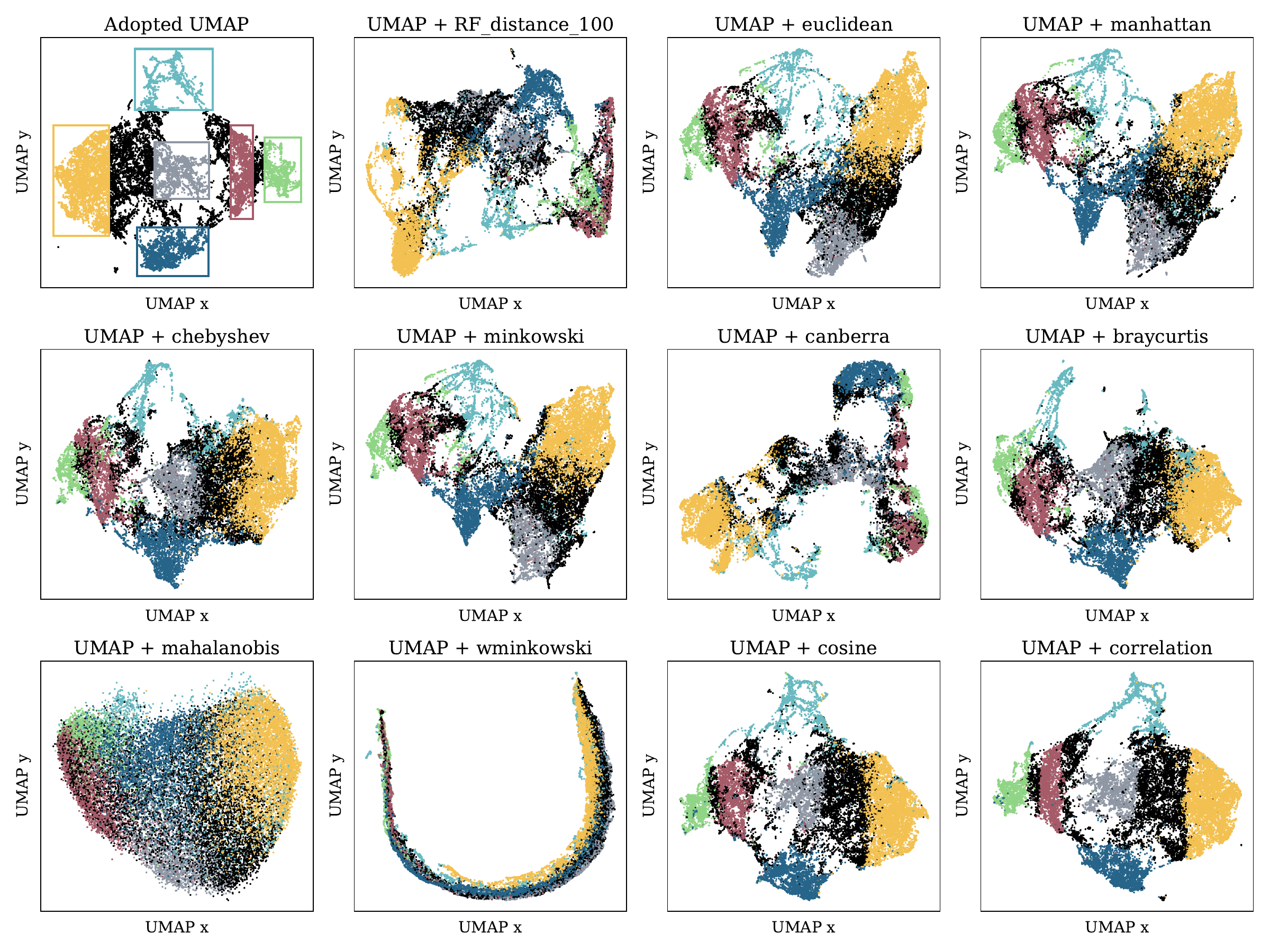}
\caption{\textbf{Comparison of {\sc umap} two-dimensional embeddings for different distance metrics.} The top left panel shows the {\sc umap} two-dimensional embedding adopted in this study, which was obtained using the following hyper-parameters: \texttt{metric=correlation}, \texttt{n\_neighbors=10}, and \texttt{min\_dist=0}. The 6 rectangles identify neighborhoods in the two-dimensional space that might be considered as different clusters. The other panels show the two-dimensional embedding obtained when varying the distance metric, assuming \texttt{n\_neighbors=25} and \texttt{min\_dist=0}. The points are color-coded according to the designated clusters of the pixels in our adopted two-dimensional embedding on the top left. These panels show that objects that were identified to be in the same cluster using one metric, will also be identified within the same cluster using another metric. This shows that the global structure of the data in the two-dimensional embedding does not depend significantly on the assumed metric.}
\label{f:UMAP_metric_variation}
\end{figure*}

Figure \ref{f:UMAP_metric_variation} shows the two-dimensional embeddings obtained with {\sc umap} for different distance metrics, and using \texttt{n\_neighbors=25} and \texttt{min\_dist=0}. These are compared to the embedding adopted in this paper (hyper-parameters: \texttt{metric=correlation}, \texttt{n\_neighbors=10}, and \texttt{min\_dist=0}). The points are color-coded according to their location in our adopted embedding, using rectangles we defined manually (top left panel). The figure shows that objects that are within a given neighborhood in one of the two-dimensional embeddings are also in the same neighborhood in another. This suggests that the global structure of the data does not depend significantly on the distance metric, and that the clusters that would have been identified using different metrics should be roughly similar to those we identify using our adopted embedding. 

\begin{figure*}
	\centering
\includegraphics[width=0.9\textwidth]{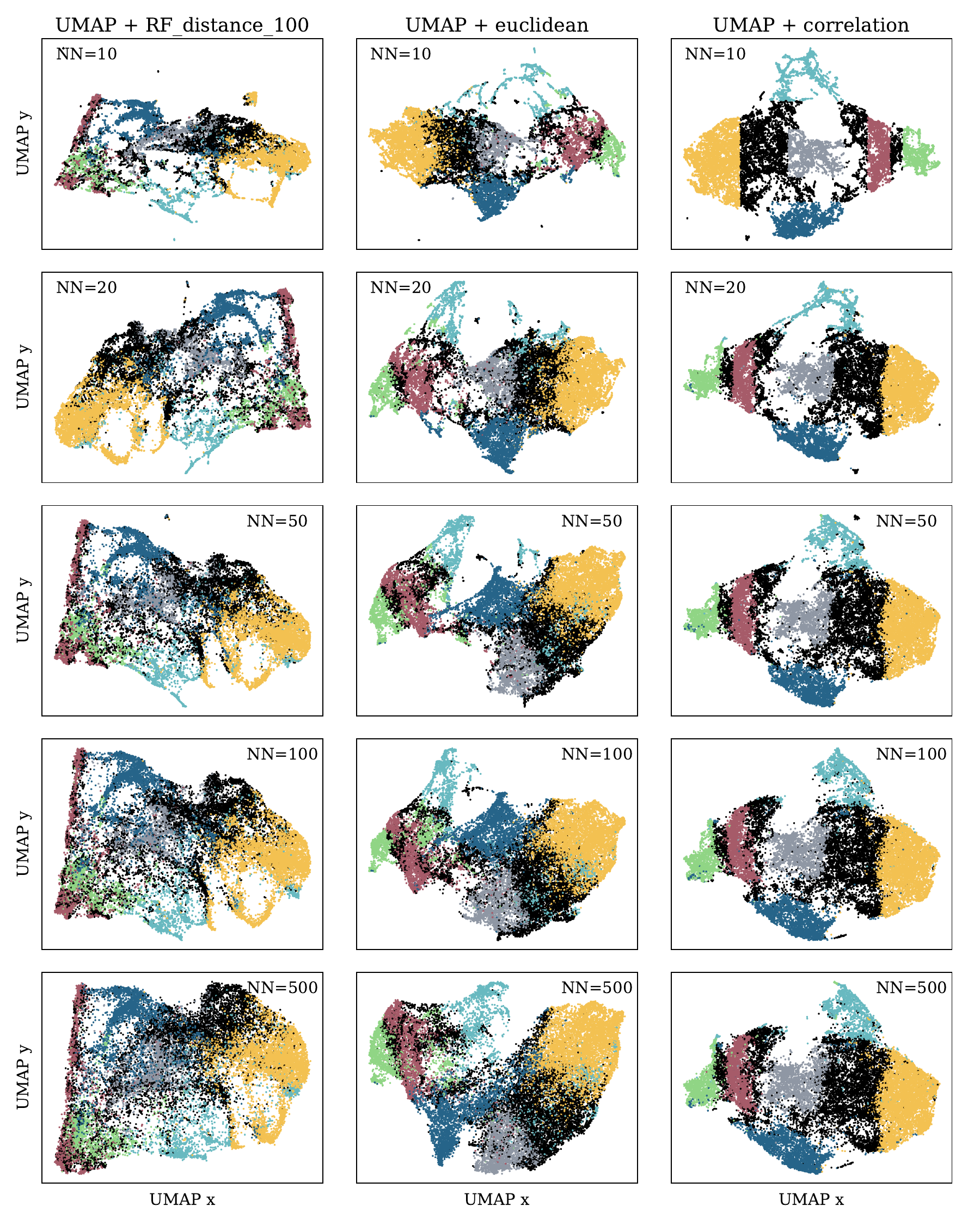}
\caption{\textbf{Comparison of {\sc umap} two-dimensional embeddings for different \texttt{n\_neighbors} parameters.} The panels show the resulting two-dimensional embedding assuming different metrics and different values of \texttt{n\_neighbors}. In each panel, the points are colored according to the designated clusters of the pixels in our adopted two-dimensional embedding (top left in figure \ref{f:UMAP_metric_variation}). For a given distance metric (given column), increasing the \texttt{n\_neighbors} parameter from 10 to 500 (top to bottom) does not change the general structure of the data in the two-dimensional space significantly, suggesting that comparable clusters would have been identified for different choices of hyper-parameters.}
\label{f:UMAP_NN_and_metrics}
\end{figure*}

Figure \ref{f:UMAP_NN_and_metrics} compares the two-dimensional embeddings by {\sc umap} for different \texttt{n\_neighbors} parameters ranging from 10 to 500. The points are color-coded according to their designated clusters we defined manually in figure \ref{f:UMAP_metric_variation} (top left panel). As expected, the general structure in the two-dimensional embedding becomes more connected and less clustered with increasing \texttt{n\_neighbors}. However, similarly to the previous figure, the figure shows that objects remain in the same rough neighborhood for different choices of \texttt{n\_neighbors}, suggesting that roughly the same clusters would have been identified using different values of \texttt{n\_neighbors}. 

%%%%%%%%%%%%%%%%%%%%%%%%%%%%%%%%%%%%%%%%%%%%%%%%%%%%%%%%%%%%%%%%%%%%%%%%%%%%%%%%%%%%%%%%%%%%%%%%%%%%%%%%%%%%%%%%%%%%%%%%%%%%%%%%%%%%%%%%%%%%%%%%%%%%%%%%%%%

\section{UMAP and clustering interpretation}\label{app:umap_interp}

%Figure \ref{f:UMAP_colorcoded_by_features} shows our adopted two-dimensional embedding, color-coded by some of the features in the input dataset. We used this figure, along with the clustering results from figure \ref{f:clustering_alg_variation}, to select our adopted clustering algorithm and its resulting clusters.

\begin{figure*}
	\centering
\includegraphics[width=1\textwidth]{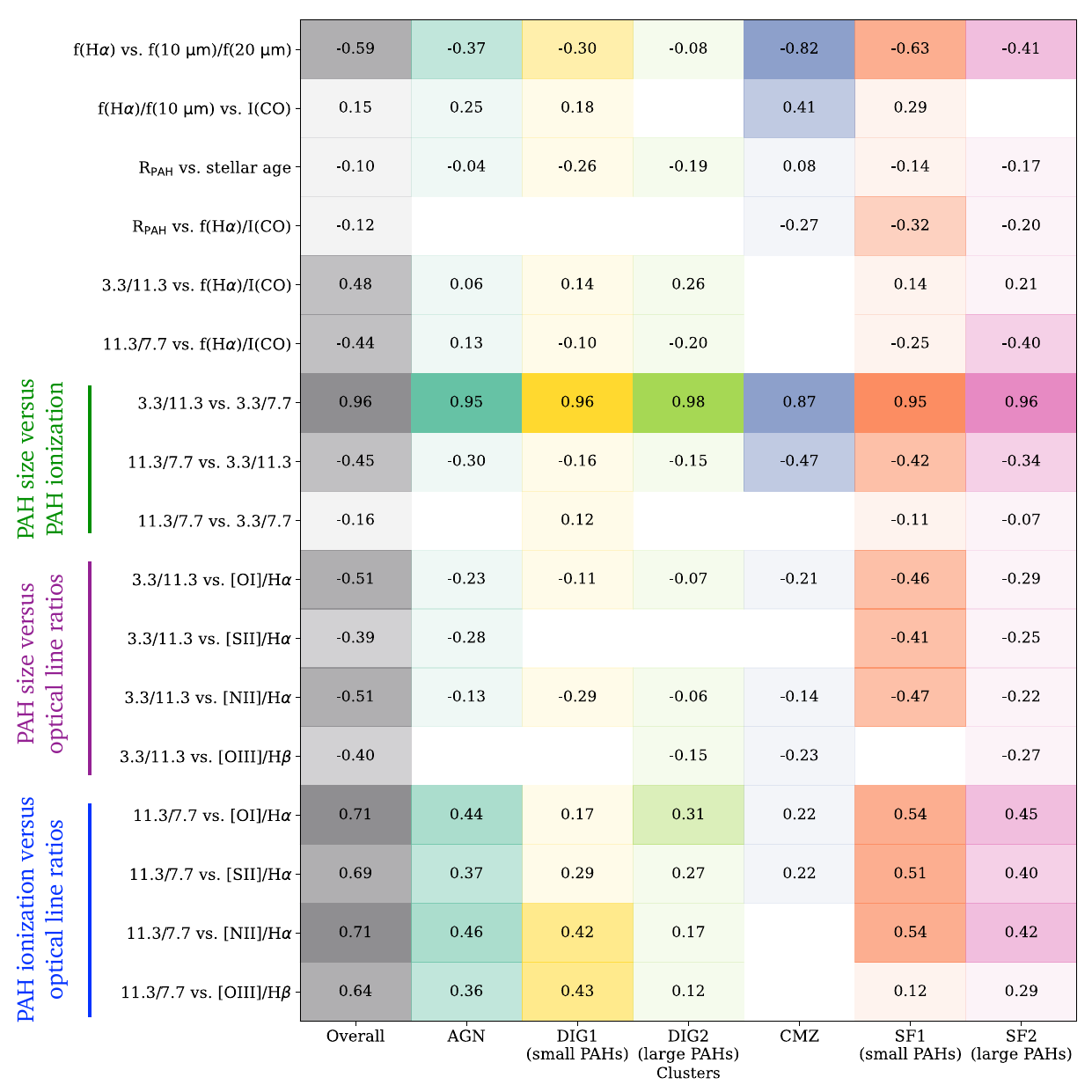}
\caption{\textbf{Correlations between various features considered in our analysis.} The matrix shows the Pearson correlation coefficient between pairs of features as indicated on the left, where empty cells correspond to correlations with p-values equal or larger than 0.01. Each row corresponds to a pair of features we consider. Each column marks the set of pixels considered for the correlation estimation, where the first column shows the correlations using all the 24\,007 pixels we considered, and the rest correspond to pixels in the different clusters we identified. High correlation coefficients are marked with brighter colors, and low coefficients with fainter colors. We identify significant and tight correlations between PAH band and optical line ratios, suggesting a strong connection between them. }
\label{f:correlation_matrix_with_labels}
\end{figure*}

\begin{figure*}
	\centering
\includegraphics[width=1\textwidth]{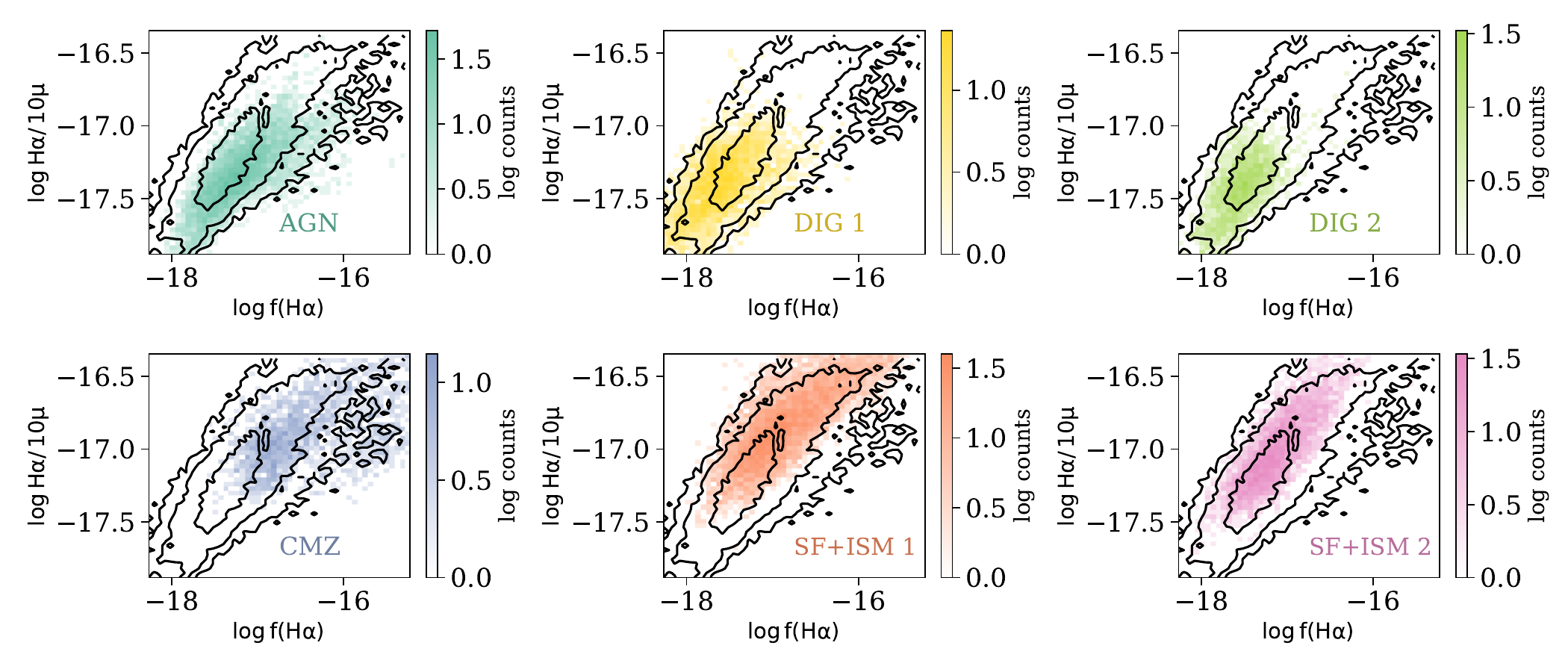}
\caption{$\log \mathrm{H\alpha/10\,\mu m}$ versus $\log \mathrm{f(H\alpha)}$ for the identified clusters. The black contours represent the 2D distribution of all the pixels in the dataset, and the colored colormaps represent the distribution in each cluster.}
\label{f:log_Ha_10mic_versus_f_Ha}
\end{figure*}

\begin{figure*}
	\centering
\includegraphics[width=1\textwidth]{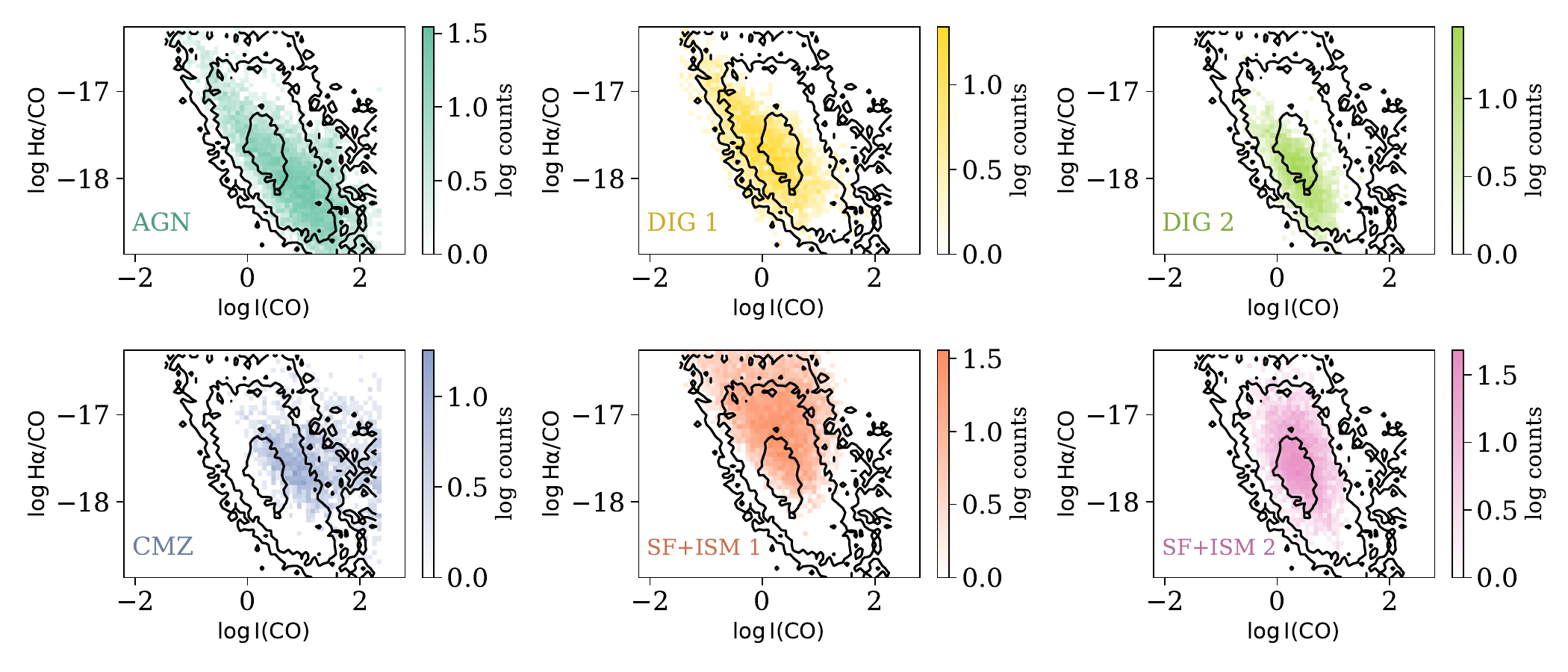}
\caption{$\log \mathrm{H\alpha/CO}$ versus $\log \mathrm{I(CO)}$ for the identified clusters. The black contours represent the 2D distribution of all the pixels in the dataset, and the colored colormaps represent the distribution in each cluster.}
\label{f:log_Ha_CO_versus_I_CO}
\end{figure*}

\begin{figure*}
	\centering
\includegraphics[width=0.8\textwidth]{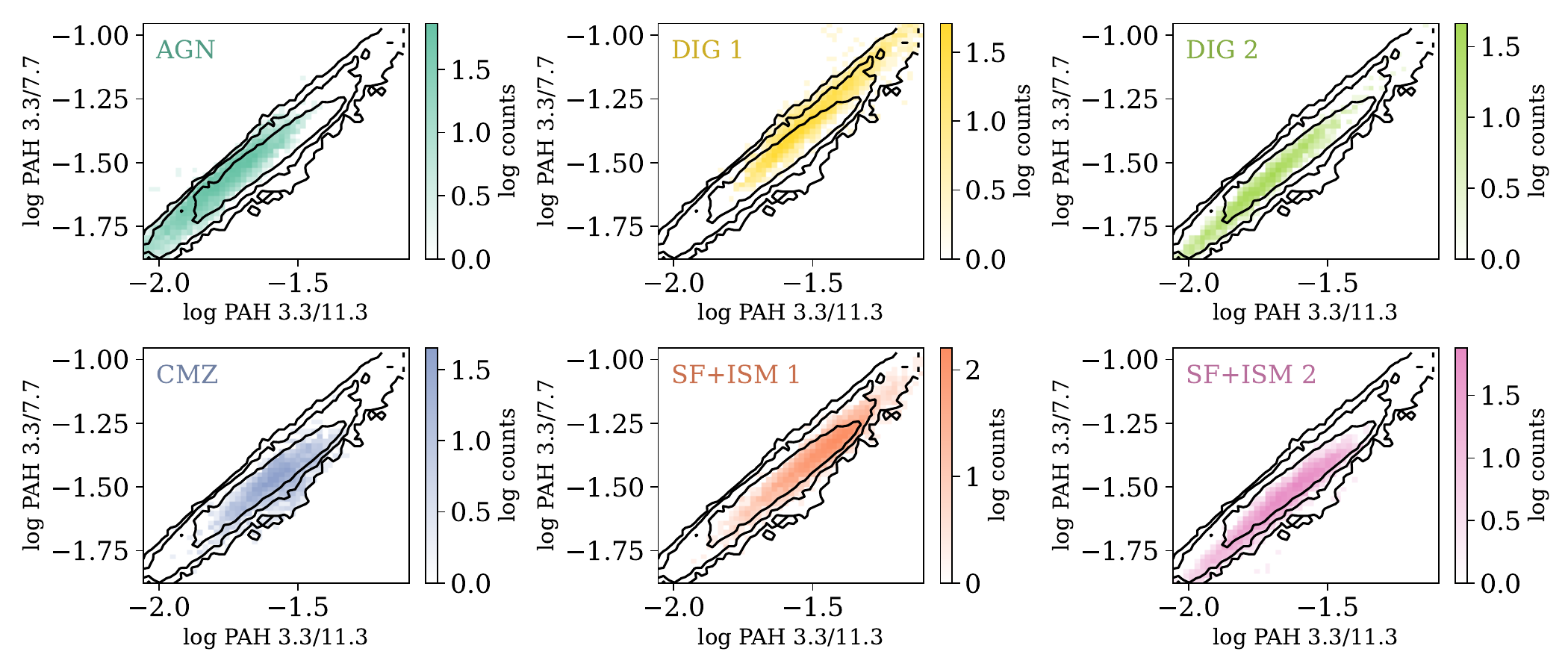}
\caption{PAH band 3.3/7.7 \mic~ versus 3.3/11.3 \mic~ ratios for the identified clusters. The black contours represent the 2D distribution of all the pixels in the dataset, and the colored colormaps represent the distribution in each cluster. }
\label{f:log_PAH_3p35-7p7_versus_PAH_3p35-11p3}
\end{figure*}

\begin{figure*}
	\centering
\includegraphics[width=0.8\textwidth]{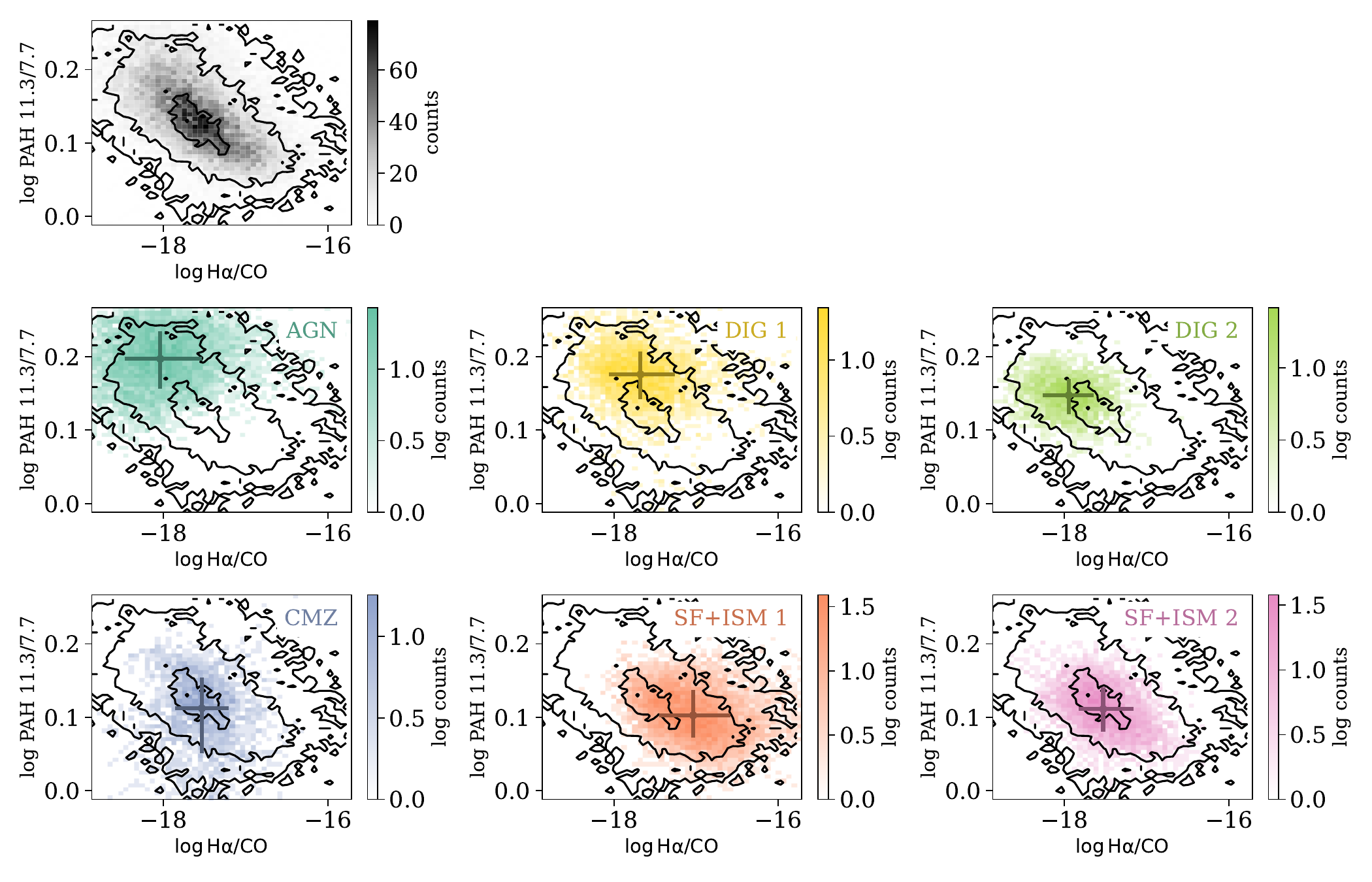}
\caption{PAH band ratio 11.3/7.7 \mic~ versus the \halpha/CO ratio. The top left panel shows the distribution for all the PHANGS pixels, and the rest show the distribution for each individual cluster. The black contours represent the distribution of all the pixels in the dataset.}
\label{f:log_PAH_11p3-7p7_versus_log_Ha-CO}
\end{figure*}

\begin{figure*}
	\centering
\includegraphics[width=0.8\textwidth]{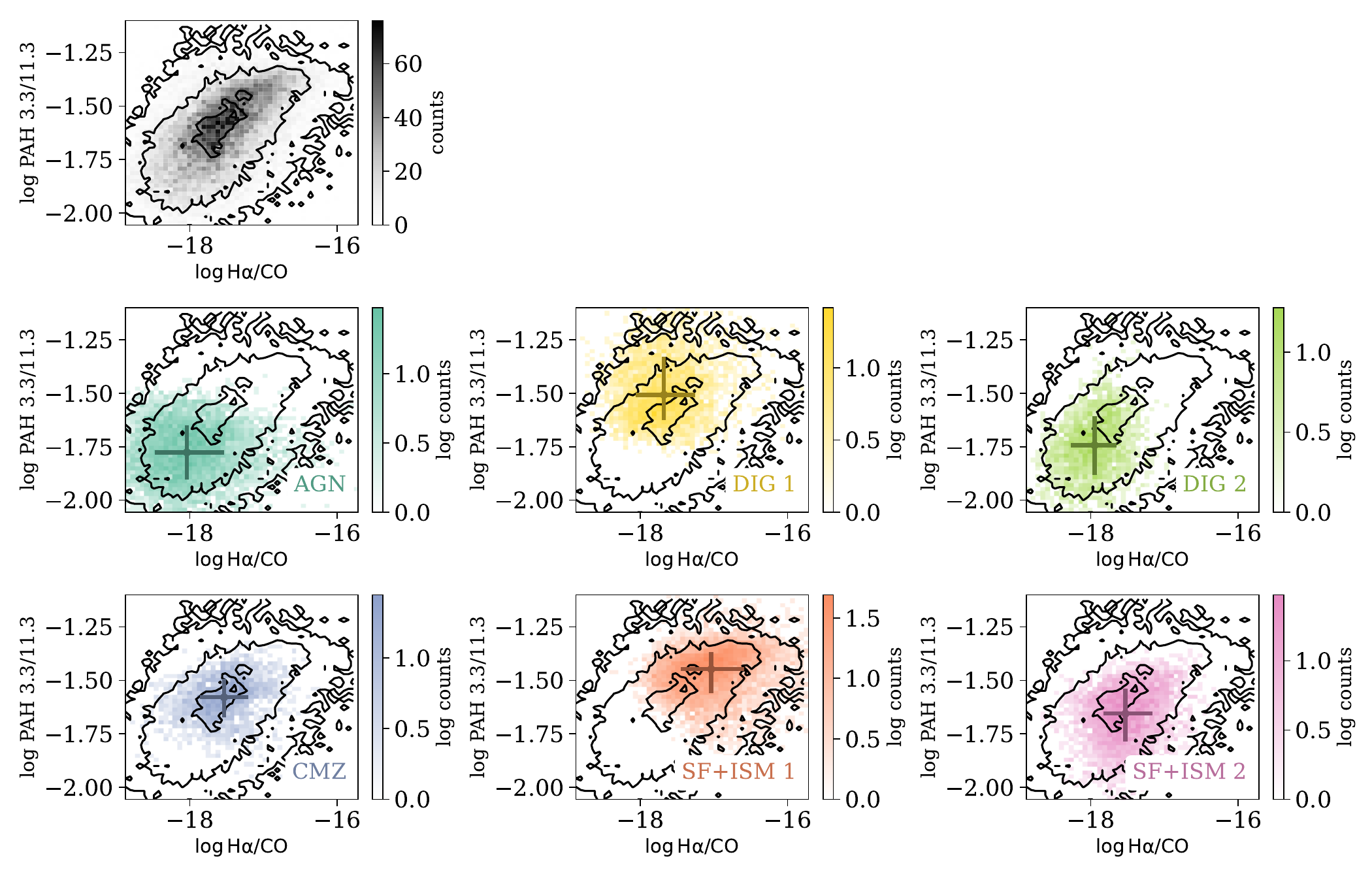}
\caption{PAH band ratio 3.3/11.3 \mic~ versus the \halpha/CO ratio. The top left panel shows the distribution for all the PHANGS pixels, and the rest show the distribution for each individual cluster. The black contours represent the distribution of all the pixels in the dataset.}
\label{f:log_PAH_3p35-11p3_versus_log_Ha-CO}
\end{figure*}

Figure \ref{f:correlation_matrix_with_labels} shows the Pearson correlation coefficient between different pairs of features. It includes the correlations calculated considering all the 24\,007 PHANGS pixels, as well as those obtained when considering pixels from individual clusters. The correlation coefficients are listed in the figure only if their associated p-value is smaller than 0.01. The figure shows significant correlations between different PAH band and optical line ratios. In figures \ref{f:log_Ha_10mic_versus_f_Ha}, \ref{f:log_Ha_CO_versus_I_CO}, \ref{f:log_PAH_3p35-7p7_versus_PAH_3p35-11p3}, \ref{f:log_PAH_11p3-7p7_versus_log_Ha-CO}, and \ref{f:log_PAH_3p35-11p3_versus_log_Ha-CO} we show the distribution of the clusters in some of the features we considered in our analysis. We used these figures to interpret the clusters we identified.

%%%%%%%%%%%%%%%%%%%%%%%%%%%%%%%%%%%%%%%%%%%%%%%%%%%%%%%%%%%%%%%%%%%%%%%%%%%%%%%%%%%%%%%%%%%%%%%%%%%%%%%%%%%%%%%%%%%%%%%%%%%%%%%%%%%%%%%%%%%%%%%%%%%%%%%%%%%

\section{Models used to interpret the PAH-ionized gas connection}\label{app:PAH_gas_models_interp}

\subsection{PAH models}\label{app:pah_models}

We make use of the models presented in \citet{draine21}. These models predict the infrared emission from dust under a wide range of assumptions, including incident radiation with varying SEDs, PAH size distributions, and PAH charge distributions. We use models with solar metallicity. For the SED, we use two sets of single-age stellar population models: BC03 \citep{bruzual03} and BPASS \citep{eldridge17}, with stellar ages of 3 Myr, 10 Myr, 100 Myr, 300 Myr, and 1 Gyr. We also consider the models calculated with the \texttt{mMMP} and \texttt{m31bulge} SEDs (see \citealt{draine21} for additional details). Therefore, we consider 12 different SEDs in our analysis. We consider the three different options for the PAH size distribution: small, standard, and large, and the three options for the PAH charge distribution: low, standard, and high. 

\begin{figure*}
	\centering
\includegraphics[width=1\textwidth]{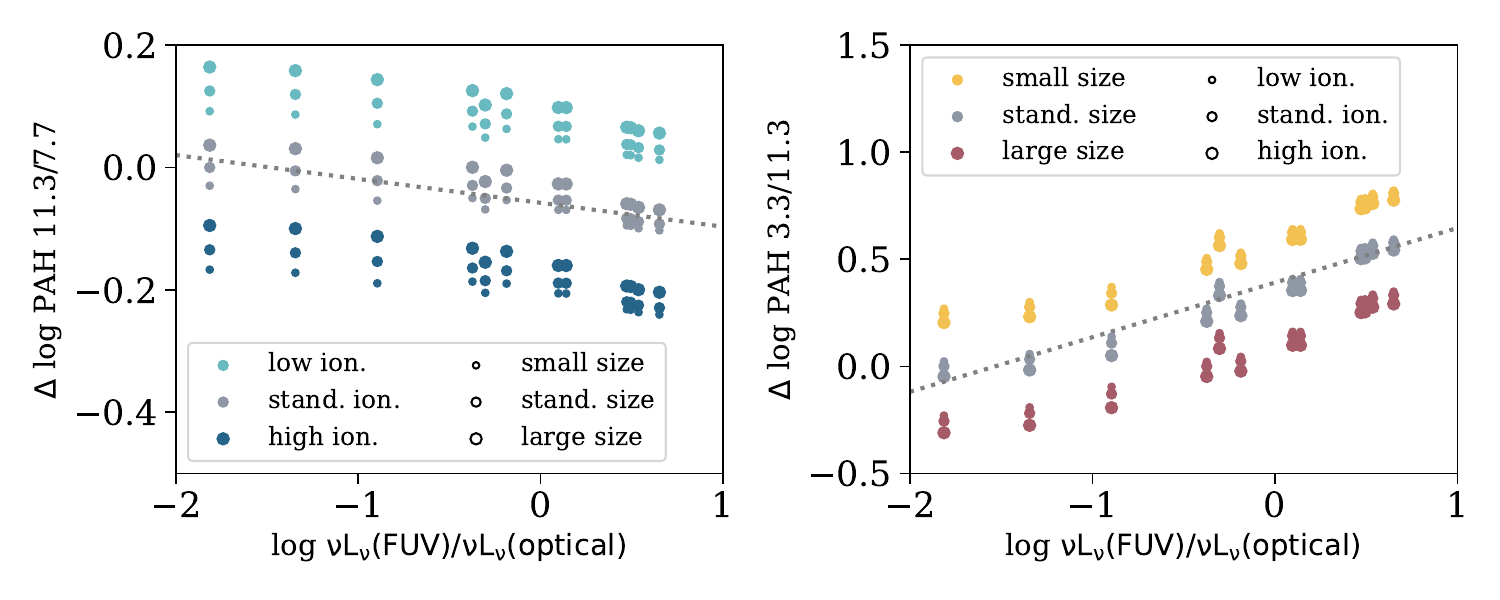}
\caption{\textbf{Relation between PAH band ratios and the FUV-optical slope for the \citet{draine21} models.} The panels show the PAH band ratios 11.3/7.7 \mic~ and 3.3/11.3 \mic~ versus the FUV-to-optical luminosity ratio of the SEDs the PAHs were exposed to. The figures show different PAH size and charge distributions. The 11.3/7.7 \mic~ band ratio is primarily sensitive to the PAH charge distribution, and increases for more neutral PAHs. It is also sensitive to the FUV-optical slope, and decreases for softer FUV-optical slopes.  The ratio is only weakly related to the PAH size distribution, and can change by $\sim0.03$ dex when the PAH sizes change more standard to small, or from large to standard. The 3.3/11.3 \mic~ ratio is primarily sensitive to the PAH size distribution, and increases for smaller PAHs. It is also sensitive to the FUV-optical slope, and increases for softer SEDs. It only weakly depends on the PAH size distribution, and can change by $\sim$0.05 dex when changing the PAH charge distribution from low to standard, or from standard to high.}
\label{f:Draine_models_FUVopt_ratios}
\end{figure*}

To estimate the 11.3/7.7 and 3.3/11.3 PAH band ratios, we follow the procedure outlined in \citet{draine21}. We parametrize the radiation field SEDs using their FUV-to-optical luminosity ratio, where the FUV luminosity was defined between $\lambda = 1350$ \AA\, and $\lambda = 1780$ \AA, and the optical between $\lambda = 3000$ \AA\, and $\lambda = 4000$ \AA. Figure \ref{f:Draine_models_FUVopt_ratios} shows how the PAH band ratios 11.3/7.7 and 3.3/11.3 change with the FUV-to-optical luminosity ratio, the PAH size distribution, and the PAH charge distribution. 

We estimate the expected change in the PAH band ratios, $\Delta \log (11.3/7.7)$ and $\Delta \log (3.3/11.3)$, for varying PAH size and charge distributions. These are represented as red and blue arrows in figures \ref{f:PAH_optical_interpretation_oiii_hbeta} and \ref{f:PAH_optical_interpretation_sii_halpha} in the main text.

For the variation with the PAH charge distribution, we use models with the standard PAH size distribution, and for each FUV-to-optical ratio, calculate $\Delta \log (11.3/7.7)$ and $\Delta \log (3.3/11.3)$ when changing the PAH ionization from low to standard and from standard to high. The resulting $\Delta \log (11.3/7.7)$ and $\Delta \log (3.3/11.3)$ do not vary significantly for different FUV-optical slopes and when considering low-standard versus standard-high changes. Therefore, we take the median over these values, with the final adopted values $\Delta \log (11.3/7.7) = -0.11$ dex and $\Delta \log (3.3/11.3) = -0.04$ dex. These represent the expected changes in the PAH band ratios when changing the PAH charge distribution from low to standard or from standard to high, and they are marked with blue arrows in figures \ref{f:PAH_optical_interpretation_oiii_hbeta} and \ref{f:PAH_optical_interpretation_sii_halpha}. A change of the PAH charge distribution from low to high would result in twice the change.

For the dependence on the PAH size distribution, we use models calculated with the standard PAH ionization distribution, and estimate $\Delta \log (11.3/7.7)$ and $\Delta \log (3.3/11.3)$ when changing the PAH size distribution from large to standard and from standard to small. These values also do not vary significantly with the FUV-to-optical luminosity ratio and when considering large to standard versus standard to small changes. The final adopted values, which are marked with red arrows in figures \ref{f:PAH_optical_interpretation_oiii_hbeta} and \ref{f:PAH_optical_interpretation_sii_halpha}, are $\Delta \log (11.3/7.7) = -0.022$ dex and $\Delta (3.3/11.3) \log = 0.24$ dex. They represent the change in the PAH band ratios when changing the PAH size distribution from large to standard and from standard to small. A change of the PAH size distribution from large to small would result in twice the change.

The dashed gray lines in figure \ref{f:Draine_models_FUVopt_ratios} are the best-fitting linear relations between the PAH band ratios $\log (11.3/7.7)$ and $\log (3.3/11.3)$ and the FUV-to-optical luminosity ratio. These lines are used to translate the observed relation between $\log (11.3/7.7)$ and \oiiihbeta~ to the expected relation between $\log (3.3/11.3)$ and \oiiihbeta, assuming that the ratios vary only due to varying SED shape. While it is clear from figure \ref{f:Draine_models_FUVopt_ratios} that the relations have some curvature, we choose to fit linear relations for simplicity.

\subsection{Spectral energy distributions}\label{app:seds}

To examine the relation between the FUV-optical slope and the hardness of the ionizing radiation, we considered several stellar population models, all of which are shown in figure \ref{f:SED_models} below. They include the stellar populations used by \citet{draine21} for the PAH models described above. We used the flexible stellar population synthesis (FSPS) code by \citet{conroy09} to simulate single stellar populations. We used the MIST isochrones \citep{choi16, dotter16} and the MILES stellar library \citep{vazdekis10}. We considered stellar population ages of 2 Myr, 100 Myr, 300 Myr, 1 Gyr, and 10 Gyr, and considered solar metallicity, as well as half and twice solar. We also considered the \citet{bruzual03} models, calculated with the Padova 1994 tracks, using two different stellar libraries: MILES and BaSeL (see \citealt{bruzual03} for details). We considered stellar population ages of 2 Myr, 100 Myr, 300 Myr, 1 Gyr, and 10 Gyr, assuming solar metallicity. We also considered the BPASS models that include a treatment of binary stars (e.g., \citealt{eldridge17}), considering the same stellar ages and a solar metallicity.

We constructed the HOLMES mixing sequence described by \citet{belfiore22}, which was used to interpret the optical line ratios seen in the diffuse ionized gas in PHANGS galaxies. The spectra are a combination of a young stellar spectrum (2 Myr), which represents the radiation leaking from HII regions, with an old stellar spectrum (10 Gyr), which represents the radiation of hot and evolved stars. The mixing between these spectra is determined through $f_{ion}$, which represents the ratio of the ionizing fluxes from the young and old stars. For example, $\log f_{ion}= 1$ represents a case where the ionizing flux of the young stars is 10 times larger than the ionizing flux from the old stars. $\log f_{ion}= 0$ represents the case that the ionizing flux from the young stars equals to the ionizing radiation from the old stars. To construct this sequence, we used the FSPS models with ages 2 Myr and 10 Gyr, assuming solar metallicity, and considering mixing fractions in the range $\log f_{ion}$ between -2 and 2. 

Finally, we constructed two AGN+SF mixing sequences. For the AGN accretion disk, the SED consists of a combination of an optical-UV continuum emitted by an optically-thick geometrically-thin accretion disk, and an additional X-ray power-law source that extends to 50 keV with a photon index of $\Gamma = 1.9$. The normalization of the UV (2500\AA) to X-ray (2 keV) flux is defined by $\alpha_{OX}$, which we take to be 1.37. We consider two models that differ in the mean energy of an ionizing photon (2.65 and 4.17 Ryd), which correspond to different different black hole masses, accretion rates, and spins (see table A1 in \citealt{baron19b}). We then construct a mixing sequence with a young stellar spectrum (2 Myr), considering mixing fractions in the range $\log f_{ion}$ between -2 and 2. 

For each of these SEDs, we estimate the FUV-to-optical luminosity ratio by integrating over the spectra in the ranges 1350--1780 \AA\, (FUV) and 3000--4000 \AA\, (optical). To probe the hardness of the ionizing SED, we estimate $\nu L_{\nu}$ at 353 and 912 \AA, which correspond to photon energies that are required to ionize Oxygen twice (probed by the \oiii~ line) and Hydrogen (probed by the \halpha~ line). Figure \ref{f:SED_models} shows the luminosity ratio $\nu L_{\nu}$(O$^{++}$)/$\nu L_{\nu}$(H$^{+}$), which traces the slope of the ionizing radiation, versus the FUV-to-optical luminosity ratio, which traces the hardness of the FUV-optical part of the SED. The figure includes all the models described in this section. The figure shows that harder ionizing radiation is typically associated with softer FUV-optical slope for a wide range of SEDs, including single stellar populations, mixing of young and old stars, and mixing of stellar and AGN SEDs. The slopes of the gray arrow in figures \ref{f:PAH_optical_interpretation_oiii_hbeta} and \ref{f:PAH_optical_interpretation_sii_halpha} was estimated by combining the observed slopes in figure \ref{f:Draine_models_FUVopt_ratios} and \ref{f:SED_models}.

\begin{figure*}
	\centering
\includegraphics[width=0.9\textwidth]{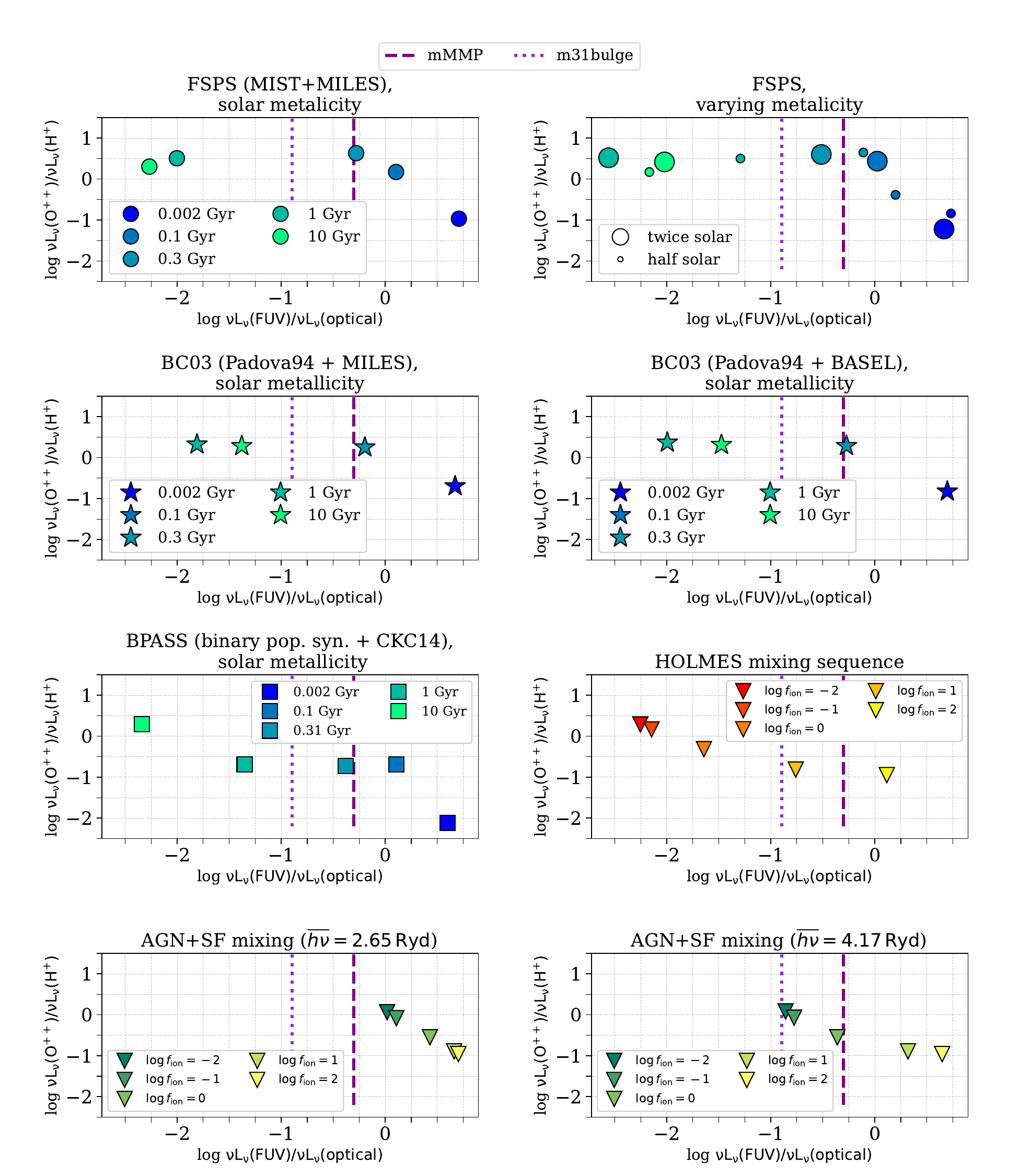}
\caption{$\nu L_{\nu}$(O$^{++}$)/$\nu L_{\nu}$(H$^{+}$) versus the FUV-to-optical luminosity ratio for a range of different assumed SEDs. Each panel shows the ratios for different sets of SEDs, where we examined single stellar populations estimated with different codes, isochrones, and stellar libraries, as well as assuming different metallicities and stellar ages. We also considered SEDs that are a combination of young and old stellar populations (HOLMES mixing sequence), which are used to interpret the optical line ratios seen in the diffuse ionized gas, and SEDs that are a combination of stellar and AGN radiation. The figure shows that harder ionizing radiation (increasing y-axis) is typically associated with softer FUV-optical slope (decreasing x-axis), for a wide range of assumptions of the SED.}
\label{f:SED_models}
\end{figure*}

\end{document}